\documentclass[11pt]{isuthesis}

\usepackage{flushend}
\usepackage{float}
\usepackage{graphicx}
\usepackage{amsfonts}
\usepackage{color}

\usepackage{algorithmic}
\usepackage{algorithm}

\usepackage{multirow}
\usepackage{subfigure}

\usepackage[hyphens]{url}

\makeatletter
\def\url@leostyle{%
  \@ifundefined{selectfont}{\def\UrlFont{\sf}}{\def\UrlFont{\small\ttfamily}}}
\makeatother
\usepackage{amsmath}

\usepackage{isutraditional}   \chaptertitle
\alternate
\usepackage{rotating}

\begin{document}

\title{Dynamic cache reconfiguration based techniques for improving cache energy efficiency}
\author{Sparsh Mittal}
\degree{DOCTOR OF PHILOSOPHY}
\major{Computer Engineering}
\level{doctoral}
\format{dissertation}
\mprof{Zhao Zhang}
\committee{4}
\members{Joseph Zambreno \\ Ahmed Kamal \\ Akhilesh Tyagi \\ David Fernandez-Baca }
\notice
\submit{the graduate faculty}
\maketitle

\chapter*{DEDICATION}

This thesis is dedicated to my teacher Dr. P. V. Krishnan who has motivated me to pursue research and given inspiration to use it for the cause of education.  



\tableofcontents
\addtocontents{toc}{\def\protect\@chapapp{}} \cleardoublepage \phantomsection

\addcontentsline{toc}{chapter}{LIST OF TABLES}
\listoftables

\cleardoublepage \phantomsection \addcontentsline{toc}{chapter}{LIST OF FIGURES}
\listoffigures
\addtocontents{toc}{\def\protect\@chapapp{CHAPTER\ }}

\cleardoublepage \phantomsection
\specialchapt{ACKNOWLEDGEMENTS}

I would like to thank my major professor Dr. Zhao Zhang for his support and guidance throughout my study.  He is always considerate of welfare of his students. He is  professional at work and I deeply value his research expertise. He has given me freedom to pursue the research ideas and develop my research skills. On numerous occasions, he has provided time, support and given constructive inputs and suggestions. I would like to remember my time with him as very memorable and enriching in my life.   
   
   I would like to thank Dr. Akhilesh Tyagi, Dr. Joseph Zambreno, Dr. Ahmed Kamal and Dr. David Fernandez-Baca for their time, discussions and valuable suggestions.
   
I wish to deeply thank my teacher Dr. P. V. Krishnan for his unflinching support and guidance in all aspects of my life. He has extended himself and given immense support especially at difficult times. His guidance has saved me from getting distracted from the goal. He has helped me in realizing the responsibility that comes with education.    
   
   I would like to thank my parents for their moral support and encouragement. My friends and well-wishers have greatly helped me and without their help my work would not have been possible. I would thank Dr. Rangan and Dr. Siddharth for their affection, encouragement and support. I would also like to heartily thank my friends,  Ankit Agrawal, Amit Pande, Venkat Krishnan, Abhisek Mudgal, Sandeep Krishnan and Vikram S. Koundinya (all PhDs) for their tremendous support to me which hardly few students may be fortunate to get. My thanks are also due to Shiva, Ganesh and Srikant.  
      
 I am grateful to God for arranging everything beyond my expectations and capabilities and wish to use these gifts properly for the purpose they are given.      
\addtocontents{toc}{\def\protect\@chapapp{CHAPTER\ }}
\cleardoublepage \phantomsection
\specialchapt{ABSTRACT} 
Modern multicore processors are employing large last-level caches, for example Intel's E7-8800 processor uses 24MB L3 cache. Further, with each CMOS technology generation, leakage energy has been dramatically increasing  and hence, leakage energy is expected to become a major source of energy dissipation, especially in last-level caches (LLCs). The conventional schemes of cache energy saving either aim at saving dynamic energy or are based on properties specific to first-level caches, and thus these schemes have limited utility for last-level caches. Further, several other techniques require offline profiling or per-application tuning and hence are not suitable for product systems.

In this research, we propose novel cache leakage energy saving schemes for single-core and multicore systems; desktop, QoS, real-time and server systems.
We propose  software-controlled, hardware-assisted techniques which use dynamic cache reconfiguration to configure the cache to the most energy efficient configuration while keeping the performance loss bounded. To profile and test a large number of potential configurations, we utilize  low-overhead, micro-architecture components, which can be easily integrated into modern processor chips. We adopt a system-wide approach to save energy to ensure that cache reconfiguration does not increase energy consumption of other components of the processor. We have compared our techniques with the state-of-art techniques and have found that our techniques outperform them in their energy efficiency. This research has important applications in improving energy-efficiency of higher-end embedded, desktop, server processors and multitasking systems. We have also proposed performance estimation approach for efficient design space exploration and have implemented time-sampling based simulation acceleration approach for full-system architectural simulators.



\addtocontents{toc}{\def\protect\@chapapp{CHAPTER\ }}
\pagenumbering{arabic}

\chapter{INTRODUCTION}\label{chap:introduction}
\section{Motivation for Present Research}

Power consumption has now become a primary design constraint for nearly all computer systems and if left un-managed, may lead to end of multicore scaling \cite{esmaeilzadeh2011dark}. In mobile and embedded computing, the amount of power consumed directly affects the battery lifetime. In desktop systems, excessive power has been one of the important reasons for the halt of clock frequency increases and wide-scale adoption
of chip multiprocessors (CMPs) since they allow high-throughput computing within cost-effective power and thermal envelopes. In supercomputers and internet data-centers also, power consumption has been on rise. For example, each of the 10 most powerful supercomputers on the TOP500 List \cite{top500list} require up to 10 megawatts of peak power \cite{feng2007green500}. This amount of power is enough to sustain a city of 40,000.  For this reason, the issue of power consumption drives major design decisions in big companies. 

Among different on-chip components, caches contribute to a large fraction of chip-power consumption. Caches occupy more than 50\% of the total area of the processor \cite{ShiKec03_CacheArea} and their size is increasing to bridge the widening gap between the speed of main memory and processor core. The number of cores on a single chip is continuously increasing; for example, IBM's POWER7 \cite{refIBMPower7}, Intel's E7-8800 Series \cite{refIntel} and AMD’s Opteron 6000 Series \cite{refAMD}  use 8 to 16 cores on a single chip; and future processor chips are expected to have much larger number of cores. To cater to the demands of the large number of cores and to bridge the widening gap between the speed of processor core and DRAM memory, large sized shared LLCs (last level caches) are being used; for example, Intel's E7-8800 processor uses 24MB L3 cache. Further, with each CMOS technology generation, leakage power has been increasing dramatically \cite{semiconductor2011international,rodriguez2006energy}.  Thus, power consumption of caches is increasingly becoming a concern in modern processor design. In this research, we propose algorithms and architectures for saving cache energy in single-core and multicore systems; desktop, QoS, real-time and server systems. 

\section{Limitations of State of The Art Techniques}
Recently, several techniques have been proposed to save cache leakage power. However, these existing techniques have several drawbacks.
\begin{enumerate}
\item Several techniques (e.g. \cite{FlaKim02_DrowsyCache,KaxHuz01_CacheDecay}) have been designed by utilizing the properties which are suited to single-core workloads. However, these techniques fail to scale for multi-core workloads, which are prevalent today. 
\item Several techniques (e.g. \cite{WanMis09_VLSI_ReconfigCache,YanPow01_IcacheResize}) require offline profiling of individual programs and hence cannot be easily used with modern servers, which run trillions of instructions of arbitrary combinations of multicore workloads.  
\item  The hardware-based techniques cannot fully exercise the trade-off between performance and energy efficiency. These techniques may cause severe cache thrashing and therefore may dramatically increase program execution time and power consumption
 in the processor core and DRAM memories. This increase may even offset the leakage power savings in cache. Thus, it is very difficult, if not impractical, for non-adaptive hardware-based schemes of cache energy saving to also take into account the components other than the cache.
 \item Most existing techniques use control mechanisms which depend on arbitrary parameters (e.g miss-bound, decay interval e.g. \cite{KaxHuz01_CacheDecay,YanPow01_IcacheResize}) that must be tuned per application. The presence of large intra-program variations and the differences between the profiled runs and actual programs make the approach of per-application tuning highly ineffective and difficult-to-scale.
\end{enumerate}
 
Thus, there is a need of novel techniques for runtime power management of caches. In this research, we seek to address this issue.  

\section{Research Statement and Approach}
\textbf{The aim of this research is to develop architectures and efficient algorithms for enabling energy-efficient operation of cache hierarchies of both single-core and multi-core systems and both single-tasking and multi-tasking systems. This research proposes specific techniques to fulfill the needs of  QoS, real-time, desktop and server systems.} 
  
In this research, we propose novel cache energy saving schemes. We present software-controlled, hardware-assisted techniques which use dynamic cache reconfiguration to configure the cache to the most energy efficient configuration. To profile and test a large number of potential configurations, we utilize  low-overhead, micro-architecture components, which can be easily integrated into modern processor chips. We focus on a system-wide approach to save energy, while keeping the performance loss bounded.

The key idea in our techniques is as follows. Since programs show large intra- and inter- program variations in their cache requirements, processor designers have to use cache with average case in mind. This, however, leads to a large wastage of energy (in the form of leakage energy) for the applications with small working set size (WSS), or cache thrashing for the applications with large WSS. Hence, at any time, by allocating an appropriate amount of LLC space to the application so that its working set can fit, the rest of the L2 cache can be turned off with little impact on performance. Thus, we employ intelligent cache reconfiguration to turnoff the parts of the cache to save large amount of energy, such that the execution time of the program is minimally affected. 
 
 We propose a low-overhead, \textbf{multi-level profiling cache}, which can profile multiple cache configurations (e.g. 32 configurations) of LLC, which include multiple number of cache ways and sets (Section \ref{sec:profdesign}). Experimental results have shown that the multi-level profiling cache produces highly accurate estimates, with an average error of 0.26 MPKI (miss-per-Kilo-instruction) in predicting the cache miss rates for 100 combinations of applications/configurations (Section \ref{sec:profvalidation}). This is extremely useful for estimating program performance for the purpose of design space exploration (Chapter \ref{chap:esto}, \cite{Mit2013_DesignContest}) and cache energy saving (Chapter \ref{chap:encache}, \cite{MitZha12_EnCache}).

For further improving the granularity of configurations profiled using multi-level profiling cache, we have proposed a \textbf{reconfigurable cache emulator} (RCE), which allows profiling at fine reconfiguration granularity (e.g. $1/128$ of the original cache size) (Section \ref{sec:palette_rce}). For multicore processors, we employ RCE to individually monitor the cache demand of each processor (Section \ref{sec:reconfigurablecacheemulator} and \ref{sec:manager_rce}). Using this information, cache can be intelligently partitioned between multiple cores and the rest of the cache can be turned off for saving cache energy with little effect on the performance (Chapter \ref{chap:palette}, \ref{chap:master}, \ref{chap:manager}, \cite{MittalPalettePaper2013}). Further, we have proposed cache reconfiguration based techniques for real-time and QoS systems (Chapter \ref{chap:cashier} and \ref{chap:manager}). For a comparison and overview of the techniques proposed in this thesis, please see Chapters \ref{chap:contributions} and \ref{chap:conclusionfuturework}.

Apart from cache reconfiguration, we also propose approaches for accelerating full-system simulation (Chapter \ref{chap:simulation}). Simulation is a vital approach for validating proposed techniques and gaining insights into the working of them. Currently, the extremely slow simulation speed of full-system simulators remains a critical bottleneck restricting  their widespread use. Although several simulation acceleration
techniques have been proposed, they have generally been limited to only few simulators or platforms. In our research, we propose integrating sampling-based simulation acceleration technique into full-system simulator \cite{MitZha12_Simulation}. Our integration approach enables the researchers to fully utilize the potential of full-system simulator and also validates the simulation acceleration technique over another platform. Results have shown that our approach leads to an average speed-up of 28$\times$ (geometric mean) over detailed full-system simulation; with an average error of only 0.73\% in estimating CPI (cycle per instruction).


%
%



\chapter{CONTRIBUTIONS OF THE WORK}\label{chap:contributions}
The contributions of our work are as follows. 
\begin{enumerate}

\item We have proposed a low-overhead, \textbf{multi-level profiling cache}, which can profile multiple cache configurations (e.g. 32 configurations) of LLC, which include multiple number of cache ways and sets (\cite{MitZha12_EnCache}, Section \ref{sec:profdesign}). For further improving the granularity of configurations profiled using multi-level profiling cache, we have proposed a \textbf{reconfigurable cache emulator} (RCE), which allows profiling cache at fine granularity (e.g. $1/128$), and is hence very useful for multicore caches.     

 
\item We have proposed ESTO, a simulation-based approach for estimating application performance (execution time and energy) under multiple last level cache (LLC) configurations (\cite{Mit2013_DesignContest}, Chapter \ref{chap:esto}). ESTO uses multi-level profiling cache which provides low-cost and non-intrusive dynamic profiling. A unique feature of ESTO is its ability to estimate performance of a cache of higher size than the baseline cache present. Experiments performed using a state-of-art simulator and benchmarks from SPEC2006 suite have shown that using ESTO, the average error in estimating execution time and memory subsystem energy are only 3.7\% and 3.3\%, respectively.
 
\item We have presented \textbf{EnCache} (Energy saving approach for Caches), a software-based approach on top of lightweight hardware support (\cite{MitZha12_EnCache}, Chapter \ref{chap:encache}).  We have compared EnCache with a well-known leakage energy saving technique, named Hybrid Dynamic Cache Resizing and have found that, EnCache outperforms the HDRI technique. For example, for a 2MB L2 cache, the average saving in EDP (energy delay product) by using EnCache and a highly optimized version of HDRI were 28.8\% and 20.6\% respectively.

\item We have presented \textbf{Palette}, a cache energy saving technique using cache coloring method. This work has been accepted \cite{MittalPalettePaper2013} and discussed in Chapter \ref{chap:palette}.    Palette uses dynamic profiling and does not require offline profiling. By virtue of using cache coloring, Palette provides fine grain cache reconfiguration. Simulations performed with SPEC2006 benchmarks show the superiority of Palette over a well-known technique, named DCT (decay cache technique). With a 2MB baseline cache, the average saving in memory sub-system energy and EDP are 31.7\% and 29.5\%, respectively.  In contrast, DCT provides only 21.3\% saving in energy and 10.9\% saving in EDP.

\item We have presented \textbf{CASHIER}, a \underline{Ca}che energy \underline{s}aving tec\underline{h}n\underline{i}qu\underline{e} for quality-of-se\underline{r}vice (QoS) systems (\cite{MitZha13_Cashier}, Chapter \ref{chap:cashier}). This technique is also useful for real-time systems. For example, for 2MB L2 cache with 5\% allowed performance slack, the average saving in memory subsystem energy using CASHIER is 23.6\%.
 
\item We have presented \textbf{MASTER}, an RCE based approach to save energy in multicore server systems (Chapter \ref{chap:master}). MASTER outperforms DCT and WAC (way-adaptable cache technique). For 2 and 4-core simulations,
the average savings in memory subsystem (which includes LLC and main memory)
energy  over shared baseline LLC are 15\% and 11\%, respectively. Also,
the average values of weighted speedup and fair speedup are close to one ($\ge$0.98).


\item We have presented \textbf{MANAGER}, a \underline{m}ulticore shared c\underline{a}che e\underline{n}ergy s\underline{a}vin\underline{g} techniqu\underline{e} for quality-of-se\underline{r}vice systems (Chapter \ref{chap:manager}). Using dynamic profiling, MANAGER periodically predicts cache access activity for different configurations. Then, cache is partitioned among running programs to fulfill the QoS requirement while saving memory subsystem (LLC+ DRAM) energy. Out-of-order simulations performed using dual-core workloads from SPEC2006 suite show that for 4MB LLC, MANAGER saves 13.5\% memory subsystem energy, over  a statically, equally-partitioned baseline cache.
  
\item We have demonstrated integration of SMARTS sampling-based simulation acceleration technique \cite{smarts} into GEMS full-system simulator \cite{gems05}. This work has been accepted (see \cite{MitZha12_Simulation}) and discussed in Chapter \ref{chap:simulation}.  Our integration approach enables the researchers
to fully utilize the potential of full-system simulator and also
validates the simulation acceleration technique over another
platform. The experiments performed over benchmarks from
SPEC2K show that using our approach leads to an average
speed-up of 28$\times$ (geometric mean) over detailed full-system
simulation; with an average error of only 0.73\% in estimating CPI (cycle per instruction).

\end{enumerate}

Our research will improve power efficiency of cache hierarchies in higher-end embedded, desktop and server processors. The algorithms proposed in this research will enable low power operation of QoS and real-time systems. Further, by virtue of using dynamic profiling, the techniques proposed here will benefit multitasking systems also.   


\chapter{ESTO: A PERFORMANCE ESTIMATION APPROACH FOR EFFICIENT DESIGN SPACE EXPLORATION}\label{chap:esto}

\section{Introduction}

Recent advancements in the field of processor architecture and chip design have opened new horizons for both architects and end-users. While these architectures promise high performance, they also pose significant challenges to the designers, due to the increasing number of design options (e.g. cache configurations) and design constraints (e.g. energy). Further, to bridge the widening gap between DRAM speed and processor speed, modern processors are using increasingly large LLCs and hence, LLCs have a significant influence on their performance. Over several years of CPU evolution, the size of L1 cache has stayed at 16KB or 32KB, while the size of the LLC has grown from nearly 256KB to 1, 2 or 4 MB in modern day processors, with future processors expected to have even larger LLC sizes. Hence, while translating a design from concept phase to a working chip, a designer must choose a suitable LLC size, based on the application requirements and also meet the constraints posed by chip power budget and real-life timing requirements. Proper choice of architectural parameters is crucial for meeting the needs of several data-critical applications \cite{MitGup08_BioinQA}. For this purpose, designers generally use detailed simulators for evaluating different design options, however, the high simulation time of these simulators makes it infeasible to use them for testing all possible configurations in the design space. This forces the designers to take decisions without considering all the design constraints or fully exploring the design space.

To address this challenge, several techniques have been proposed for performance estimation and fast design space exploration. However, existing techniques of performance estimation have several drawbacks. Superscalar out-of-order processors use speculative execution and hence, the possible overlap between execution and different miss events such as cache misses and branch mispredictions etc. make it challenging to estimate performance under multiple design options.  For this reason, several techniques use simplistic platforms or require offline profiling or multiple runs (e.g. \cite{puzak2007pipeline,chou2004microarchitecture}) and hence these techniques are difficult to scale to real-world processors and applications, which execute trillions of instructions. Many performance estimation techniques use intrusive methods which have a large space/time overhead. A few other techniques have a large error of estimation and hence, the conclusions derived from them could be very misleading. Thus, an efficient and accurate performance estimation method is required for design space exploration and making crucial design decisions. 

 In this chapter, we present ESTO, a dynamic profiling based technique for estimating the performance of an application program under a range of possible last level cache (LLC) sizes\footnote{In this chapter, we use the term performance to refer to execution time (ET) and energy consumption together.}. The key idea behind our approach is the use of a small \textit{profiling cache}, to estimate the number of LLC misses under different cache configurations and to compute their effect on program performance. Profiling cache is a data-less cache which is based on the idea of set sampling \cite{MitZha12_EnCache} and has an energy overhead of less than 1\% of that L2 cache. ESTO uses memory stall cycle model to take into account the possible overlap between different miss events and thus ESTO can be used in out-of-order processors with speculative execution support.

For a system with L2 cache size of $X$, we define any cache with size $\leq$$X$ as \textit{sub}-sized cache and any cache of size $\geq$$X$ as \textit{super}-sized cache. A unique feature of ESTO is its capability to estimate execution time and energy of both super-sized caches and sub-sized caches. Thus, for example, using a 4MB L2 cache, a designer can estimate the performance of 8MB L2, as well as 2MB, 1MB, 512KB and 256KB caches. This feature is extremely useful for making projections about a future configuration which may be presently unavailable.  Thus, ESTO helps a designer in choosing most suitable LLC configuration and fulfill the design constraints.

ESTO addresses several limitations of the existing approaches. Firstly, ESTO uses non-intrusive dynamic profiling and hence, does not require any changes to application source code or binaries. The profiling cache works in parallel with L2 and hence does not affect the access latency of the L2 cache. ESTO provides online estimates of performance and does not require offline profiling or any separate runs.  To evaluate ESTO, simulations were performed using Sniper \cite{CarHei2011_Sniper},  simulator and benchmark programs from SPEC2006 suite. Across 80 combinations of benchmarks and configurations, the average error in execution time (ET) estimation is 3.7\%. Further, the average error in memory subsystem energy (L2 cache+ main memory energy) is 3.3\%.  These results confirm the effectiveness of ESTO. 

As computer systems are becoming increasingly power constrained, workload optimized system design is expected to become even more prominent, as seen through example of Intel’s Many Integrated Core (MIC) architecture and IBM's BlueGene processor. Hence, our approach is likely to become even more important in the design of future computer chips. Profiling cache can be easily used for saving cache energy, thus helping the designers in realizing the goals of sustainable and green IT.

 The rest of the chapter is organized as follows. Section \ref{sec:esto_motivation} and \ref{sec:esto_relatedwork} present the motivation and scope of the work and the related work. Section \ref{sec:esto_methodology} discusses the ESTO methodology and Section \ref{sec:overheadesto} computes the overhead of ESTO. Section \ref{sec:experimentalplatform} and \ref{sec:esto_results} discuss the experimental platforms and presents results. Finally, Section \ref{sec:esto_conclusion} presents the conclusion and future work.

\section{Motivation and Scope of The Work}\label{sec:esto_motivation}
 
 We present the motivation for using ESTO with a typical design scenario. Modern portable devices such as personal digital assistants, phones, laptops and iPODs etc are powered by the battery which supplies limited energy. Thus, the amount of battery dissipation which is induced by program execution becomes an important factor in assessing battery life and gives valuable information to take decision about recharging or replacement. This is especially important in situations such as traveling in flight etc. To address such needs, ESTO enables an architect to use a suitable cache size, taking into account the energy budget, usage scenario and quality of service (QoS) requirement. For example, if a certain delay in response is acceptable, the architect can use a smaller sized cache if that is more energy efficient. Similarly, within a same energy budget, an architect can use a larger sized cache if that is more performance efficient.

Our objective in this chapter is to propose and experiment with the methods which enable exact program execution time estimation for a given input and hardware for different configurations of L2 cache sizes.  The WCET analysis approach is different from our work. Worst-Case Execution Time (WCET) prediction approach seeks to estimate the upper bound of the program execution time under different program inputs or hardware platforms or system resources. Given the large number of possible inputs, only a range or bound is estimated for WCET. Moreover, such analysis has been done by assuming simplified/idealized platforms (e.g. perfect processor pipeline with no stalls \cite{LixMit03_AccurateET} etc). In contrast, we estimate exact execution time, using a detailed out-of-order superscalar processor which presents challenges of its own.

 
\section{Related Work}\label{sec:esto_relatedwork}
    
Recently, several methods have been proposed for estimating cache miss rate, execution time and energy of a program. In the following, we review them briefly.
 
\textbf{Miss Rate Estimation: } Tam et al. \cite{TamAzi09_RapidMRC} present a software based L2 miss rate prediction approach. This technique works by recording data addresses of memory accesses to a data address register and later feeding the log of addresses  to an LRU stack simulator to  generate the miss rate curve (MRC) using the Mattson stack algorithm. This technique only takes into account L1 data cache misses and does not take into account  L1 instruction cache misses and L1 data write-backs. This, however, leads to loss of accuracy and hence the miss rate curve generated using this approach need to be vertically shifted to better match real MRC. Moreover, this approach only works for fully-associative caches, while the modern processors use set-associative caches with finite (e.g. 8 or 16 way) associativity.

Qureshi et al. \cite{QurPat06_UtilityBasedCP} propose Utility Monitors (UMONs) for tracking miss rate of L2 caches for different ways of an LRU cache, using Mattson stack algorithm.  However, due to the high cost of implementation of true-LRU technique, most real-world processors use an approximation of LRU (e.g. pseudo-LRU ). Hence, true-LRU based miss rate prediction approaches are not suitable for real-world processors. In contrast, ESTO uses set-based profiling, and hence, it can easily work with different ``approximate-LRU'' replacement policies.
    
\textbf{Execution Time Estimation: } Techniques for estimation of execution time is especially important for high-performance computing applications. 
Yamamoto et al. \cite{YamIsh06_ExecTimeAnalysis} propose an execution time prediction method which combines measurement-based execution time analysis and simulation-based memory access analysis. As for memory access analysis, the memory access latency value is estimated in terms of the memory access pattern of a function level and the properties of the target processor cache architecture.  However, the authors observe an error up to 64\% in ET estimation on Pentium-M processor. 

 Most methods of computation of L2 cache latency require running the program twice (e.g \cite{puzak2007pipeline,chou2004microarchitecture}). Once the program is run, with the assumption of infinite cache and then with finite (real) cache. This method, however, introduces large  overhead and is not suitable for real-time applications.

 Because of their dynamic behaviors, caches present several challenges in WCET analysis. Several studies have focused on addressing this issue. Li et al.  \cite{LiyMal96_CacheModeling} build an Integer Linear Programming solution for WCET estimation problem for direct mapped and set-associative caches, while Ferdinand et al. \cite{FerMar99_AbsInt} use abstract interpretation to model the instruction cache behavior for WCET analysis.

\textbf{Energy Estimation: }
Dhouib et al. \cite{EstimateEnergy_Embedded} propose a multi-layer power and energy estimation approach for embedded systems. Their approach works by first estimating energy and power consumption of standalone tasks and then adding energy overheads of operating system services such as timer interrupt, inter process communications etc. Zhao et al. \cite{zhao2008fine} present a microarchitectural approach to estimate the energy consumption of embedded operating systems by taking into account the energy spent in system calls and kernel execution paths etc. Our approach is different from these, since we estimate memory sub-system energy under many configurations in a single run.

\begin{figure*}[htp]
 \centering
  \includegraphics [scale=0.30] {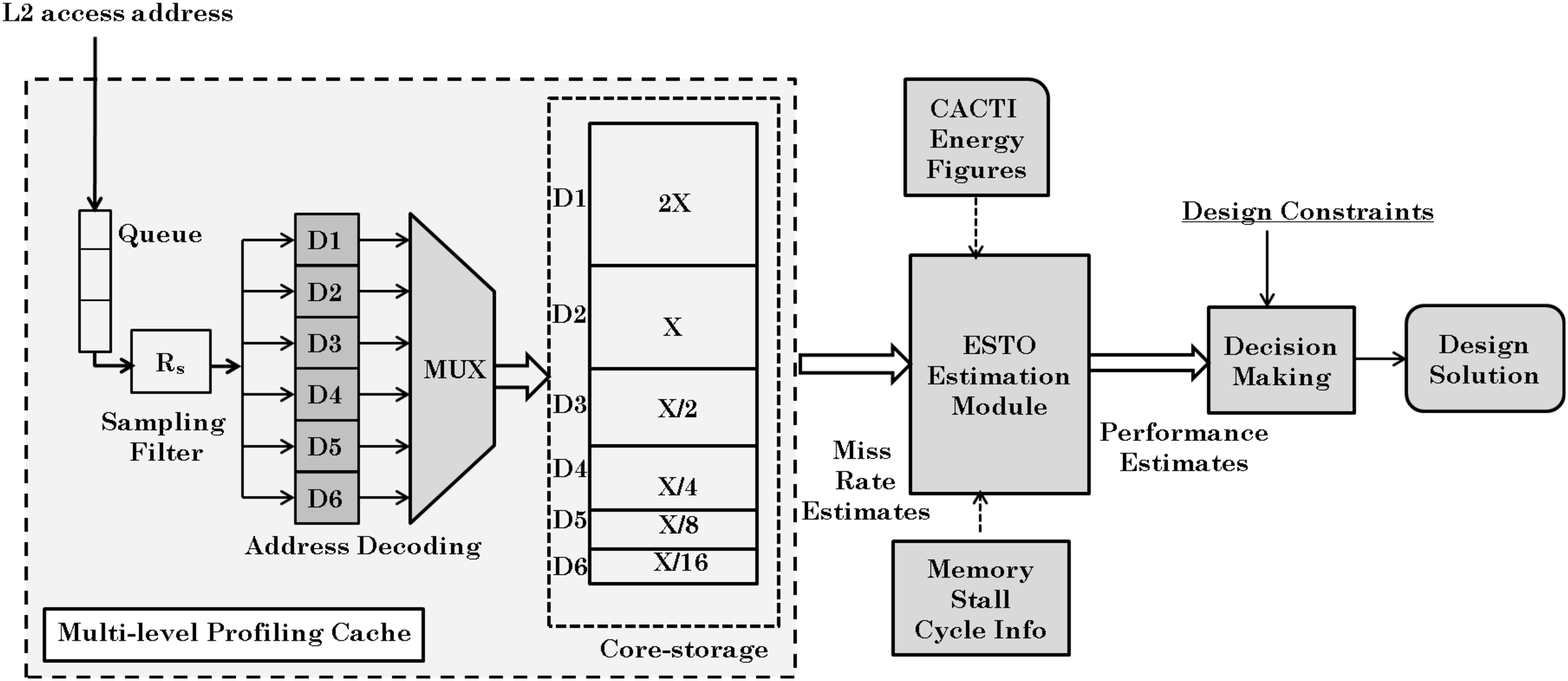}
 \isucaption{ESTO flow diagram }\label{fig:esto_flowdiagram}
 \end{figure*}

\section{Methodology}\label{sec:esto_methodology}

It is well-known that under different cache configurations, program applications show different number of cache misses, and hence different performance. Hence, to estimate the impact of multiple L2 cache configurations on performance, ESTO uses \textit{profiling cache} to predict L2 misses under those configurations (Section \ref{sec:profilingcache}). Using these estimates, along with CPI stack model, ESTO estimates execution time of the application under those configurations (Section \ref{sec:executiontimeestimation}). Finally, using these estimates, ESTO estimates both dynamic and leakage energy component of memory subsystem energy (Section \ref{sec:energyestimation}). Based on these estimates and the domain knowledge of design constraints, a designer can take suitable design decisions. Figure \ref{fig:esto_flowdiagram} shows the overall flow diagram of ESTO. In what follows, we explain each of these components in detail.

\subsection{Profiling cache} \label{sec:profilingcache}
Profiling cache is a small, dataless (tag only) cache, which is designed based on the well-known set sampling technique \cite{MitZha13_Cashier}, which states that the miss rate characteristics of a set associative cache can be estimated by sampling only a few of its sets. The ratio of set count of L2 and that of a profiling cache is termed as sampling ratio ($R_s$). Profiling cache emulates L2 and thus, has same associativity, block size and replacement policy as L2. On an access to profiling cache, a hit or miss is decided and corresponding counters are updated. Note that, it does not store or communicate data and hence does not generate traffic. On a miss, the tag of missed address is copied and the victim is evicted. Thus, profiling cache is decoupled from L2 cache and as shown in Section \ref{sec:overheadesto}, the size of this `single level' profiling cache is only 0.10\% of L2 cache size. 


We use the above mentioned properties to extend profiling cache, such that it profiles multiple cache sizes in parallel; each size is referred to as a level. For our experiments, we choose six levels, each level profiling a cache of size 2X, 1X, X/2, X/4, X/8, X/16 respectively. These levels, also referred to as configurations, are, in general, shown as $C$ and the baseline (1X) configuration is shown as $C^{\star}$. Also, note the unique capability of profiling cache: because of its decoupled operation with L2, it can also profile a cache of 2X size (double the baseline cache size) with reasonable accuracy, as we will see in the results section (Section \ref{sec:esto_results}). This feature is an important improvement over previous works based on profiling and it allows a designer to estimate program performance for a cache size which may be currently unavailable.
 
As shown in Section \ref{sec:overheadesto}, even with this extension, the size of multilevel profiling cache is only 0.40\% of L2 cache size. Thus, the multilevel profiling cache has a small size and access latency and since it does not lie on the critical access path, its latency is easily hidden. In what follows, we use the word profiling cache to refer to a multilevel profiling cache, unless otherwise mentioned.

The profiling cache works as follows (ref. Fig. \ref{fig:esto_flowdiagram}). The L2 access addresses are passed through a small queue and then sampled using a sampling filter. Then these sampled addresses are passed through address decoding region for calculating the set (index) and tag values. Then these addresses are sent to the core storage component through a multiplexer (MUX). We mention that even though profiling cache is accessed multiple times for each sampled address, the presence of the queue and use of a large sampling ratio avoids the possibility of any congestion.


 \subsection{ Execution time Estimation} \label{sec:executiontimeestimation}
 For estimating both execution time and leakage energy under different cache configurations, we need to estimate memory stall cycles under those configurations as a function of L2 misses. However, modern out-of-order processors use several features for hiding latency (e.g. overlap between miss events such as branch misprediction and L2 miss), and hence the memory stall cycles cannot be computed as a linear function of the number of L2 misses. 

To address this issue, ESTO uses a well known technique, called CPI stack model \cite{CarHei2011_Sniper}. CPI stack shows the contribution of base execution along with different miss events, (such as branch mispredictions, cache misses) in the overall CPI of the program.   For example, in any interval $i$, the memory stall cycle component of CPI stack (termed as StallCPI$_i(C^{\star})$ ) shows the net contribution of memory stall cycles on overall cycles, \textit{after} taking into account the overlap with other miss events.
  Let LoadMisses$_i(C^{\star})$ show the number of load misses in interval $i$. Now, since memory stall cycles are primarily due to L2 load misses \cite{MitZha12_EnCache}, we define K$_i(C^{\star})$ as follows.
 \begin{equation}
\text{K}_i(C^{\star}) = \dfrac{\text{StallCPI}_i(C^{\star})} {\text{LoadMisses}_i(C^{\star})}
\end{equation}

Here K$_i(C^{\star})$ shows memory stall CPI per load miss. We assume that K$_i(C^{\star})$ value is independent of the number of load misses and hence remains same for different cache configurations, thus K$_i$=K$_i(C^{\star})$, for all configurations. Further, we also use extra counters in profiling cache to record load misses, along with total misses, for different L2 configurations. Then, StallCPI$_i(C)$ for any configuration ($C$) can be computed, using  
  \begin{equation}
\text{StallCPI}_i(C) = \text{K}_i \times \text{LoadMisses}_i(C)
\end{equation}
    
Then, using StallCPI$_i(C)$ and other components of CPI stack, total CPI value at any configuration $C$ can be computed. Using total CPI, along with given frequency value and number of instructions, execution time under $C$ can be easily estimated.

 \subsection{Energy Estimation}\label{sec:energyestimation}
 We now discuss the energy model used in ESTO and also show the procedure for estimating program energy value under any configuration using the estimates of miss rates and execution time. Since other components of processor are minimally affected by change in L2 cache size, we only consider memory subsystem energy, which is given as the sum of L2 and memory energy. 
\begin{equation}\label{eq:esto_Etot}Energy= E_{L2}+E_{mem}\end{equation}

 We use the symbols $E^{dyn}_{L2}$ and $P^{leak}_{L2}$ to show the dynamic energy \emph{per access} and leakage energy \emph{per second}, respectively, consumed in L2 cache. For memory, these parameters are shown by $E^{dyn}_{mem}$ and  $P^{leak}_{mem}$ respectively.  

To calculate L2 energy, we assume that an L2 miss consumes twice the energy as that of an L2 hit \cite{MitZha13_Cashier}.  Thus,
\begin{equation} \label{eq:cacheenergy}E_{L2}= E^{dyn}_{L2}\times(2M_{L2}+H_{L2}) + P^{leak}_{L2}\times Time\end{equation}
 
Here, for any configuration, we have corresponding $M_{L2}$=L2 misses, $H_{L2}$=L2 hits, $Time$=execution time. The L2 energy values are obtained using CACTI 5.3 (http://quid.hpl.hp.com:9081/cacti/) for 4 bank, 8-way caches with 64 byte block size at 45nm. These values are shown in Table \ref{tab:esto_energyvalues}.
\begin{table}[htbp]  \small
  \centering
  \isucaption{L2 cache Energy Values}
    \begin{tabular}{|r|r|r|}
    \hline
   & $E^{dyn}_{L2} (nJ/access)$& $P^{leak}_{L2}(Watt)$ \\\hline
    
    8MB       & 1.525          & 5.588      \\ \hline
    4MB       & 1.148          & 2.848      \\\hline
    2MB       & 0.985          & 1.568      \\\hline
    1MB       & 0.912          & 0.966      \\\hline
    512KB     & 0.872          & 0.664      \\\hline
    256KB     & 0.848          & 0.500      \\\hline
    
    \end{tabular}
    \label{tab:esto_energyvalues}
\end{table}

To calculate memory energy, we note that $ P^{leak}_{mem}$=0.18 Watts and $ E^{dyn}_{mem}$=70 nJ \cite{MitZha12_EnCache}. Thus, we get,
\begin{equation} \label{eq:memenergy} E_{mem}= E^{dyn}_{mem}\times A_{mem}+ P^{leak}_{mem}\times Time \end{equation}
where $A_{mem}$ denotes the number of memory accesses. From Eq. \ref{eq:cacheenergy} and \ref{eq:memenergy}, it is clear that using miss rate and execution time estimates, program energy under any configuration can be estimated.

\section{Overhead of ESTO}\label{sec:overheadesto}
ESTO uses profiling cache and computations for performance estimation, and hence the overhead of ESTO comes from these two components. ESTO does computations for ET and energy only at the end of a large interval length (e.g. 5M instructions). Thus, the cost of these calculations is amortized over interval length. In remainder of this section, we first compute the size of single level and multilevel profiling cache and then compute the energy consumption of multilevel profiling cache, to show that the overhead of ESTO is extremely small. We use the subscripts $Single$ and $Multi$ to represent any quantity (e.g. size) for single level and multilevel profiling cache respectively.   
  
For a $W$ way L2 cache having $Q$ sets, $B$ byte cache block and $G$ bit tag, the total cache size in bits is
\begin{equation}
\text{Size}_{L2} = Q\times W\times (B\times 8 + G)
\end{equation}
Since profiling cache is a dataless cache, its size is
\begin{equation}
\text{Size}_{Single} = \dfrac{Q}{R_s}\times W\times G
\end{equation}
If $\Theta_{Single}$ shows the size of single level profiling cache as a percentage of L2 size, we get
\begin{equation}
\Theta_{Single} = \dfrac{G}{R_s(G+B\times 8)}\times 100
\end{equation}
For $R_s$=64, $B$=64 and $G$=36 we get $\Theta_{Single}$=0.10\%.

For computing size of multilevel profiling cache, we first compute the number of sets ($\text{Sets}_{Multi}$) in it, as follows. 
\begin{equation}
\text{Sets}_{Multi} = \dfrac{2Q}{R_s}+\dfrac{Q}{R_s}+\dfrac{Q}{2R_s}+\dfrac{Q}{4R_s}+\dfrac{Q}{8R_s}+\dfrac{Q}{16R_s}
\end{equation}
\begin{equation}\label{eq:multilevel}
\text{Sets}_{Multi} = \dfrac{63Q}{16R_s} < \dfrac{4Q}{R_s}
\end{equation}
Using above equations, we compute the size of multilevel profiling cache as a percentage of L2 size ($\Theta_{Multi}$) as follows.
\begin{equation}
\Theta_{Multi} = \dfrac{4G}{R_s(G+B\times 8)}\times 100
\end{equation}

Thus, for $R_s$=64, $B$=64 and $G$=36 we get $\Theta_{Multi}$=0.40\%.  To cross-check, we have computed the area of L2 and multilevel profiling cache using CACTI, for the cache sizes used in our experiments (Section \ref{sec:experimentalplatform}). Since multilevel profiling cache is a tag only structure, we take 8B block size, which is smallest allowed block size in CACTI and only take the area values for tag arrays. From these values, we compute  $\Theta_{Multi}$  and find that $\Theta_{Multi}$=0.29\%, which is in the same range as that obtained above.   

To compute the energy values for (multilevel) profiling cache, we take $R_s$=64 and use CACTI 5.3. As explained above, we only take the energy figures for tag arrays. For a profiling cache corresponding to a baseline L2 of 2MB, we get the energy values as $E^{dyn}_{Multi}$= 0.004 nJ/access and $P^{leak}_{Multi}$=0.007 Watt. Noting that, profiling cache is accessed only 6 times for every 64 L2 accesses, we find that profiling cache energy consumption is a very small fraction of L2 cache energy consumption. Thus the overhead of ESTO is indeed very small. Moreover, by taking large value of sampling ratio (e.g. $R_s$=128), the overhead of ESTO can be even further reduced.

\section{Experimental Platform }\label{sec:experimentalplatform}
For evaluating ESTO, we have used Sniper \cite{CarHei2011_Sniper},  which has been validated against the real hardware. We model 4-way processor with 1GHz frequency. L1I and L1D are 32KB, 4-way caches with 4 cycle latency. L2 is 4MB, 8-way cache with 12 cycle latency. All caches use LRU and 64B block size. Memory has 90 cycle latency, 6GB/s peak bandwidth and memory request queue is also modeled. The performance estimates are collected after every 5M instructions. Our workload consists of 16 benchmark programs from SPEC2006 (astar, bwaves, cactusADM, gamess, gemsFDTD, gobmk, h264ref, hmmer, lbm, leslie, libquantum, mcf, perlbench, sjeng, sphinx and tonto), which represent a wide range of cache usage characteristics. Each benchmark program was fast forwarded for 10B instructions and then simulated for 100M instructions.

\section{Results} \label{sec:esto_results}
In this section, we present the results on accuracy of estimation of program execution time and memory subsystem energy.  Further, to be strict in evaluation, we compare execution time and energy values only for cache sizes other than 1X, since, for 1X size (i.e. baseline), these values are easily predicted with high accuracy.  ESTO provides performance and energy estimates for five cache sizes (other than baseline) and with 4 MB cache as baseline, these caches have the size of 8MB, 2MB, 1MB, 512KB and 256KB (4MB itself is baseline and is skipped). Hence, using 4MB cache, a single run was performed for each benchmark, and performance estimates were obtained using ESTO. These estimates were compared with the corresponding actual values obtained using 8MB, 2MB, 1MB, 512KB and 256KB caches and percentage errors were computed with respect to baseline values.

\begin{figure*}[htbp]
  \begin{center}
     \subfigure{\label{fig:time}\includegraphics[scale=0.53]{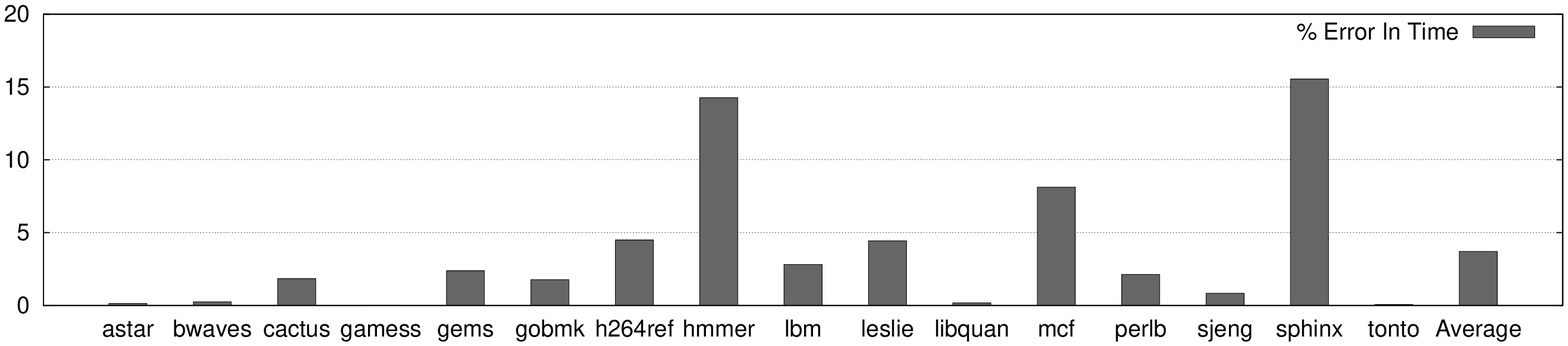}}
     \subfigure{\label{fig:energy}\includegraphics[scale=0.53]{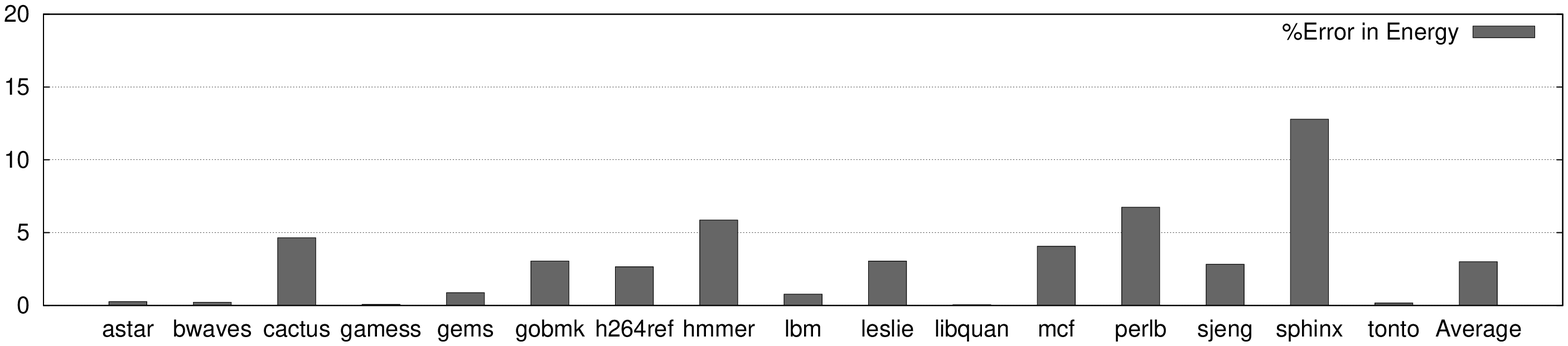}}
  \end{center}
\isucaption{Percentage Error in Execution Time and Energy Estimation} \label{fig:esto_results}

\end{figure*}

Figure \ref{fig:esto_results} shows the average error for each benchmark, across all cache sizes.  Across all benchmark/configuration combinations, the average errors in execution time estimates and energy estimates are 3.7\% and 3.3\% respectively.

\begin{figure}[htp]
   \centering     
     \subfigure{\label{fig:time64Size}\includegraphics[scale=0.72]{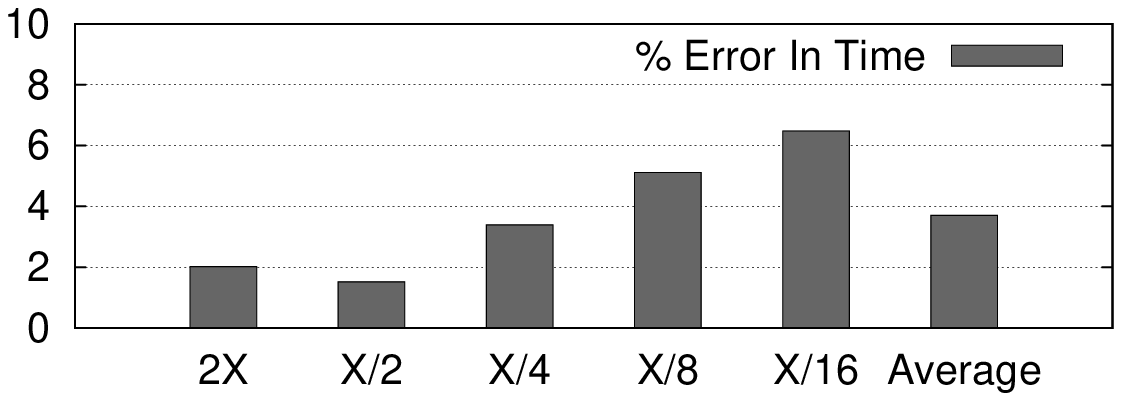}}
     \subfigure{\label{fig:energy64Size}\includegraphics[scale=0.72]{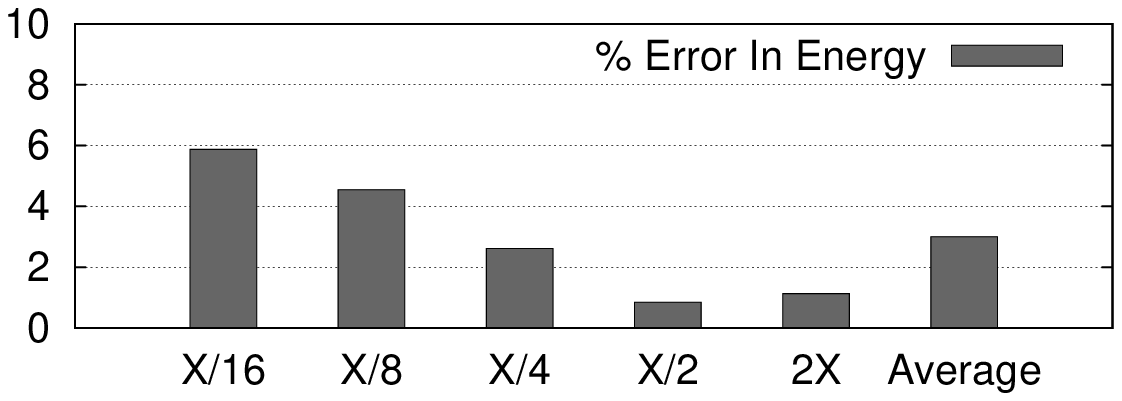}}
\isucaption{Percentage Error in Execution Time and Energy Estimation} \label{fig:esto_results64Size}
\end{figure}

Figure \ref{fig:esto_results64Size} presents the same result; this time for each cache size, across all benchmarks. Clearly, for 2X (8MB) and X/2 (2MB), the accuracy is the highest, which decreases gradually as we move to cache sizes farther from 1X.

We have also tested ESTO for  sampling ratio value of 128 and observed that ESTO still provides high estimation accuracy. Further, for approximate LRU schemes, such as round-robin replacement policy also, ESTO provides high accuracy, which implies that ESTO does not require implementation of true-LRU policy.

We have shown the effectiveness of ESTO in execution time and energy estimation. ESTO can be easily extended to estimate total system energy by simply including the energy model of processor core in total energy equations. Also, since ESTO predicts both energy and execution time; using these estimates the energy delay product (EDP) of the program can also be estimated, although with higher error.

%
%
%

\section{Conclusion}\label{sec:esto_conclusion}

In this chapter, we presented ESTO, a dynamic profiling based approach for estimating application performance and energy consumption under different LLC configurations. We have shown the utility of ESTO for the case when the LLC is an L2 cache, although our approach can can also be applied to an L3 cache. Our future work will focus on making more accurate prediction of impact of cache miss on execution time. This will improve the accuracy of execution time and energy estimation.


\chapter{EnCache: A  CACHE ENERGY SAVING APPROACH FOR DESKTOP SYSTEMS} \label{chap:encache}
\section{Introduction}

In this chapter, we present \textbf{EnCache} (Energy saving approach for Caches), a new software-based approach on top of lightweight
hardware support.  The key component of the
hardware support is a simple \textbf{profiling cache}. It is tag-only cache and uses set-sampling to predict cache miss rates of multiple cache configurations of much larger sizes in an {\em online} manner. It works non-intrusively and due to its decoupled and parallel operation and small-size, its latency is easily hidden. Profiling cache is not a part of the cache hierarchy and it does not lie at the critical access path of the cache.

The previous approaches such as \cite{QurPat06_UtilityBasedCP} utilize sampling only to profile different associativities of the current size of the cache, while EnCache provisions a separate cache structure which can profile different associativities at different cache sizes. Thus, EnCache considerably expands upon the potential of sampling. This is a significant difference, which enables the prediction of energy efficiency of multiple cache sizes and thus, guide reconfiguration. Profiling cache has an energy overhead of less than 0.5\% of L2 cache energy. Our simulation results show that a profiling cache is highly accurate, with an average error of 0.26MPKI (miss-per-Kilo-instruction) in predicting the cache miss rates for 100 benchmark/configuration combinations.

Our profiling cache is designed to also estimate the impact of cache miss-rates on performance, in terms of memory stall cycles. Using these estimates and other performance counters, an OS component periodically predicts the memory-subsystem (which includes LLC and main memory) energy for multiple cache configurations. Then, the cache configuration with the minimum estimated energy is chosen for the next interval and, if necessary, the cache is reconfigured to that configuration.

EnCache addresses the aforementioned shortcomings of the hardware-based
approaches. It optimizes for memory-subsystem energy rather than merely cache energy. It optimizes directly for \textit{energy}, unlike previous approaches which work by trying to control \textit{miss-rate} and thus optimizing cache energy indirectly. Furthermore, EnCache uses dynamic performance monitoring and regulation and thus, does not require offline profiling or per-application tuning. A comparison with a popular technique named Hybrid Dynamic ResIzing (HDRI) cache~\cite{YanPow02_HybridCache,YanPow01_IcacheResize} shows the superiority of EnCache approach. 


The rest of the chapter is organized as follows.  Section~\ref{literatureReview} discusses related work and Section~\ref{sec:profdesign}, \ref{sec:DPMR} and ~\ref{sec:EnergySave} explain the design and algorithms in more detail. Section~\ref{sec:LastLevel} discusses the hardware implementation.  Section \ref{sec:profvalidation} and \ref{sec:encache_results} present the results on profiling cache accuracy and energy saving. Finally, we conclude in Section~\ref{conclusion}.

\section{Related Work}\label{literatureReview}
We employ ``Multi-Level Profiling cache'', which is based on the idea of set-sampling, which states that the behavior of the cache can be estimated by sampling only a small subset of cache sets. Kessler et al. discuss set-sampling and time-sampling techniques and the conditions under which those techniques may be used \cite{KesHil94_SamplingSetTime}.  Qureshi and Patt employ sampling idea
for estimating hit-miss information about possible cases, when the L2 cache
used by them contains 1 to 16 ways, which equals associativity of L2 cache \cite{QurPat06_UtilityBasedCP}.
Profiling cache's ability to estimate performance of multiple cache sizes is a significant improvement over these works, where set-sampling is used to predict the performance of only the current size cache. This difference is critical for the purpose of improving energy efficiency.

Many studies have been done to save power consumption by caches and main memories \cite{Mit_DRAMsurvey}.  Several existing  techniques (e.g. \cite{DroBuy02_Accounting,selectiveCache} are aimed at saving dynamic energy of the cache. Leakage energy forms a large fraction of energy spent in last-level caches and hence, these techniques are not so useful for saving energy in LLCs.

Some researchers have proposed statically reconfiguring cache characteristics such as cache size and cache active ways to save energy~\cite{Albonesi99_Selective, ZhaVah03_Configurable}. The work by Kaxiras et al. reduces leakage energy by turning off the cache lines which have not been accessed for a certain number of cycles, called decay interval \cite{KaxHuz01_CacheDecay}. However, the techniques based on a fixed decay-interval are shown to be less effective for L2 than for L1~\cite{AbeGon05_IATAC}. Apart from this, the optimal value of decay interval varies widely for different benchmarks \cite{ZhoTob03_AMC}.  Thus, for real-world applications, the utility of these approaches is limited.

  Flautner et al. use the technique of placing idle cache lines in state-preserving mode and thus reduce static power consumption \cite{FlaKim02_DrowsyCache}. Similarly, Hanson et al.  use the technique of dynamically changing the threshold voltage to place the cache lines into low-leakage mode, without destroying the contents of the cache line \cite{HanHri02_TVLSI}. However, these techniques require two supply voltages for each cache line. This increases the probability of soft-errors in the cache.

Unlike some techniques (e.g. \cite{ZhoTob03_AMC,AbeGon05_IATAC}) in which only data is turned off and tag fields are always kept on, our technique can have both tag and data arrays turned off. Most of the techniques proposed in literature (e.g. \cite{AbeGon05_IATAC}) have been evaluated by considering their effect on cache energy only. However, we include both LLC energy and memory energy in our energy equations for providing a more comprehensive evaluation.
Simulation holds a vital role in computer architecture research to model, study and experiment with any hardware design proposed. 

\section{Design of Profiling Cache}\label{sec:profdesign}
 
For estimation of program response for multiple configurations, we use ``\emph{profiling cache}'', which  employs set-sampling to estimate cache miss rate. Profiling cache is a data-less cache and gives accurate predictions even for sampling ratios ($R_S$) as high as 32 or more; thus its storage size is very small compared with L2 cache. Furthermore, it is \emph{decoupled from L2 cache} and works non-intrusively. These properties of profiling cache enable us to further extend it to a multi-level
profiling cache: each level emulating a cache of 1X, 0.5X, 0.25X and 0.125X size of the L2 cache. Here all the four L2 caches are assumed to have same block-size and associativity and differ only in number of sets. This extension still keeps the overhead of profiling cache small. Note that such approach has also been used in other fields \cite{Mittal123}.

The L2 cache in our experiments uses LRU replacement policy and for such cases, the profiling cache uses extra counters to provide miss rates for configurations having different number of ways as well. This is based on the Mattson stack algorithm~\cite{Mat70_LRUStack} for caches with LRU replacement policy, which states that an access that hits in an N-way cache also hits in an M-way
cache with the same number of sets, if $M > N$. Thus, with merely $(2M)/R_s$ sets, a profiling cache can simultaneously emulate many caches of much large sizes. This feature is especially useful for miss-rate curve generation. For the purpose of energy saving, we provision the configuration to only four levels, since this gives a large saving in energy, with a small performance loss. Thus, our cache reconfiguration technique chooses a suitable configuration from a large configuration space of 32 configurations of L2 (four states with eight ways each). These configurations are shown as an ordered 2-tuple ($S$, $W$), where $S$ and $W$ denote the L2 state and active ways respectively.

\begin{figure}[htp]
\centering
 \includegraphics [scale=0.50] {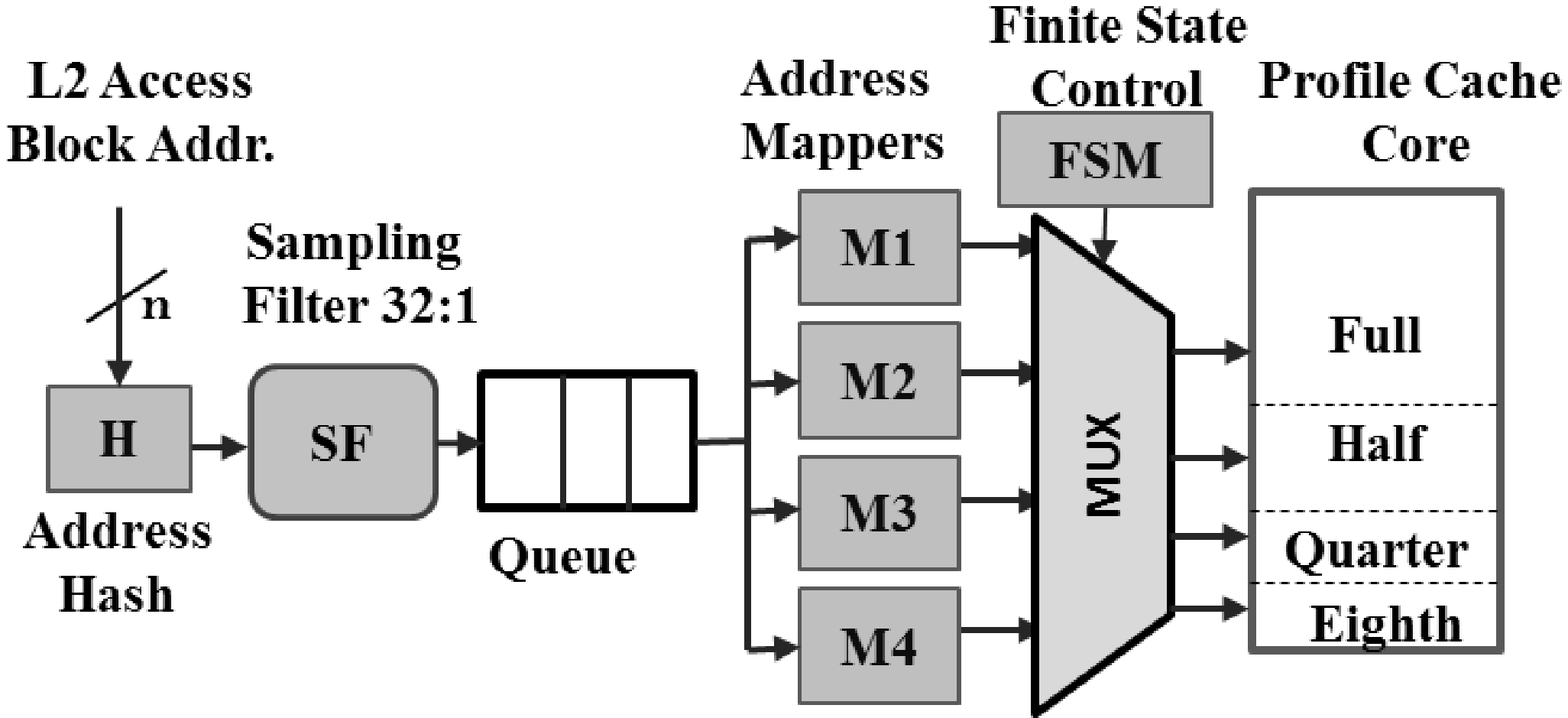} 
\isucaption{The Design of Profiling Cache.}
\label{fig:profdesign}
\end{figure}

Figure~\ref{fig:profdesign} shows the details of the profiling cache design. Its
core storage is a tag-only cache, which has the same set-associative structure and replacement policy as
the L2 and thus, it emulates normal cache accesses. A simple frontend logic component is shown in the left part of Figure~\ref{fig:profdesign}.  Each L2 cache access block address first
passes a hashing logic (for randomization) and then goes through to a sampling filter.  The sampling ratio is chosen at design time and has been taken as 32 in our experiments, which means that only 1 out of 32 of memory
block addresses in the physical address space will pass the filter. Sampling is implemented by merely a bit-shifting operation. Then, those addresses are sent through a small queue to the profiling cache core.

The profiling cache core storage is split into four regions, called ``$Full$'',
``$Half$'', ``$Quarter$'' and ``$Eighth$'', respectively.  Each region represents an
emulated cache size (also known as L2 cache state), i.e. ``$Full$'' for full size, ``$Half$'' for half size, and so on.  Each hashed address from the head of the
queue is sent to four address mappers (M1, M2, M3 and M4). Each mapper is a simple logic that
removes a subset of the address bits (decided by $R_S$) and then
inserts a subset of bits that is the offset of each region in the profiling
cache core. Thus it maps
the addresses onto a unique cache set in one of those four regions (M1 for
``$Full$'', M2 for ``$Half$'' and so on). The four mapped addresses
are sent to a multiplexer (MUX), from where they are sequentially sent to the profiling cache core
 by the control of a small finite state machine. A ``miss" in profiling cache does not generate any request for other caches or memory; rather, the LRU block is evicted and the tag of the address missed is simply copied in its place.

 The profiling
cache core is accessed four times for each address that passes the filter. Note that this does not cause congestion even in the case of bursty last-level cache accesses, because of a large value of $R_S$, the presence of queue and lack of any data-transfer operation. Due to its smaller size and parallel operation, the latency of profiling cache is small and easily
hidden. Moreover, it does
not lie at the critical access path of the cache and does not affect L2 cache access time.

\section{Dynamic Performance Monitoring and Regulation (DPMR)} 
\label{sec:DPMR} 

Cache reconfiguration-based energy minimization
involves performance trade-off. To control the aggressiveness of cache
reconfiguration, while still making performance-efficient choices,
EnCache employs dynamic performance regulation, which works as follows.
Let $\text{Time}_{i}(S,W)$ denote the estimate of execution time for
interval $i$. Then for any configuration
$(S,W)$, we define, \[\Delta
\text{Time}_{i}{(S,W)}=\dfrac{\text{Time}_{i}{(S,W)}-\text{Time}_{i}(Full,Assoc)}{\text{Time}_{i}{\text(Full,Assoc)}}
\times 100\]  Note that $\text{Time}_{i}(Full,Assoc)$ is also obtained
at runtime (i.e. not offline) with the help of profiling cache, even
though the actual configuration in interval $i$ may be different.
$\Delta \text{Time}_{i}{(S,W)}$ gives an estimate of the time overhead of a
configuration $(S,W)$ compared to the baseline $(Full,Assoc)$. Then, in interval
$i+1$, EnCache only searches from those configurations that satisfy
the criterion $\Delta \text{Time}_{i}{(S,W)}$ $\leq$
$\lambda$. The parameter $\lambda$ is an
\textit{application-independent} constant and is set to be $3\%$ in our
experiments. In summary, DPMR dynamically adjusts the configuration space available for the EnCache
energy saving algorithm.  The dynamic performance regulation approach of EnCache is
suitable for real-world applications and is a considerable enhancement
over the static approaches used in previous studies.

\section{Energy Saving Algorithm}
\label{sec:EnergySave} 

It is well-known that different applications and even different phases of the
same application may have different active working set size (WSS). In
any interval, by allocating just minimum LLC space to an application so
that its working set can fit, the rest of the L2 cache can be turned off
to save leakage energy with little impact on performance. Based on this
observation, at the end of each interval, the system software (which could
be a kernel module) is designed to use the following algorithm to choose a
configuration with minimum estimated energy.  If a configuration $(S, W)$ has been rejected by DPMR, the system software does not compute its energy value to reduce
computations. Initially, the cache configuration is $(Full,Assoc)$.

\begin{algorithm}

\isucaption{EnCache: Algorithm For Energy Saving}
\label{L2SwitchOnOff}

 \begin{algorithmic}[1]
{
\small

    \INPUT $Misses$ and $Time$ estimates (for all configurations)
   \OUTPUT Best \emph{$State $} and \emph{$Ways$} for interval $i+1$

         \STATE $Energy^{\star}=\infty$, $S^{\star}$= -1, $W^{\star}$= -1

       \FOR{ $S$ = \{$Full$,$Half$,$Quarter$, $Eighth$\} }
         \FOR{ $W$ = 1 to Assoc}
            \STATE Estimate  $PPM_{i}$($S$,$W)$, $\Delta Time_{i}$($S$,$W$)

	            \IF {$\Delta Time_{i}(S,W)$ $\geq$ $\lambda$ }
                 \STATE  Disregard ($S$,$W$) for interval $i+1$; Continue to next configuration
                \ENDIF
             \STATE Estimate $Energy_{i}$($S$,$W$)
        \\
	\IF {$Energy_{i}$($S$,$W$) $<$ $Energy^{\star}$}
	\STATE $Energy^{\star}$=$Energy_{i}$($S$,$W$), $S^{\star}$=$S$, $W^{\star}$=$W$
	
	\ENDIF

\ENDFOR
\ENDFOR
       
        \STATE RETURN ($S^{\star},W^{\star}$) for interval $i+1$
\\
}

 \end{algorithmic}
\end{algorithm}

\section{Hardware Implementation}\label{sec:LastLevel}

\begin{figure}[htp]
\centering
 \includegraphics [scale=0.31] {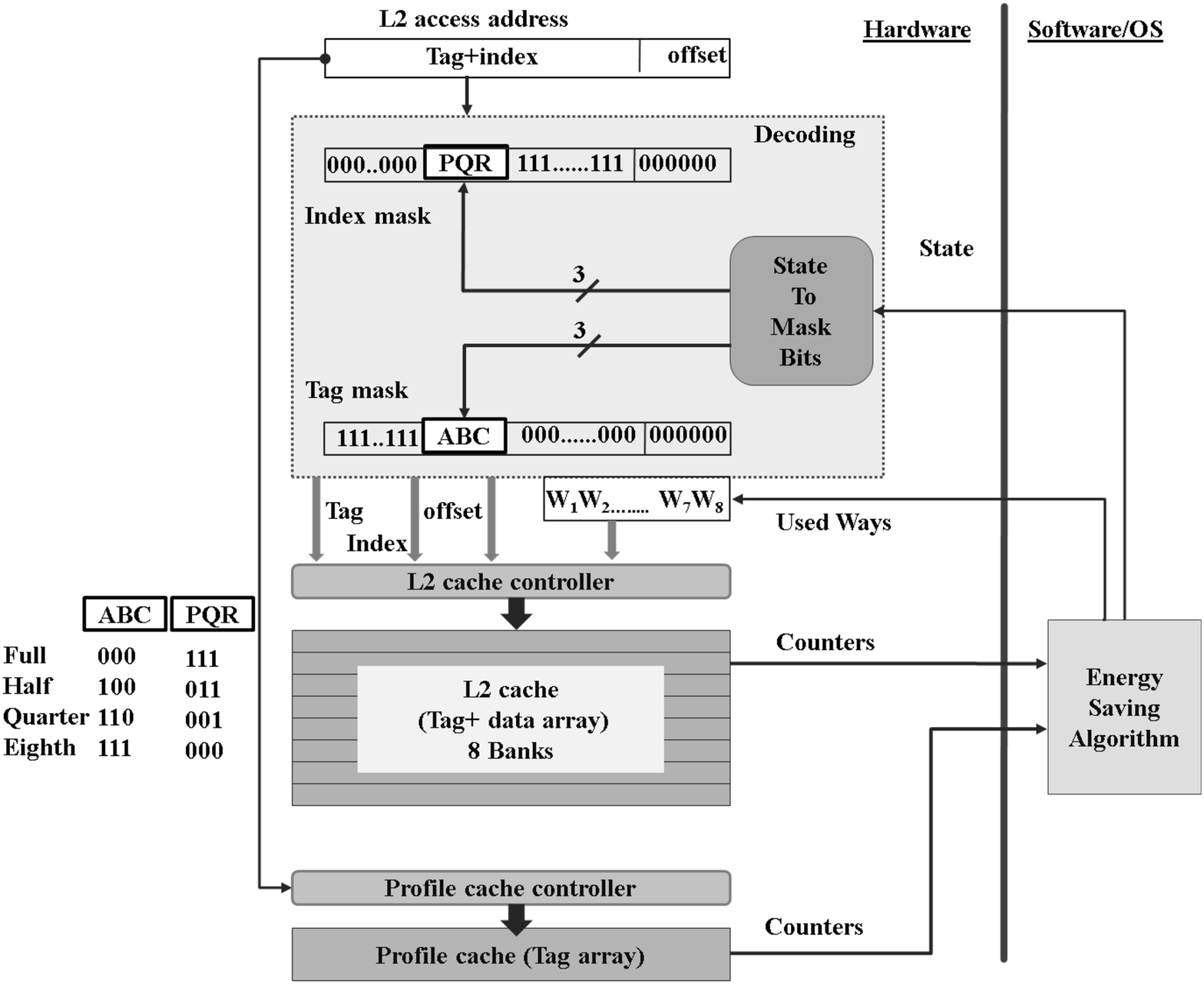}
\isucaption{L2 cache controller in EnCache}
\label{fig:L2cachework}
\end{figure}

Figure~\ref{fig:L2cachework} shows the L2 cache controller design. For
$W$-way cache ($W=8$ in our case), the controller uses a $W$-bit mask
called way-selection mask. By controlling a particular bit $W_k$ (for
$k$=\{1, 2, ...8\}), the corresponding way $k$  can be turned on or
turned off respectively.  The L2 cache has an eight-bank structure. For
accomplishing switching to $Half$, $Quarter$ and $Eighth$ states, cache
controller keeps four, two and one bank of the cache turned-on
(respectively) and turns off the rest of the banks. This is achieved by
a simple logic controlled by a set-selection mask (not shown in the
Figure \ref{fig:L2cachework}). Note that the approach of turning off cache banks to save cache leakage has been used in other studies~\cite{WanMak10_BankTurnOff,RamYal07_Resizing} also.

We define ActiveRatio as the average fraction of L2 cache lines which
are turned on over the execution of the program. Mathematically
\begin{eqnarray*}
  \text{ActiveRatio} & = & \displaystyle\dfrac{\sum_{i=1}^{N}Fraction(S_i^{\star})\times W_i^{\star}}{N\times Assoc}\times 100 \\
   S_i^{\star}  & = & \{\text{Full}, \text{Half}, \text{Quarter},  \text{Eighth} \} \\
   Fraction(S_i^{\star})  & =  & \{1, 0.5, 0.25, 0.125\}
\end{eqnarray*}

Here $N$ is the number of intervals, $Assoc$ is the associativity of
L2, and $(S_i^{\star}, W_i^{\star})$ denotes the actual configuration used in an
interval $i$.

The L2 cache controller uses suitable tag and index (set) masks to handle
the change in set and tag decoding resulting from change in L2 state
(Figure~\ref{fig:L2cachework}). The calculation for these masks for 2MB,
8-way cache with block size of 64-byte is done as follows. $Full$ state has
4,096 sets and hence for index mask, a total of 12 bits are required.
Since $Eighth$ state has 512 sets, 9 least-significant-bits out of 12
bits are always set to 1. The three most-significant-bits are calculated
as: $a_2a_1a_0= \text{Binary}(8\times Fraction(S_i)-1)$. In
Figure~\ref{fig:L2cachework}, these bits are shown as PQR. For a 45-bit
address and 6 bits of block offset, the maximum number of bits in
the tag-mask is $45-6-9= 30$, as required for $Eighth$ state. Out of these,
27 most-significant-bits  are always set to 1 since a minimum of
$45-6-12=27$ bits are required for $Full$ state. The three
least-significant-bits are simply
$\overline{a_2}\overline{a_1}\overline{a_0}$. In
Figure~\ref{fig:L2cachework}, these bits are shown as ABC. Since the
index and tag masks are modified only at most once at the end of an
interval, the address decoding can be optimized to hide the extra
latency caused by the change in decoding.

For handling reconfigurations, we use the following approach. When only
the number of $Ways$ is decreased, the clean blocks of the disabled ways
are discarded and the dirty blocks are written back. On a change in L2 state,
the new set (index) and tag values for cache blocks are computed and the
blocks are re-located to their new set-locations. Out of the blocks not
fitting the available cache space, the clean blocks are discarded and
the dirty blocks are written back.  Such an approach may incur a ``one
time'' high overhead but is simple and requires small state storage.
Since, the reconfigurations take place at a fixed interval boundary,
block transitions do not lie at the critical path of cache access.
Further, on reconfigurations involving an increase in only active ways
or active sets, writebacks to memory are not required.  As shown in
Section~\ref{sec:encache_results}, EnCache keeps reconfiguration overhead small,
which is easily amortized over the phase length.

\section{Profiling Cache Prediction Accuracy Verification} \label{sec:profvalidation}
\begin{figure*}[htbp]
  \begin{center}
     \subfigure{\label{fig:Prof2}\includegraphics[scale=0.29]{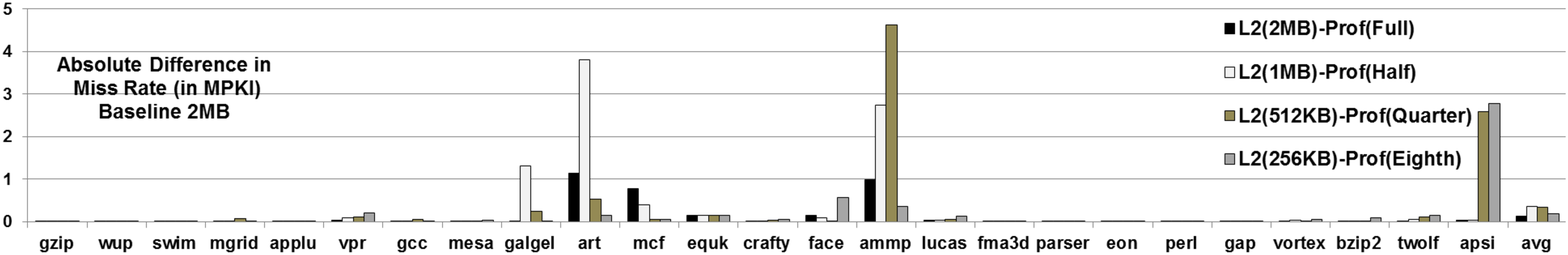}}
     \subfigure{\label{fig:Prof4}\includegraphics[scale=0.29]{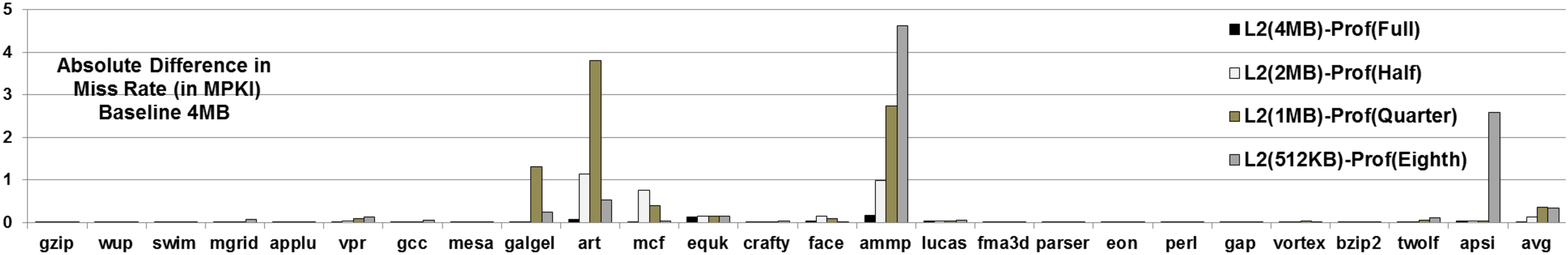}}
     \subfigure{\label{fig:Prof8}\includegraphics[scale=0.29]{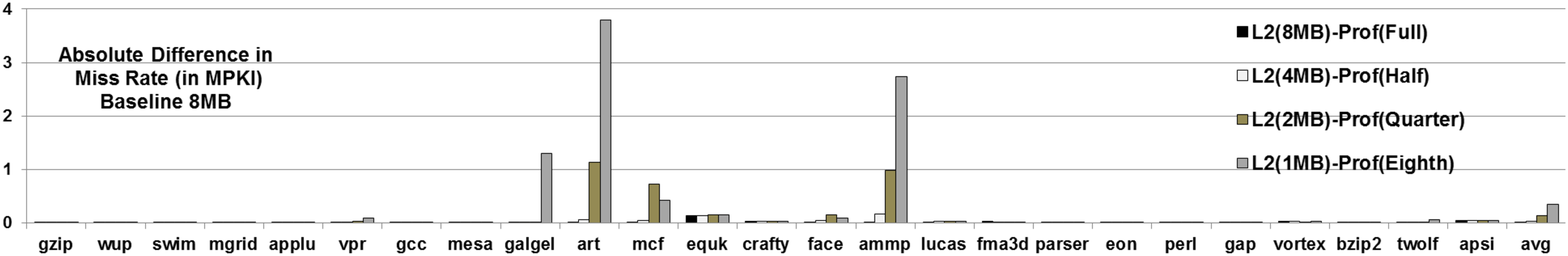}}
  \end{center}
\isucaption{Profiling Cache Prediction Accuracy Verification.} \label{fig:profwork}

\end{figure*}

We present the results of the experiments performed for verifying profiling cache accuracy. We explain the procedure for the case, when the baseline L2 cache has a size of 2MB. The profiling cache predicts miss-rates for four states and when the L2 cache has a maximum size of 2MB, these states profile L2 cache sizes of 2MB, 1MB, 512KB and 256KB. Hence, for each benchmark in our workload, experiments were carried out using baseline cache configuration of size 2MB, 1MB, 512KB and 256KB (each having 8-way and 64B block size) and miss per Kilo instructions (MPKI) were recorded. These values were compared with corresponding estimates obtained from a four-level profiling cache. For example, the miss-rate obtained from 2MB cache was compared with the miss-rate estimate obtained from profiling cache region that emulates ``$Full$'' size cache; the miss-rate with 1MB cache was compared with the miss-rate estimate obtained from profiling cache region that emulates ``$Half$'' size cache and so on. The results are shown in Figure~\ref{fig:profwork}. 

Across 100 combinations of benchmark/configurations (25 SPEC2000 benchmarks with 4 states each), the average absolute difference in miss-rates estimated from profiling cache and that obtained from corresponding size actual L2 cache is merely 0.26 misses/Kilo instructions. The average of percentage absolute difference in miss-rates is 5.91\%. Figure~\ref{fig:Prof4} and Figure~\ref{fig:Prof8} show these values when baseline cache has maximum size of 4MB and 8MB respectively and the values of average absolute difference in miss-rates for these cases are 0.22 MPKI and 0.13 MPKI respectively. Further, the average of percentage absolute difference in miss-rates for 4MB and 8MB baseline caches are 5.34\% and 4.15\% respectively. These results confirm the high accuracy of the multi-level profiling cache.

\section{Energy Saving Results} \label{sec:encache_results}
The experiments performed over sim-outorder simulator and the
comparisons made with a well-known technique named Hybrid Dynamic
ResIzing (HDRI) cache~\cite{YanPow02_HybridCache,YanPow01_IcacheResize}
show the effectiveness of EnCache approach in saving memory-subsystem
energy.
 
\begin{figure*}[htbp]
  \begin{center}
     \subfigure{\label{fig:EDP2}\includegraphics[scale=0.29]{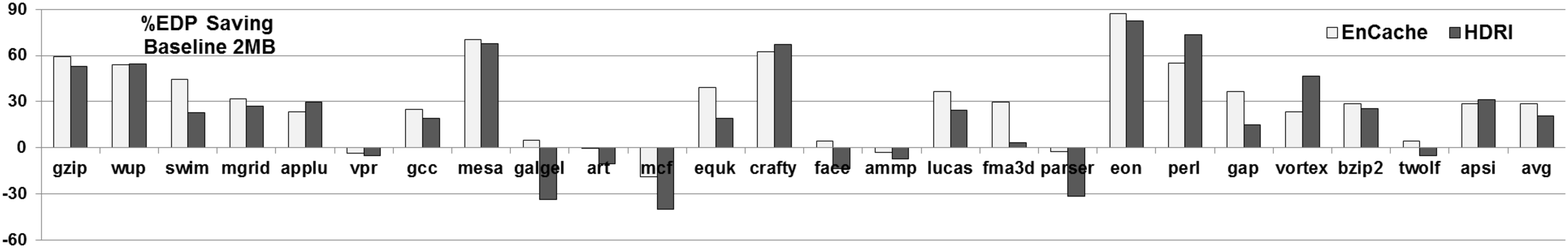}}
     \subfigure{\label{fig:Energy2}\includegraphics[scale=0.29]{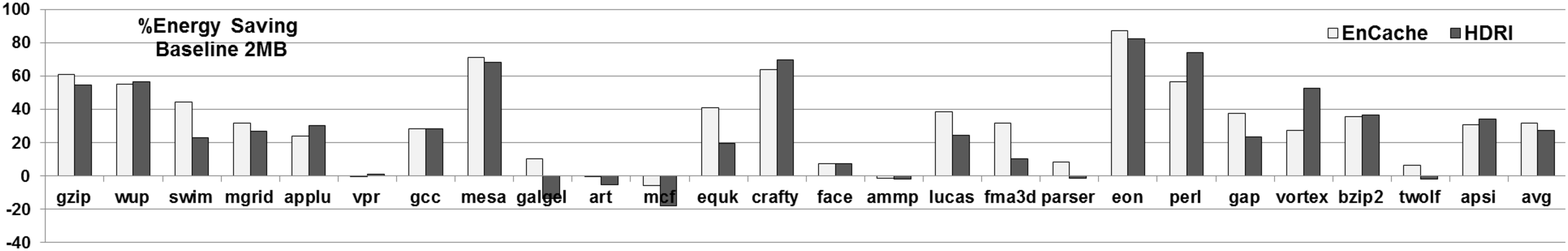}}
     \subfigure{\label{fig:Cycle2}\includegraphics[scale=0.29]{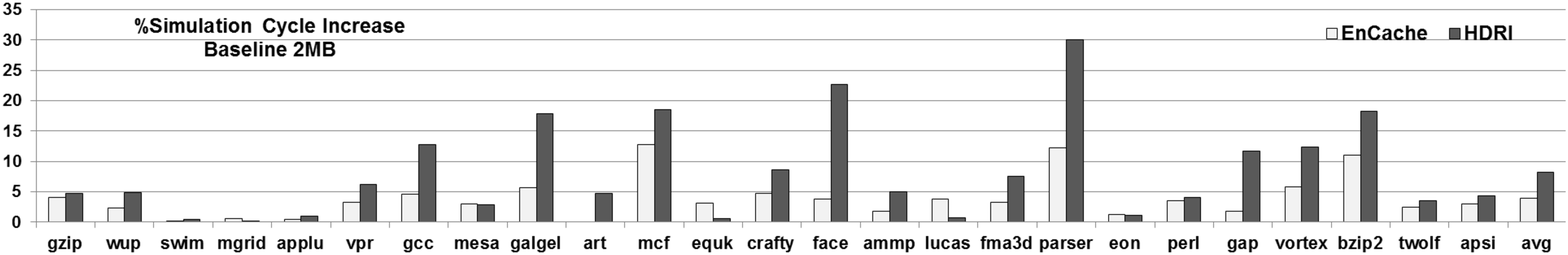}}
  \end{center}
    \isucaption{ EnCache: Experimental Results with 2MB Baseline Cache }
  \label{fig:result2}
\end{figure*}

\begin{figure*}[htbp]
  \begin{center}
     \subfigure{\label{fig:EDP4}\includegraphics[scale=0.29]{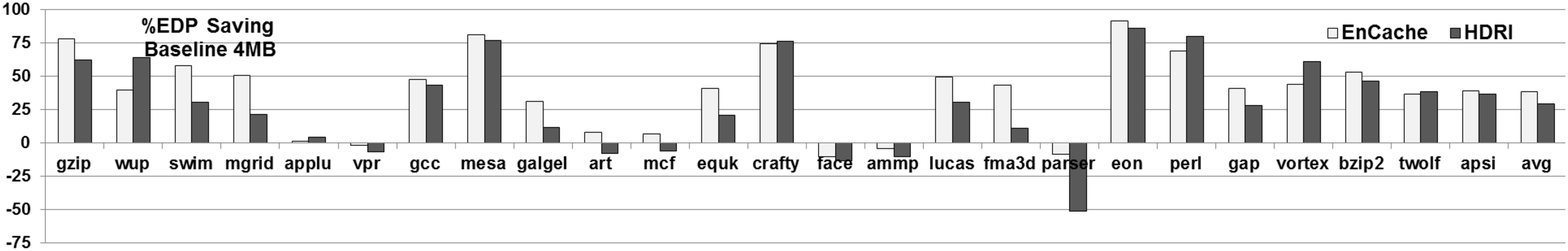}}
     \subfigure{\label{fig:Energy4}\includegraphics[scale=0.29]{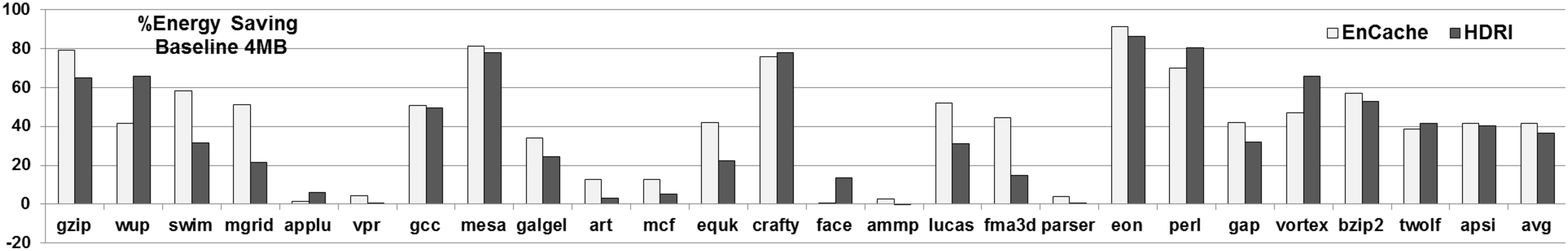}}
     \subfigure{\label{fig:Cycle4}\includegraphics[scale=0.29]{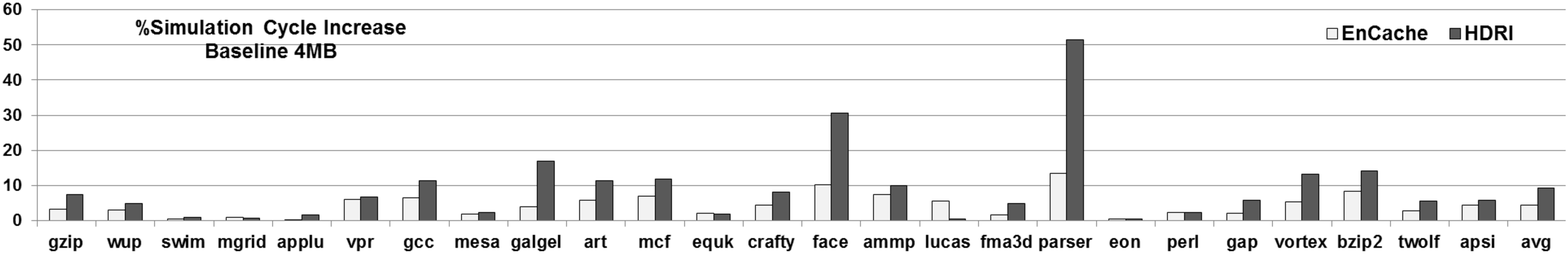}}
  \end{center}
    \isucaption{ EnCache: Experimental Results with 4MB Baseline Cache }
  \label{fig:result4}
\end{figure*}

\begin{figure*}[htbp]
  \begin{center}
     \subfigure{\label{fig:EDP8}\includegraphics[scale=0.29]{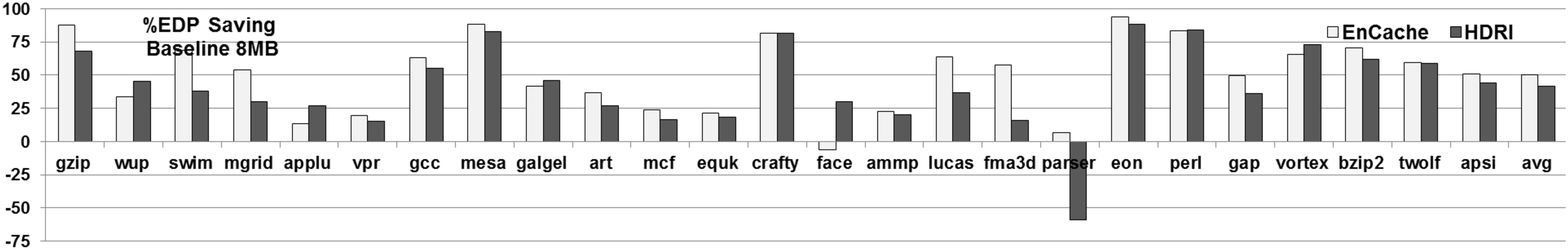}}
     \subfigure{\label{fig:Energy8}\includegraphics[scale=0.29]{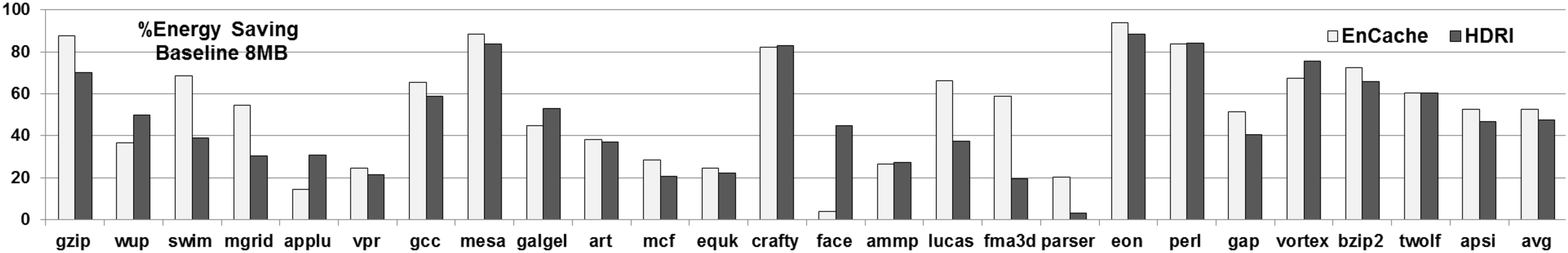}}
     \subfigure{\label{fig:Cycle8}\includegraphics[scale=0.29]{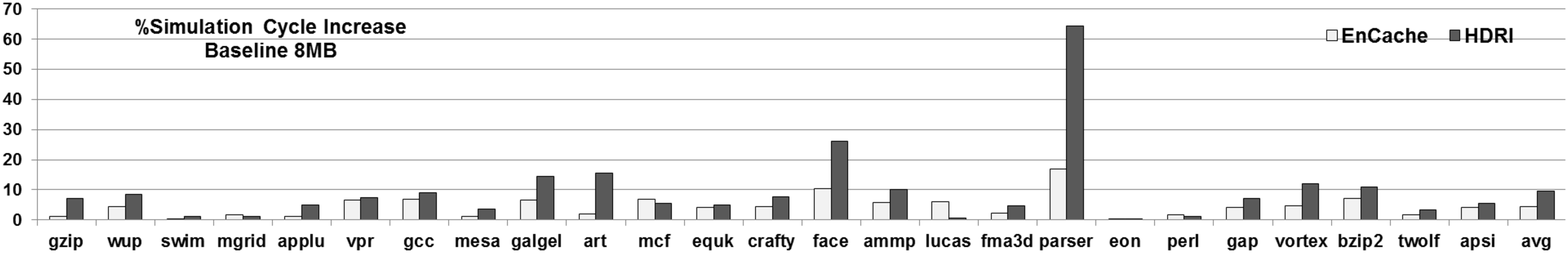}}
  \end{center}
    \isucaption{ EnCache: Experimental Results with 8MB Baseline Cache }
  \label{fig:result8}
\end{figure*}

 Fig.~\ref{fig:result2}, \ref{fig:result4} and \ref{fig:result8}  show the saving in memory subsystem energy for the case when
baseline cache has 2MB, 4MB and 8MB size, respectively. 

The average saving in memory subsystem energy for EnCache and HDRI are
31.7\% and 27.4\% respectively. Average increase in simulation cycles
for EnCache and HDRI are 3.93\% and 8.2\%, respectively; and the average
saving in EDP are 28.8\% and 20.6\%, respectively. The average
ActiveRatio with EnCache and HDRI are 49.5\% and 47.1\%, respectively; and
the average increase in MPKI are 0.45 misses and 0.62 misses, respectively.
Out of 100 intervals, for EnCache, reconfigurations occur about 26 times
on average, with about 16 for associativity changes and the other for
set changes. For HDRI, those values are 42 and 33 respectively. The
figures for other quantities have been omitted for brevity.

The adaptive nature of both the algorithms especially benefits
benchmarks such as \textit{eon}, \textit{gzip}, \textit{mesa},
\textit{crafty}, \textit{wupwise},\textit{ perlbmk} etc, where a large
saving in energy is achieved. The worst-case performance of HDRI is very
poor; as \textit{mcf} shows loss in EDP of 39\%. Similarly
\textit{galgel} shows loss of energy of 13\% and \textit{parser} shows
simulation cycle increase of 30\%. For EnCache, the worst-case
performance happens on \textit{mcf}, where loss in EDP is 19\%.  For
\textit{art}, EnCache does not choose to reconfigure the cache at all,
since the extra misses generated by reconfiguration would have offset
energy saved in cache.  On the other hand, HDRI performs poorly for
\textit{art} and shows loss in energy. A negligibly small (0.2\%) loss
in energy, observed with EnCache arises due to the use of profiling
cache.

Firstly, for both techniques, the saving in cache energy is large enough
to offset the energy cost of the algorithm ($E_{algo}$). At all the three
cache sizes, for both energy and EDP saving, EnCache performs superior
to HDRI in terms of best-case, average-case and worst-case behavior.

With HDRI technique, for different applications, the best (i.e. lowest) value of EDP is observed at different values of $MissBound$. Further, some benchmarks show large variation in EDP saving with change in $MissBound$.
For example, with the 8MB baseline cache, the saving in EDP in
\textit{wupwise} increases from 4.5\% to 45.5\%  when going from $\eta
+200$ to $\eta + 400$.   Also, intra-program variations make the HDRI
approach of using fixed value of $MissBound$ highly ineffective.  This
is evident from \textit{parser} benchmark at 8MB baseline, where the
loss in EDP is 59\% even at $\eta+200$ and even worse at other
$MissBound$ values. Thus even a small offset of $200$ misses leads to
severe cache thrashing.

HDRI is generally more aggressive in turning off L2 cache.
Despite this, the large increase in number of misses and execution time
offset the saving achieved in L2 cache energy. This highlights the
importance of dynamic performance regulation (DPMR), which EnCache uses.

EnCache allows a direct change to one state from any other state without
having to go through intermediate state (e.g. from $Full$ to $Eight$ without
going through $Quarter$). Thus, whenever L2 WSS changes drastically, the
EnCache algorithm directly reconfigures the cache to the most
appropriate size. On the other hand, the HDRI approach must go through all
the intermediate configurations before reaching a desired configuration;
and thus it incurs a large reconfiguration overhead.

For different benchmarks, the impact of increased cache misses on energy
is different. The HDRI approach fails to capture this relationship since it
works by trying to keep number of extra misses small and thus, it does
not directly work to choose an energy-efficient configuration. On the
other hand, EnCache optimizes directly for energy and captures the
effect of increased misses on energy consumption.

EnCache uses a profiling cache to provide online profiling results for
guiding reconfiguration, while the choice of suitable $MissBound$ in
HDRI scheme requires multiple simulation-runs in offline profiling.
Moreover, with HDRI, changing the simulation length/parameters (e.g.
simulating 500M instructions or using L1 cache of 32KB) would require
completely new offline profiling, since benchmark behavior may vary a
lot between different configurations. Given that the $TotL2Miss$ for
different benchmarks varies over three orders of magnitude, choosing a
benchmark-specific $MissBound$ is absolutely necessary with the HDRI
technique. Finally, EnCache can optimize based on the energy consumption
in other components (such as main-memory) also, while the HDRI scheme is
insensitive to the overall energy picture.

\section{Conclusion} \label{sec:encache_conclusion}
In this chapter, we discussed EnCache, a novel scheme for saving leakage power
consumption of last-level caches. It uses a system-level approach with
lightweight hardware support. Using a novel, low-cost hardware component called profiling cache, system software can accurately predict memory-subsystem energy of a program for multiple cache-configurations. The dynamic performance monitoring allows  controlling aggressiveness of reconfiguration and strike suitable balance between energy minimization and performance loss. The experiments performed show the superiority of EnCache over conventional energy-saving scheme.

\chapter{PALETTE: A CACHE ENERGY SAVING USING CACHE COLORING} \label{chap:palette}

\section{Introduction}
 In this chapter, we present \textbf{Palette}, a cache coloring based leakage energy saving 
technique using dynamic cache reconfiguration. Palette uses a small hardware component called ``reconfigurable cache emulator'' (RCE) which provides miss rate estimates for multiple cache sizes. Using this, along with the memory stall cycle estimation model, Palette estimates program execution time under multiple possible cache configurations. Then, for these configurations, memory sub-system energy is estimated. Further, using the energy saving algorithm, the cache is reconfigured to the most energy efficient  configuration and the unused colors are turned off for saving leakage energy. For switching (i.e. turning on/off) cache blocks, Palette uses the gated V$_{\text{dd}}$ scheme \cite{PowSeh00_GatedVdd}. 
 
Palette has several salient features which address the limitations of previous techniques. Palette uses dynamic profiling and not offline profiling and hence, it can be easily used in product systems.  Palette optimizes for energy \textit{directly}, unlike existing techniques, which control other parameters (e.g. miss rate, number of dead blocks \cite{YanPow01_IcacheResize,KaxHuz01_CacheDecay}) to save energy in an \textit{indirect} manner. By virtue of this feature, Palette can optimize for system (or subsystem e.g. memory sub-system) energy, and not merely cache energy and hence, it can easily detect the case when saving cache energy may increase the energy consumption of other components of the processor. Palette takes into account the benefit (i.e. utility) from cache allocation and not access intensity. Hence, it saves large amount of energy for most programs, including streaming programs.   

We perform microarchitectural simulations using out-of-order core model from Sniper simulator \cite{CarHei2011_Sniper} and benchmark programs from SPEC2006 suite. Further, we compare Palette with a well-known cache leakage saving technique, called ``decay cache technique'' (DCT) \cite{KaxHuz01_CacheDecay}. The experimental results show that Palette is effective in saving energy and outperforms the conventional energy saving technique. Using Palette, the average saving in
memory sub-system energy and EDP, compared to a 2MB baseline cache are 31.7\% and 29.5\%, respectively. In contrast, using DCT, the saving in energy and EDP are only 21.3\% and 10.9\%, respectively.

The rest of the chapter is organized as follows. Section \ref{sec:relatedwork} discusses the  related work and Section \ref{sec:systemimplementation} explains the design of Palette. Section \ref{sec:makealgorobust} presents the energy saving algorithm and Section \ref{sec:palette_implementation} discusses the hardware implementation of Palette.  Section \ref{sec:simulationmethodology} discusses the simulation environment, workload, and energy model. Section \ref{sec:palette_results} presents results on energy saving. Finally, Section \ref{sec:palette_conclusion} concludes the work.

\section {Background and Related Work}\label{sec:relatedwork}
 Recent advances in high performance computing has made several applications computationally amenable . High performance computing platforms provision large cache resources to bridge the gap between the speed of processor and main memory. This however, also brings the issue of managing power consumption of caches. In this chapter, we address this issue using cache dynamic reconfiguration approach.  
 
In literature, several techniques have been proposed for saving cache energy. A few techniques aim to save cache dynamic energy \cite{selectiveCache}. However, a large fraction on energy dissipated in LLCs is in the form of leakage energy \cite{li2002leakage}, and hence, cache dynamic energy saving techniques have only limited utility in saving energy in LLCs.  Palette aims at saving leakage energy of the cache and hence, it is useful for saving energy in LLCs.

Several techniques use \textit{static} cache reconfiguration \cite{ZhaVah03_Configurable,Albonesi99_Selective}, however, programs show a large variation in their cache demands over different phases and hence, dynamic cache reconfiguration is important to achieve large energy savings.  Some leakage energy saving techniques always keep the tag fields turned on and only turn off only the selected regions of the data-array, e.g. \cite{ZhoTob03_AMC}. In contrast, Palette turns off both tag and data arrays of the inactive region.     

Different energy saving techniques turn off cache at different granularity, such as cache ways \cite{Albonesi99_Selective}, cache sets \cite{YanPow01_IcacheResize}, hybrid (sets and ways) \cite{YanPow02_HybridCache,MitZha12_EnCache} and cache blocks  \cite{KaxHuz01_CacheDecay,ZhoTob03_AMC}. Selective ways approach incurs low reconfiguration overhead; however, its cache allocation granularity is limited by the number of cache ways. Selective sets and hybrid approaches generally incur higher reconfiguration overhead, since on a change in the set-counts, the set-decoding scheme changes and whole cache needs to flushed. In contrast, the cache coloring scheme used in Palette incurs smaller reconfiguration overhead than selective sets or hybrid approaches, since on a change in the number of active colors, the set-locations of only the affected cache colors are changed.

As for circuit-level leakage control mechanism,  both state-destroying \cite{PowSeh00_GatedVdd} and state-preserving \cite{FlaKim02_DrowsyCache,HanHri02_TVLSI} techniques have been used. The state-preserving techniques typically save less power in low-leakage than the state-destroying techniques and also increase the noise susceptibility of the memory cell \cite{ayala2007energy}. For this reason, Palette uses state-destroying leakage control using gated V$_{\text{dd}}$ mechanism \cite{PowSeh00_GatedVdd}.

%


\section{Palette Design and Architecture}\label{sec:systemimplementation}

\begin{figure*}[htp]
 \centering
  \includegraphics [scale=0.46] {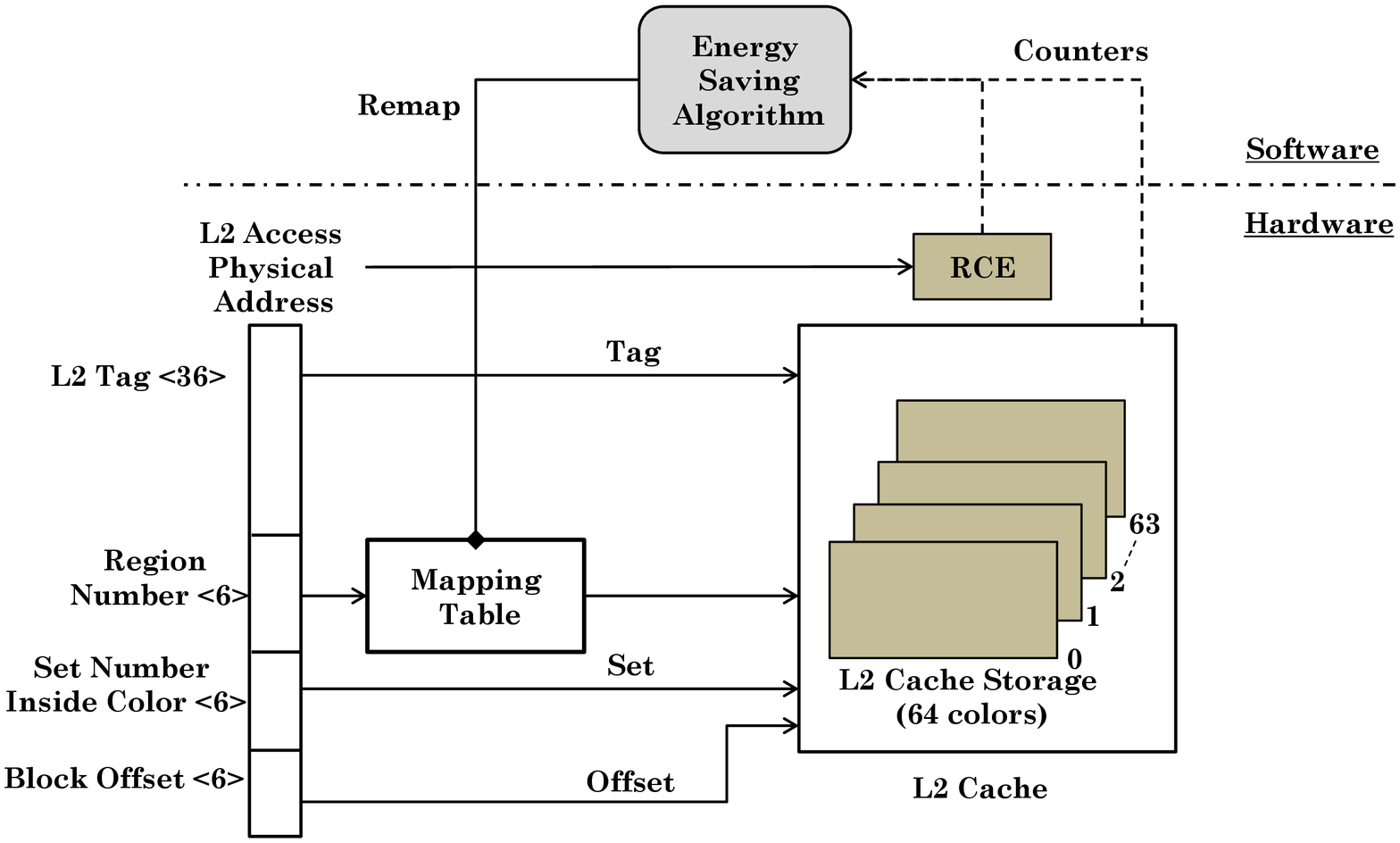} 
 \isucaption{Palette Flow Diagram}\label{fig:algoflow}
 \end{figure*}

It is well-known that there exists large intra-application and inter-application variations in the cache requirements of different applications. Since several applications executed on the modern processors are performance-critical and hence, designers use an LLC size which meets the requirements of such performance-critical applications. However, this leads to wastage of energy in the form of cache leakage energy.  Palette works on the intuition, that in any interval, a suitable amount of cache can be allocated to a program, while the rest of the cache can be turned-off for saving leakage energy. Figure \ref{fig:algoflow} shows the overall flow of Palette. In this section, we discuss each of the components of Palette in detail. We assume that the LLC is the L2 cache, and the discussion can be easily extended to the case when the LLC is an L3 cache.  

\subsection{Coloring Scheme}
To selectively reconfigure the cache, Palette uses cache coloring technique \cite{LinLuq09_softwarecp,KesHil92_PageColoring}.   Firstly, the cache is divided into $N$ non-overlapping bins, called cache-colors. Let $B$ denote the L2 block size; $H$ denote the physical page size and  Size$_{L2}$ denote the number of sets in L2. Then, $N$ is given by
\begin{equation}
N = \dfrac{\text{Size}_{\text{L2}}\times B}{H}
\end{equation}

In modern memory management, physical memory is divided into physical pages. We logically group these pages into $N$ \textit{memory regions}.  A memory region refers to a group of physical pages that share the $\log_2(N)$ least significant bits of the page number. Cache coloring works by controlling the mapping from memory regions to cache colors such that all the physical pages in a memory region are mapped to the same color in the cache.

To enable flexible cache indexing and also avoid the cost of page migration (as in \cite{LinLuq08_hpca}),  we use a small \textit{mapping-table} (MT), which stores the region-to-color mapping. Thus MT has $N$ entries. To see the typical value of $N$, we note that for a page size ($H$) of 4KB and L2 block size ($B$) of 64 byte (or 512 bits), for 8-way, 2MB L2 cache, $N=64$. Hence, the size of MT is 384 ($=64\times \lg_2(64$) bits. Clearly, the size of MT is extremely small and hence, its access latency and energy consumption are negligible.

For enabling reconfiguration, the amount of cache allocated to the application is controlled by controlling the number of active cache colors.  At any point of execution, if the number of colors allocated to an application is $M$ ($\le$$N$), then the mapping-table stores the mapping of $N$ regions to $M$ colors. Note that, here $M$ can have a non-power-of-two value also and thus Palette has the flexibility to allocate any cache size to the application. Thus a cache configuration is specified in terms of the number of active colors.

 \subsection{Reconfigurable Cache Emulator}\label{sec:palette_rce}
To estimate the cache miss-rate under  various cache configurations, Palette uses a small microarchitectural component, called reconfigurable cache emulator (RCE). RCE has one or more profiling units. Each profiling unit is based on the principle of set sampling \cite{puzaksampling,KesHil94_SamplingSetTime} and thus estimates L2 miss rate by sampling only a few sets. The profiling unit is a data-less (tag-only) component and it emulates the L2 cache by having similar replacement policy and associativity. It does not store data and hence does not communicate with other caches on a hit or miss. It works in parallel to L2 and does not lie at the critical access path. We use the sampling-ratio ($R$) of 64, which implies that profiling unit samples only 1 out of 64 sets of the L2 cache.

The small size of profiling unit and parallel operation enables us to use multiple profiling units in the RCE. For our technique, we use six profiling units, each of which profiles a cache size of  $X/16$, $2X/16$, $4X/16$, $8X/16$, $12X/16$, $16X/16$, where $X$ shows the L2 cache size (or equivalently number of L2 colors). A unique feature of the RCE design is that the profiling unit can profile a cache size for which the set-counts are \emph{not} power-of-two values. This becomes possible by using cache coloring scheme (as explained above). This is a significant improvement over previous works based on cache reconfiguration (e.g. \cite{MitZha12_EnCache}).

\begin{figure}[htp]
 \centering
  \includegraphics [scale=0.55] {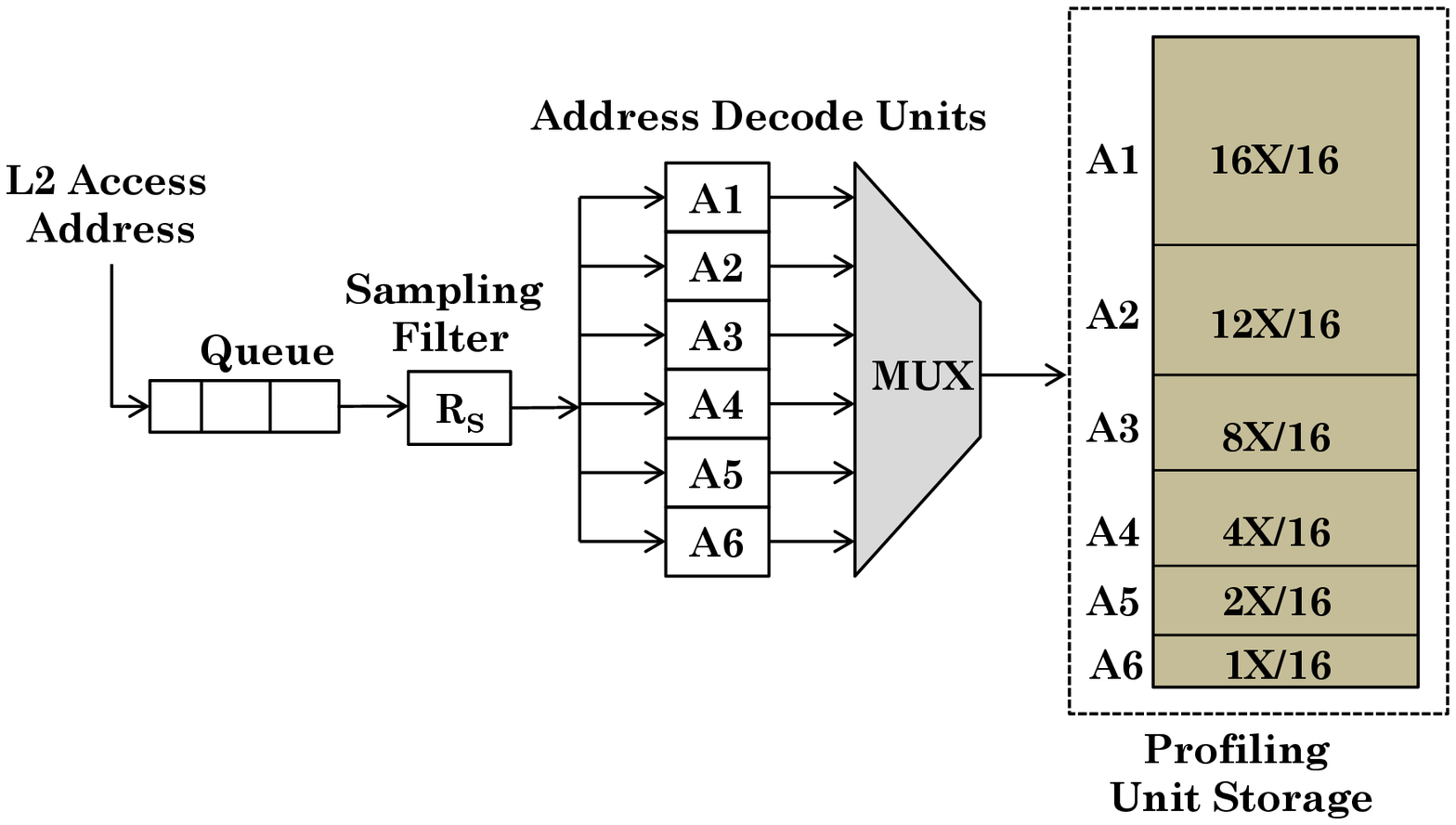}
 \isucaption{RCE block diagram}\label{fig:palette_profcache}
 \end{figure}

The RCE works as follows (Figure \ref{fig:palette_profcache}). Each L2 access address is passed through a small queue and then passed through to a sampling filter. The sampled addresses are fed to address decoding units (ADUs). Each ADU uses its own mapping table.  To compute set(index) and tag of the address, first, the region number of address is computed and then, its color is read from the mapping-table. Using this, the set(index) value of the address is computed. After ADU, the accesses are fed to the core storage using a simple MUX.

Let $P$ denote the number of sets in L2 cache and $S$ denote the total number of sets in the RCE. Then, we have  
\begin{equation}\label{eq:sumofprof}
S = \dfrac{P}{16R} + \dfrac{2P}{16R} + \dfrac{4P}{16R} + \dfrac{8P}{16R} +\dfrac{12P}{16R} + \dfrac{16P}{16R} = \dfrac{43P}{16R} 
\end{equation}

To see the overhead of RCE ($F_{prof}$) compared to L2 cache size, we assume $W$ way L2 cache, with $B$ bit block size and $T$ bit tag. Thus,
\begin{equation}
F_{prof} = \dfrac{\text{Size}_{\text{RCE}}}{\text{Size}_{\text{L2}}}=\dfrac{S\times W \times T }{P\times W\times(B+T)}  = \dfrac{43T}{16R(B+T)}
\end{equation}

For $R=64$, $T=40$, $B=512$ (i.e. 64 byte), we get $F_{prof}=$.003 or 0.3\%. Thus, the overhead of RCE is small.

\subsection{Predicting Memory Stall Cycle For Energy Estimation} \label{sec:StallCycle}
To compute the leakage energy of memory sub-system under different L2 configurations, the program execution time under those configurations needs to be estimated. This, however, presents several challenges, since modern out-of-order processors use ILP (instruction level parallelism) techniques to hide cache miss latency \cite{GenEye2010_IntervalCoreModel}. To get an estimate of program execution time under different configurations, Palette uses a hardware counter to continuously measure effective memory stall cycles, taking into account possible overlap with other miss events (e.g. branch misprediction, L1 miss). Further, extra counters are used with RCE for also measuring the number of L2 load misses under different cache configurations.

Using above hardware support, we proceed as follows. First, the total-cycles of the program is decomposed into base-cycles and stall-cycles.   We assume that in an interval $i$ with configuration $C_{\star}$, the effective stall-cycles (StallCycles$_i(C_{\star})$) are proportional to the number of load-misses (LoadMisses$_i(C_{\star}$)). Thus, their ratio (termed as stall-cycle per load-miss or SPM$_i$) is independent of the number of load-misses. Using this, the StallCycles$_i(C)$ for any configuration $C$ can be estimated as
\begin{equation}
\text{StallCycles}_i(C) = \text{SPM}_i\times \text{LoadMisses}_i(C)
\end{equation}
  where LoadMisses$_i(C)$ shows the number of load-misses under that configuration. 
   
From StallCycles$_i(C)$ value, the total-cycles (or equivalently execution time) under configuration $C$ are computed by adding base-cycles value to it. Using this, the leakage energy of the program under any configuration can be easily estimated (Section \ref{sec:palette_energymodel}).
   
A limitation of this approach is that for the programs which show significant variation in the number of load-misses with the L2 cache size, the SPM value varies with L2 cache sizes and this affects the accuracy of energy estimation. However, as shown next, Palette only searches for configurations which differ in a small-number of colors from $C_{\star}$ and hence the above assumption holds reasonably well.

%

\section{Palette Energy Saving  Algorithm}\label{sec:makealgorobust}
In each interval, Palette uses energy saving algorithm (ESA) which works by intelligently selecting a small number of candidate configurations, estimating their energy and then selecting the most energy efficient configuration from them.
Before discussing the energy saving algorithm, we first discuss the concept of marginal gain and then show its use in ESA. 
\subsection{Marginal Gain Computation}
Palette computes \textit{marginal gain} values and utilizes them to make an intelligent guess about candidate configurations.  At any configuration $C$, the value of marginal gain, $G(C)$, is defined as the reduction in cache misses on increasing a single color. Thus, $G(C)$ is a measure of utility of increasing unit cache resource of the program. We assume that between two profiling points, the  number of misses vary linearly with cache size (piecewise linear approximation) and hence, the marginal gain remains constant. For the six profiling points viz. $C_{p}^1=N/16$, $C_{p}^2=2N/16$, $\ldots$ $C_{p}^6=16N/16$ ; if the number of L2 misses at these profiling points (i.e. cache sizes) is denoted by $Miss(C_{p}^j)$ (where $j=\{1,2,3,4,5,6\}$), then the marginal-gain $G(C)$ at $C$ ($C_{p}^{1} \le C \le C_{p}^{6}$ ) is defined as

     \begin{equation}
G(C) = \begin{cases}
\dfrac{ Miss(C_{p}^j)-Miss(C_{p}^{j+1})}{C_{p}^{j+1}-C_{p}^j}  & C_{p}^{j} \le C < C_{p}^{j+1}\\
\dfrac{ Miss(C_{p}^5)-Miss(C_{p}^{6})}{C_{p}^{6}-C_{p}^5} & C=C_p^6\\

\end{cases}
\end{equation}

\subsection{ESA Description}
We now discuss the working of ESA and then present its pseudo-code. We use the following notations. Let ConfigSpace denote the set of candidate configurations, which are initially chosen in an interval. Also, let $D$ be its cardinality, i.e. the number of candidate configurations. Also, we use $C_{\star}$ to denote the actual configuration in interval $i$.

To keep the reconfiguration overhead small and avoid oscillation, ESA selects configurations in neighborhood of $C_{\star}$ using following criterion. 
\begin{enumerate}
\item The algorithm always considers the current configuration ($C_{\star}$) as one of the candidates. 
\item To keep algorithm overhead low, $D$ is set to a small value. In our experiments, $D$ is  taken as 11 which includes $C_{\star}$ itself.   
\item To avoid the possibility of thrashing/starvation of the application, ESA only selects configurations with at-least $Min$ active colors; thus, at least $Min$ colors are allotted to the application. In our experiments $Min$ is set to $N/16$.  
\item The granularity of cache allocation is taken as two colors, since this allows testing a wider range of configurations, while still keeping algorithm overhead small. Thus a configuration $C$ is `valid' if  fulfills the criterion $(N/16) \le C \le N$ and $C\pmod 2 =0$.
\item To allow for possible reduction or increase in number of active colors, the candidate configurations include both kinds of configurations, namely those with lower and higher number of active colors than $C_{\star}$.  Intuitively, for a program with low $G(C_{\star})$ value, configurations with smaller cache size are likely to be energy efficient and vice-versa. Thus, for programs with low $G(C_{\star})$ value, out of $D$ configurations, the number of candidate configurations having colors less than  $C_{\star}$ is higher than those having colors more than $C_{\star}$. Similarly, for programs with high $C_{\star}$ value, out of $D$ configurations, the number of candidate configurations having colors more than  $C_{\star}$ is higher than those having colors less than $C_{\star}$.
 \end{enumerate}

Afterwards, for each configuration in the ConfigSpace, the memory subsystem energy is computed and the configuration with the least amount of energy is selected for the next interval. Algorithm 1 shows the pseudo-code of ESA.

\begin{algorithm}
 \isucaption{Palette Energy Saving Algorithm (ESA)}
 \label{algo:palette_energysaving}
 \begin{algorithmic}[1]
{

    \INPUT Estimates of $Misses$ (from RCE),  Current Config=$C_{\star}$
   \OUTPUT Best configuration for interval $i+1$
         \STATE $BestEnergy$ = $\infty$, $BestConfig$= -1 
          \STATE $G(C_{\star})$ = marginal\_gain\_at($C_{\star}$)
         \STATE ConfigSpace= config\_space\_for($C_{\star}$, $G(C_{\star})$)
        \FOR{ each config $C_i$ in ConfigSpace }
         \STATE $E_i$= estimated\_energy\_of($C_{\star}$)
	            \IF {$E_i$ $<$ $BestEnergy$ }
                 \STATE  $BestEnergy$= $E_i$
        \STATE $BestConfig$= $C_i$ 
                \ENDIF

\ENDFOR
       
        \STATE RETURN $BestConfig$

}

 \end{algorithmic}
\end{algorithm}

\section{Hardware Implementation}\label{sec:palette_implementation}
For cache block switching (i.e. turning off and on), we use the well-known gated-$V_{\text{dd}}$ technique \cite{PowSeh00_GatedVdd}. Gated V$_{\text{dd}}$ works on the basis of transistor stacking effect \cite{ye1998new}. A gated V$_{\text{dd}}$ memory cell uses an extra transistor in the supply or ground path. For the active regions of the cache, this transistor is kept on. For deactivating a memory cell, this transistor is turned off, which drastically reduces the leakage current supply in the cell and the memory cell loses its stored value. We use a specific implementation of gated-$V_{dd}$ (NMOS with dual Vt, wide, with charge pump) which results in minimal impact on access latency but introduces a 5\% area penalty. We account for the effect of increased area on leakage energy in Section \ref{sec:simulationmethodology}.  
  
The reconfigurations are handled in the following manner. When the active cache colors are decreased, the contents of the disabled cache colors are flushed (i.e. dirty blocks are written-back to memory and other blocks are discarded). The memory regions mapped to these colors are remapped to other active colors. When the active cache colors are increased, some memory regions, which were mapped to another color, are remapped to newly active colors and the blocks of those memory regions in their previous colors are flushed. Our reconfiguration scheme is simple and requires less state storage than previous schemes \cite{RanAdv00_Recon, LinLuq08_hpca}. Further, reconfigurations only happen at the end of an interval, thus, cache color turning off/on  does not lie at critical path of cache access. Moreover, since Palette uses a large interval length (e.g. 10M instructions), reconfigurations are minimized and their cost is amortized over the phase length. Our experimental results (Section \ref{sec:palette_results}) have shown that Palette provides large saving in energy and also keeps the increase in execution time and L2 MPKI small. This confirms that the reconfiguration overhead of Palette is indeed small.
    
Palette allocates cache at the granularity of cache colors and not cache ways, and hence, Palette can easily work with caches of low-associativity, (e.g. a 4-way cache), which have low dynamic energy. Moreover, Palette uses dynamic reconfiguration and hence, it does not require storing values from offline profiling (unlike \cite{wang2010leakage}).

\section{Simulation Methodology } \label{sec:simulationmethodology}
\subsection{Platform, Workload and Evaluation Metrics}
For microarchitectural simulations, we have used Sniper \cite{CarHei2011_Sniper}, state-of-art simulator, which has been validated against real hardware \cite{CarHei2011_Sniper}. We model a 1.5 GHz, 4 wide processor with ROB size of 128. L1 data/instruction caches are 32KB, 4 way, LRU, 64B line size and have a latency of 4 cycles. The unified L2 is 2MB, 8 way, 64B line size LRU with 12 cycles latency. The DRAM memory has a latency of 105 cycles and a peak bandwidth of 6GB and the queue contention is also modeled.

To simulate the representative behavior, while still limiting the simulation time, we use  12 benchmarks from SPEC2006, which represent the behavior of entire SPEC2006 suite, as shown by Phansalkar et al. \cite{phansalkar2007subsetting},  based on their multivariate statistical data analysis.  These benchmarks are 6 each from integer point (gcc, hmmer, libquantum, mcf, sjeng, xalancbmk) and floating point (cactusADM,  lbm, milc, povray, soplex, wrf) benchmarks. We use reference inputs. Each benchmark was fast-forwarded for 10B instructions and then simulated for 1B instructions. Algorithm interval size is taken as 10M instructions.   

 Our baseline is the full size L2 cache which does not use energy saving technique. For evaluation, we show results on five metrics, which are as follows.
\begin{enumerate}
 \item Percent of energy saved over baseline.
 \item Percent of simulation cycle increase over baseline.
 \item Percent of EDP (energy delay product) saved over baseline.
 \item Active ratio, which shows the average fraction of active cache lines over entire simulation and is expressed as a percentage \cite{KaxHuz01_CacheDecay}.
 
\item Absolute increase in L2 MPKI (miss-per-kilo-instructions).  
 \end{enumerate}
 The computation of energy is shown in Section \ref{sec:energymodel}. For computation of EDP, delay is taken to be same as simulation cycles. For MPKI increase, we report \textit{absolute} increase and not \textit{relative} increase, following \cite{MitZha12_EnCache,TamAzi09_RapidMRC}, since MPKI value for some workloads can be arbitrarily small and hence, even a small change in a small value may show up as a large percentage, which misrepresents its contribution in the performance.

\subsection{Comparison With Existing Technique}\label{sec:palette_comparison}
For comparison purposes, we have implemented decay cache technique (DCT) \cite{KaxHuz01_CacheDecay}. Our choice of DCT is motivated by two reasons. Firstly, DCT, like Palette, uses state destroying leakage control. Secondly, it is a well-known technique and has been used/evaluated by several researchers (e.g. \cite{HanHri02_TVLSI,AbeGon05_IATAC,li2002leakage}).  

Decay cache technique (DCT) monitors accesses to cache blocks and turns off a block which has not been accessed for the duration of `decay interval' to save cache energy. For implementing DCT, we follow \cite{KaxHuz01_CacheDecay,LiyPar04_Skadron} and use gated V$_{\text{dd}}$ for hardware implementation and hierarchical counters for measuring access intensity. Also, the latency of waking up decayed block is assumed to be overlapped with memory access latency and to maximize energy saving, both data and tag arrays are decayed \cite{KaxHuz01_CacheDecay,LiyPar04_Skadron}. 

We compute decay interval using competitive algorithms theory  \cite{KaxHuz01_CacheDecay}. As shown in Section \ref{sec:energymodel}, the DRAM access energy ($E^{\text{dyn}}_{\text{mem}}$) is 70nJ and the leakage power consumption of 2MB, 8-way L2 cache is 1.568 Watts. Let $U$ denote the leakage energy (in nJ) per block per cycle for L2 for 1.5GHz frequency, then $U$ is given by 
\begin{equation}
 U = \dfrac{1.568}{1.5\times 32768}
\end{equation}
Then, the ratio $E^{\text{dyn}}_{\text{mem}}/U$ shows the ratio of DRAM access energy and the L2 leakage energy per block per cycle, which in our case is 2.19M cycles. This suggests the range of decay interval. To choose a suitable decay interval, we simulated DCT with five decay intervals, viz. 3M, 5M, 7M, 9M and 11M cycles. We did not choose decay intervals which are smaller than 2.19M, since for several benchmarks, even at 3M cycle decay intervals, the performance degradation becomes very high. Based on these simulations, we chose the decay interval for DCT as 7M cycles, since this gives the largest average improvement (saving) in EDP.

\subsection{Energy Model}\label{sec:palette_energymodel}
We take into account the energy spent in L2 cache ($E_{L2}$), main memory ($E_{mem}$) and in execution of the algorithm ($E_{Algo}$), since other components are minimally affected by our approach. 

\begin{equation}\label{eq:palette_Etot}Energy= E_{L2}+E_{mem}+E_{Algo}\end{equation}
where energy spent in L2 and memory is composed of both leakage and dynamic energy.

To calculate $E_{L2}$, we note that the leakage energy depends on active ratio of the cache \cite{YanPow01_IcacheResize,li2002leakage}. Also, an L2 miss is assumed to consume twice the energy of an L2 hit \cite{HanHri02_TVLSI,MitZha12_EnCache,MitZha13_Cashier}.   Thus,
\begin{equation}E_{L2}= E^{dyn}_{L2}\times(H_{L2}+2M_{L2}) + (P^{leak}_{L2}\times Time\times C_{\star})/N\end{equation}
 
Here, for any interval, we have corresponding $H_{L2}$ = L2 hits $M_{L2}$ = L2 misses, $Time=\text{Time consumed}$ and $C_{\star}$=active colors. Also $P^{leak}_{L2}$ and $E^{dyn}_{L2}$ show the dynamic energy per L2 access and L2 leakage energy per second. We use CACTI 5.3 \cite{cacti_53} to compute these values for 8-bank, 8-way caches with 64 byte block size. We obtained $P^{leak}_{L2} = $1.568 Watts and $E^{dyn}_{L2}=$0.985 nJ/access.  To account for the effect of increased area due to gated V$_{\text{dd}}$ (Section \ref{sec:palette_implementation}), we assume 5\% higher value of $P^{leak}_{L2}$ for both Palette and DCT, but not for baseline LRU cache. 

To calculate $E_{mem}$, we note that the leakage power of memory, $P^{leak}_{mem}$= 0.18 Watt and dynamic energy per access of memory $E^{dyn}_{mem}$= 70 nJ  \cite{ZheLin09_DIMM,MitZha12_EnCache}. Using $A_{mem}$ to denote the number of memory accesses, we get,
\begin{equation}E_{mem}= E^{dyn}_{mem}\times A_{mem}+ P^{leak}_{mem}\times Time \end{equation}

The overheads of RCE (for Palette) and block transitions (for both Palette and DCT) are calculated as follows. 
\begin{equation}\label{eq:palette_Ealgo}E_{Algo}= E_{dyn}^{prof}\times A_{prof} +P_{leak}^{prof}\times Time + E_{Tran}\end{equation}
Here $A_{prof}=\text{profiling cache accesses}$ and $E_{dyn}^{prof}$ and $P_{leak}^{prof}$ are dynamic energy per access and leakage energy per seconds for profiling cache. $E_{Tran}$ shows the energy consumed in block transitions. 

To calculate energy values for profiling cache, we use CACTI along with Eq.~\ref{eq:sumofprof}, with $R$=64. Since CACTI only provides values for power-of-two size caches, we take an upper bound as $S=64P/16R$. For the L2 caches used, we compute the energy values for corresponding profiling cache, by only taking tag energy values since profiling cache is a tag-only cache. For the RCE corresponding to a 2MB L2, we get $P^{leak}_{RCE}$=0.007 Watt and $E^{dyn}_{RCE}$=0.004 nJ/access.    Clearly, profiling cache consumes a negligibly small fraction of energy compared to the energy consumed by L2 cache.  

Each block-transition is assumed to take $0.002$ nJ \cite{MitZha12_EnCache}, thus  the energy spent in block transitions is
\begin{equation}E_{Tran}= 0.002\times Tran \text{  nJ}\end{equation}
where $Tran$ denotes the total number of blocks transitions.

For both DCT and Palette, we ignore the overhead of counters and algorithm execution etc., since many processors already contain counters for measuring performance etc. \cite{KaxHuz01_CacheDecay} and also because they work with a large interval size.

%
%
%
%
\section{Results and Discussion}\label{sec:palette_results}

\begin{figure*}[htp]
 \centering
  \includegraphics [scale=0.73] {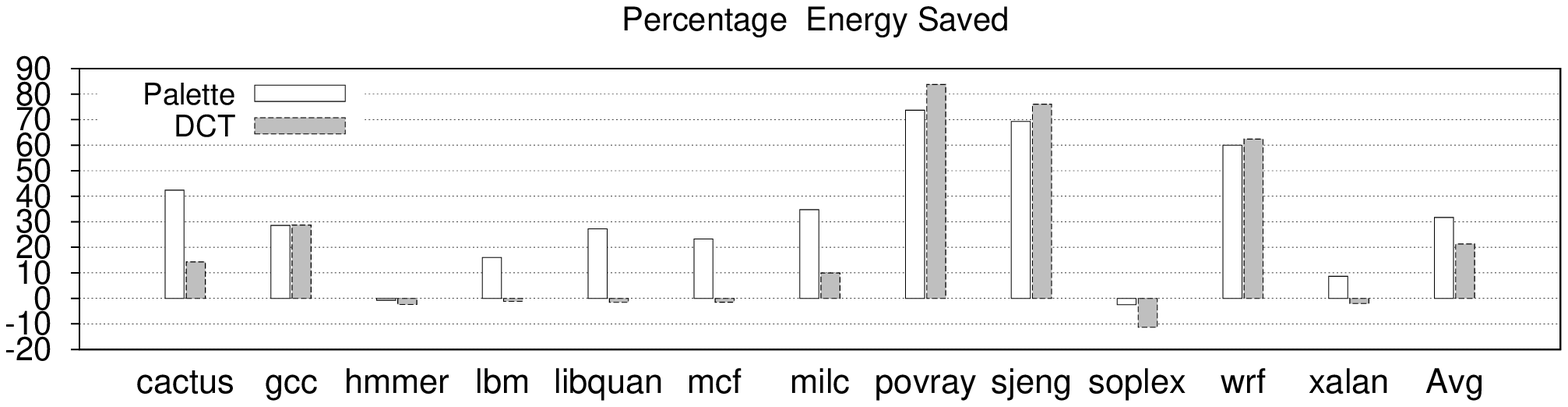}
  \includegraphics [scale=0.73] {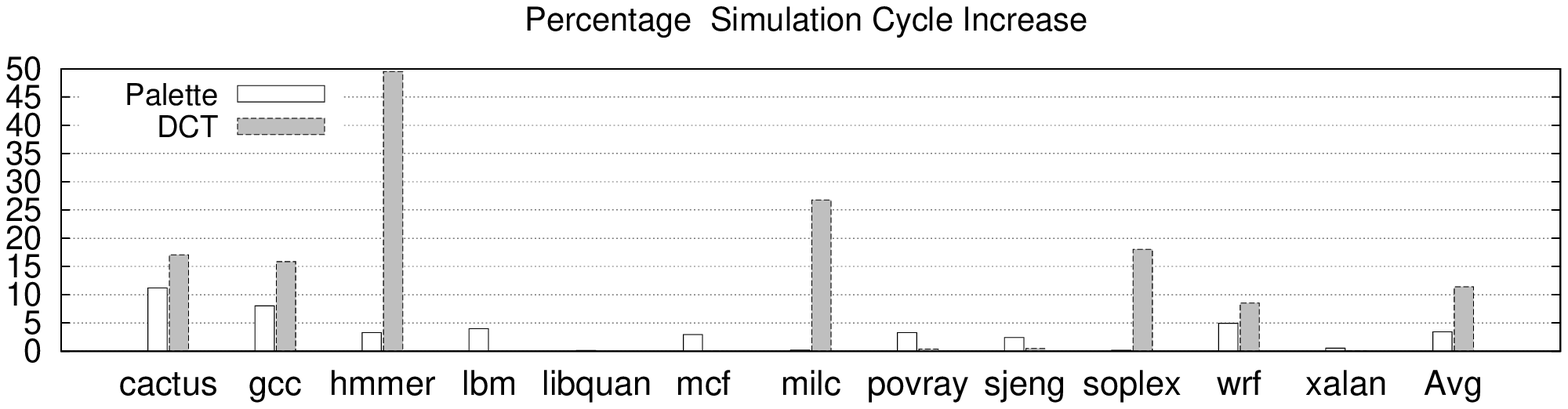}

\isucaption{Experimental Results with DCT and Palette }
 \label{fig:exptresult2MBetedp}
 \end{figure*}

\begin{figure}[htp]
 \centering  
 \includegraphics [scale=0.73] {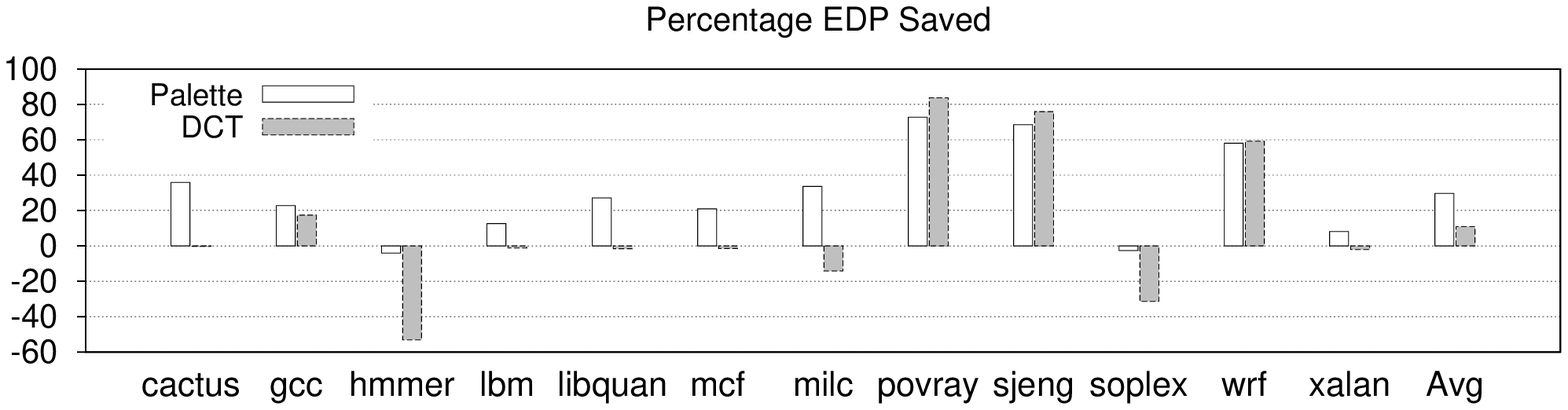} 
  \includegraphics [scale=0.73] {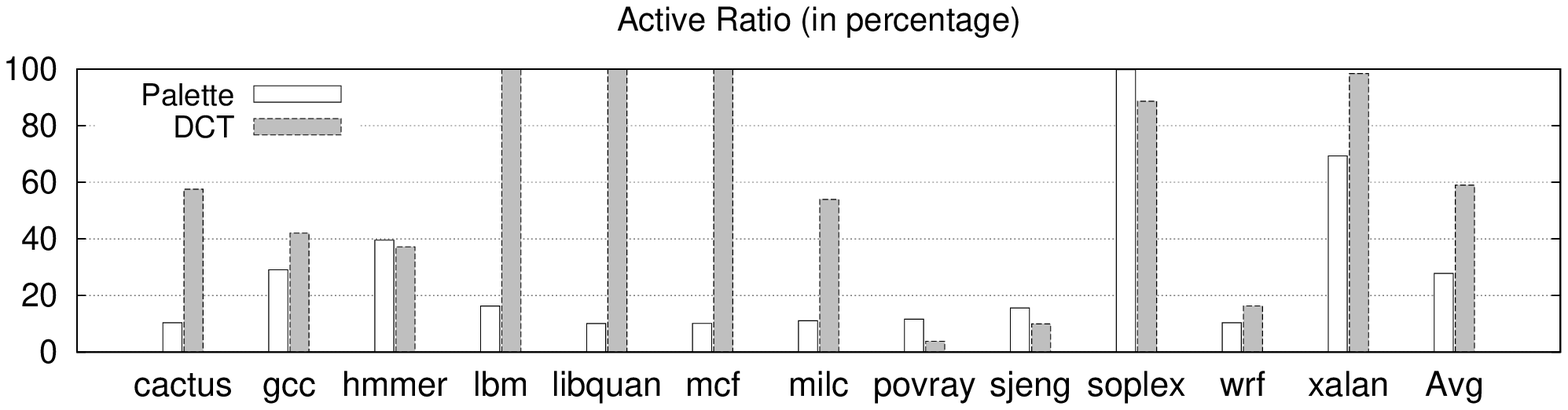}  
  \includegraphics [scale=0.73] {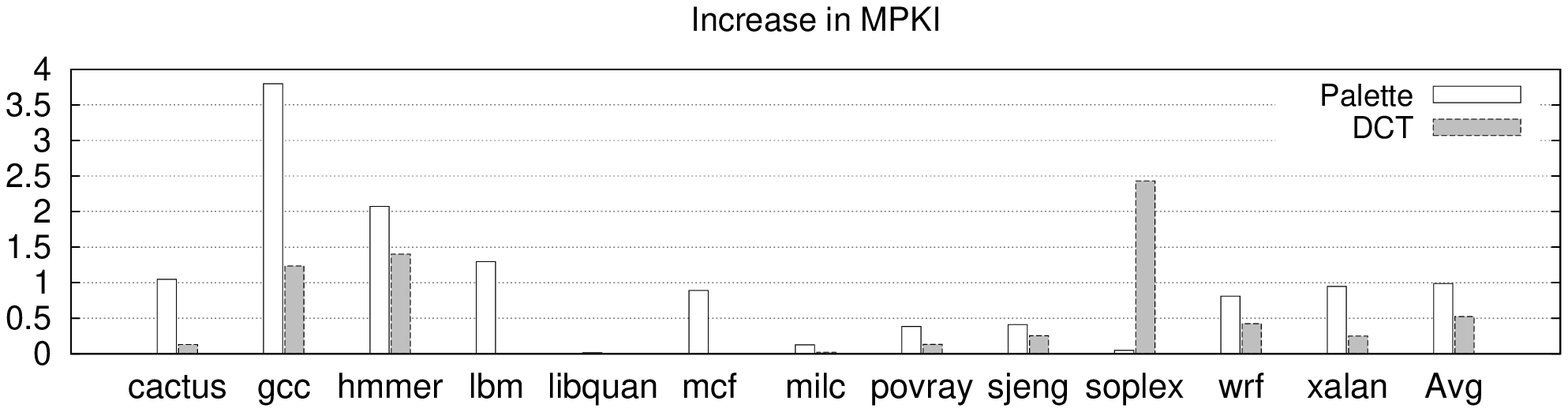} 
   
\isucaption{Experimental Results with DCT and Palette }
 \label{fig:exptresult2MBarm}
 \end{figure}

Figure \ref{fig:exptresult2MBetedp} and \ref{fig:exptresult2MBarm}  show the experimental results. On average, Palette saves  31.7\%  energy, while DCT saves  21.3\% energy. The increase in simulation cycle
 using Palette and DCT are  3.4\% and 11.4\%, respectively. The percentage saving in EDP by using Palette and DCT are 29.7\% and 10.9\%, respectively. The cache active ratio using Palette and DCT are  27.7\% and    59.0\%, respectively. Further, the increase in MPKI by using Palette and DCT are 0.99 and 0.52, respectively.
 
The results clearly show that compared to DCT, Palette saves much larger amount of energy and EDP, and keeps increase in simulation cycle smaller. The average percentage saving in EDP using Palette are nearly double of that obtained using DCT. Further, Palette turns off nearly 72\% of the cache, while DCT turns off only 41\% of the cache.

 DCT turns off a cache block based on its access intensity. However, for some benchmarks, such as mcf, lbm, and libquantum, although the access intensity is large, but the cache reuse remains very small. For such benchmarks, DCT turns off a negligible fraction of cache and thus, DCT does not save energy for these benchmarks. In contrast, Palette turns off cache based on the marginal gain from the allocation of cache, and hence, Palette saves more than 15\% energy for each of these benchmarks. 

DCT turns off cache at the granularity of a single cache block and hence, it can exercise very fine-grain cache reconfiguration. Hence, for povray and sjeng, DCT  saves larger amount of energy than Palette. Towards this, we note that in RCE, by using extra levels of profiling units, such as $X/32$, the profiling information can be obtained for even smaller cache sizes (Section \ref{sec:palette_rce}), and thus, the minimum number of colors allocated to the program can be lowered from $X/16$ to $X/32$. However, this increases the number of profiling units which are consulted in each cache access and hence, a designer can select a suitable trade-off between desired energy saving and acceptable RCE overhead.

For DCT, choice of a suitable decay interval requires significant efforts in offline profiling. Moreover, the optimal value of decay interval varies for different benchmarks \cite{ZhoTob03_AMC}.   In contrast, Palette works by using dynamic profiling and optimizing based on energy estimates. Thus, Palette can be easily used in real-world systems, which execute trillions of instructions of arbitrary applications. 
    
The hardware-based techniques such as DCT work by keeping the miss-rate increase from cache reconfiguration low. Hence, they do not \textit{directly} optimize for energy and cannot easily take the energy consumption of components into account. As an example, on including the energy model of processor core, the optimal value of decay interval will also change. In contrast, Palette directly optimizes for energy and hence, it can easily take the energy consumption of components other than cache into account.      

The average increase in MPKI by using Palette is larger than that from DCT. However, still the increase is small and the extra energy dissipation due to increased DRAM accesses is compensated by the leakage energy saving achieved in the L2 cache.
            
The results presented in this section confirm that Palette is effective in saving cache energy and also outperforms the conventional cache energy saving technique.

\section{Conclusion}\label{sec:palette_conclusion}

We have presented Palette, a cache coloring based technique for saving leakage energy in last level caches. Palette employs online profiling to estimate memory subsystem energy for multiple cache configurations and then dynamically reconfigures the cache to optimize memory subsystem energy efficiency. The experimental results have shown that Palette offers large energy savings, while keeping the performance loss small and also outperforms a conventional leakage energy saving technique. Our future work will focus on integrating cache reconfiguration scheme of Palette with dynamic voltage/frequency scaling to further increase the energy savings.


\chapter{CASHIER: A CACHE ENERGY SAVING APPROACH FOR QOS SYSTEMS} \label{chap:cashier}
\section{Introduction}

In this chapter, we present \textbf{CASHIER} (a \underline{Ca}che energy \underline{s}aving tec\underline{h}n\underline{i}qu\underline{e} for quality-of-se\underline{r}vice (QoS) systems). Several real-world applications present soft real-time resource demands \cite{ref12}. In such applications, the task deadlines are usually more relaxed than the task completion time and as long as a task is completed by its deadline, the actual completion time does not matter from user's perspective. CASHIER is designed for saving energy in such systems. CASHIER exploits the available slack by using dynamic cache reconfiguration to save leakage energy, while making best possible effort to meet the task deadline. Unlike aggressive cache energy saving techniques (e.g. \cite{KaxHuz01_CacheDecay}) which may fail to meet QoS requirements, CASHIER saves energy while fulfilling the QoS requirement.

CASHIER uses a small microarchitecture component called ``reconfigurable cache emulator'' (RCE), which uses set sampling idea to estimate program miss rate for various cache configurations in an on-line manner. Additionally, CASHIER uses CPI stacks to estimate program execution time under different LLC configurations. Using these estimates, the energy saving algorithm estimates memory subsystem energy under different cache configurations. Then, an appropriate cache configuration is chosen to strike a right balance between opportunity of energy saving and performance loss, thus making best possible efforts to not miss the deadline. CASHIER optimizes memory subsystem (which includes LLC and main memory) energy, instead of merely LLC energy.  CASHIER technique is a useful technique for state-of-the-art multimedia transmission systems which require quality-of-service \cite{pande2011quality}.

The rest of the chapter is organized as follows. Section \ref{sec:cashier_relatedwork} discusses related work. The architecture of CASHIER and  energy saving algorithm are discussed in Section \ref{sec:cashier_system} and \ref{sec:cashieralgorithm}, respectively. The energy model and energy saving results are discussed in Section \ref{sec:energymodel} and \ref{sec:cashier_energyresults}, respectively. Finally, Section \ref{sec:cashier_conclusion} concludes this work.  

\section{Related Work}\label{sec:cashier_relatedwork}
Some researchers have presented techniques for saving cache energy while meeting deadlines (e.g. \cite{chi2007cache,WanMis09_VLSI_ReconfigCache}). Wang and Mishra \cite{WanMis09_VLSI_ReconfigCache} use offline analysis to profile a large number of configurations of two-level cache hierarchy and explore these configurations during run-time for finding the best configuration.  However, due to the use of offline profiling, their technique is not suitable for product systems, which generally execute trillions of instructions of arbitrary applications. Apart from cache reconfiguration, dynamic voltage/frequency scaling (DVFS) has also been used  for saving energy while still meeting the deadlines (e.g. \cite{weissel2002process,JejGup05_SlackReclaim,PilShi01_RTDVS}). DVFS  aims to save the dynamic energy of the processor, while CASHIER aims to save the leakage energy of the processor. Thus, CASHIER can be synergistically used with DVFS  to save additional amount of energy.  
\section{CASHIER: System Architecture}\label{sec:cashier_system}
\subsection{Cache coloring } \label{sec:coloring}

To selectively and dynamically allocate cache to an application for the purpose of saving leakage energy, CASHIER uses cache coloring technique \cite{KesHil92_PageColoring,LinLuq08_hpca,LinLuq09_softwarecp}. Cache coloring is also known as page coloring and works as follows. Firstly, the cache is logically divided into multiple non-overlapping bins, called cache colors. The maximum number of colors, $N$, is given by
\begin{equation} \label{eq:Nvalue}
N = \dfrac{CacheSize}{ PageSize \times Associativity} \end{equation}

\begin{figure}[htp]
 \centering
  \includegraphics [scale=0.53] {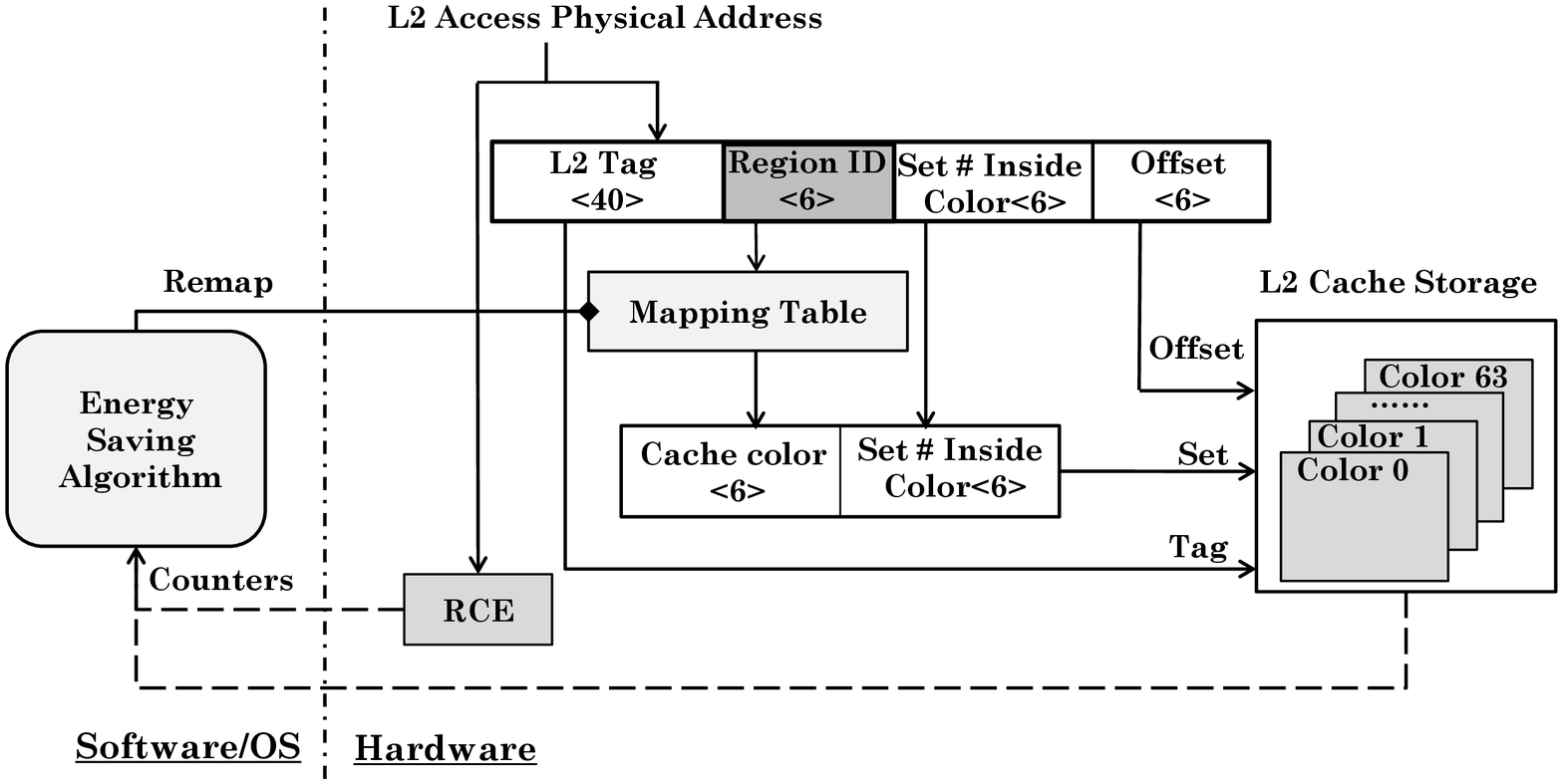}
 \isucaption{CASHIER Flow Diagram (Using example of $N$= 64) }\label{fig:flowdiagram}
 \end{figure}

Further, the physical pages are divided into $N$ \textit{memory regions} based on the least significant bits (LSBs) of their physical page number. In Fig.~\ref{fig:flowdiagram}, where page size is taken as 4KB and $N$= 64, these bits are referred to as Region ID.
Cache coloring maps a memory region to a unique color in the cache. For this purpose, CASHIER uses a small {\em mapping table} (MT) which stores the cache color assigned to each memory region. By manipulating the mapping between physical pages and cache colors, CASHIER allocates a particular cache color to a memory region and thus, all physical pages in that memory region are mapped to the same cache color. 

CASHIER works on the key idea that for restricting the amount of active cache, all memory regions can be allocated to merely few cache colors. Thus, the rest of the colors are effectively not utilized and can be turned off to save cache energy. This is implemented using the mapping table (MT). At any point of execution, if $M$ ($\le$$N$) colors are allocated to the application, the mapping table stores the mapping of $N$ regions to $M$ colors. Thus, CASHIER reconfigures the cache at the granularity of a single cache color. Also, a salient feature of this cache coloring technique is that, unlike previous approaches (e.g. \cite{LinLuq08_hpca}), it does not require a change in underlying virtual address to physical address mapping, and thus can be implemented with little overhead. We refer to ``active'' or ``turned on'' color, as one that stores data and consumes power normally. Also, an ``inactive'' color is one that has been ``turned off'' to save leakage energy and hence does not store data. 

Figure~\ref{fig:flowdiagram} shows the flow diagram of CASHIER with values from the following example. We assume a 2MB, 8-way L2 cache of 64B block size and a $PageSize$ value of 4KB. Then from Equation~\ref{eq:Nvalue}, we get $N$= 64 colors. Hence, in this case, MT has 64 entries, each 6-bits wide (Figure~\ref{fig:flowdiagram}). Also note that the size of mapping table is small and hence, its access latency and energy consumption are negligible.  

\subsection{Reconfigurable Cache Emulator (RCE)}\label{sec:cashier_rcedesign} 
The design of RCE follows similar to as explained in Section \ref{sec:palette_rce}.

\subsection{CPI Stack for Execution Time Estimation}\label{sec:cpistack}
For estimating program execution time under different L2 configurations, CASHIER uses 
the CPI stack technique \cite{EyeEec06_CPIOutOrder,CarHei2011_Sniper}. A `CPI stack' is a stacked bar that shows the different components contributing to overall performance. It presents base CPI (which represents the useful work being done) and `lost' cycle opportunities due to instruction interdependencies, cache misses etc., taking into account the possible overlaps between execution and miss events.
 
Out of various components of CPI-stack, CASHIER makes use of the memory stall cycle component, since the change in L2 configurations shows its effect on execution time in terms of change in memory stall cycles. We assume that, in an interval, memory stall cycles vary linearly with the number of load misses, and thus, their ratio, called SPM (Stall cycles Per load Miss), remains independent of the number of load misses themselves. Then, the stall cycles under any cache configuration can be computed by multiplying SPM with the number of estimated load misses with that configuration. Using these stall cycle estimates and base CPI value from the CPI stack, the total number of cycles (and hence total execution time) under that configuration can be computed. These estimates are used for computing memory subsystem energy values . Also, the execution time and energy estimates are used by the energy saving algorithm.

\section{CASHIER Energy Saving Algorithm}\label{sec:cashieralgorithm}

We now explain the energy saving algorithms of CASHIER. Throughout the chapter, we refer to `baseline cache' as the full size cache which does not use any cache reconfiguration or energy saving technique. We assume that the available slack can be specified in one of the two ways. First,  the slack can be specified as extra time itself ($T_{slack}$). For example, a $T_{slack}$ value of 100$\mu$s denotes that an application can be slowed down by 100$\mu$s, without missing the deadline. This is called \textit{Magnitude Slack Method} (MSM). Second, the slack can be specified as a percentage of extra time over baseline, denoted as $\Upsilon$. This is called \textit{Percentage Slack Method} (PSM). For example, an $\Upsilon$ value of 3\% denotes that an application can be slowed down by 3\% and still it meets its deadline. Note that both these methods have been used in previous studies \cite{LinLuq08_hpca,roy2008toward,white1997rsvp,weissel2002process}. We now discuss the algorithms for each of these methods. A salient feature of CASHIER is that \textit{neither} of these two algorithms require \textit{a priori} knowledge of the baseline execution time for their operation.

We first discuss the steps which are common to both the algorithms.
In any interval $i$ with $C_{\star}$ active colors; both the algorithms select those configurations as candidates which satisfy following two conditions. Firstly, to avoid thrashing, a configuration should have at least $N/16$ active colors. Secondly, to keep the reconfiguration overheads small, in any interval, only up to $L$ ($L=8$ in this chapter) colors can be turned ON or OFF. If $E$ denotes the set of configurations, fulfilling these conditions, we have $E=\{ C \mid (C_{\star}-L) \le C \le (C_{\star}+L) \text{ and } C \ge N/16 \}$.

For understanding the algorithms, it is useful to define a quantify $t_{i}$, as follows. Using program execution time estimates, in every interval, the algorithms estimate the extra time, which the current configuration is taking over and above the baseline configuration\footnote{Note that the execution time estimates for baseline cache configuration are also obtained in run-time using RCE and not in offline manner.}. Over all the intervals, the Algorithm accumulates these values. At the end of any interval $i$, this gives the estimate of increased execution time ($t_{i}$) due to energy saving algorithm (viz. PSM or MSM), \textit{till that interval} $i$. Thus, $t_{i}$ shows the amount of slack already exploited.

 We now explain the steps which are specific to each algorithm.
 
 \textbf{MSM Algorithm:} 
\begin{enumerate}
\item To be conservative, MSM Algorithm keeps a reserved slack of $T_{reserve}$ (which is $T_{slack}/10$ in this chapter) and assumes an effective slack of $T_{eff}$ =$T_{slack}-T_{reserve}$.
\item At the end of interval $i$, $(T_{eff}-t_i)$ shows the amount of slack remaining. Based on this, MSM Algorithm decides allowed \emph{maximum absolute slack} (MAS$_{i+1}$) for next interval $i+1$, e.g. if the remaining slack is 60$\mu$s, the Algorithm may choose to use MAS$_{i+1}$ as 2$\mu$s.
\item Then, the configurations having a slack greater than MAS$_{i+1}$ are rejected from $E$.  In effect, the configurations with number of active colors below a certain threshold color are rejected. We call this step as thresholding. 
\item If $E \neq \phi$, then the configuration from $E$ with minimum estimated energy is selected for the next interval $i+1$. 
\item If $E=\phi$  then the configuration closest to the threshold, viz. $(C_{\star}+L)$ is chosen for next interval. This is to avoid possible oscillations due to sudden change in working set size of the application. Since the algorithm aims to meet a global deadline, and not per-interval deadline; by feedback adjustment, it compensates for positive or negative deviations from the allowed slack. 
\end{enumerate}


\textbf{PSM Algorithm:} 

\begin{enumerate}

\item   If the total execution time at the end of interval $i$ is $T_{i}$, then $(T_{i}- t_{i})$ gives the estimate of baseline time till interval $i$. Using this, $\Delta _{i}$ is calculated as follows: 
\begin{equation}
\Delta _{i} = \dfrac{t_{i}\times 100}{(T_{i}- t_{i})}
\end{equation}
Clearly, $\Delta _{i}$ gives the estimate of percentage of extra time taken by the PSM Algorithm over the baseline. 
\item The PSM Algorithm always tries to conservatively keep  $\Delta _{i}$ below the actual allowed percentage slack ($\Upsilon$), by a small margin $\delta$ (0.3\% in this chapter). Thus, $\Delta _{i} \leq \Upsilon - \delta$.

\item Based on $\Delta _{i}$ and $\Upsilon$, Algorithm computes \emph{maximum percentage slack} over the baseline for $i+1$.  This is termed as  MPS$_{i+1}$ and represents the maximum percentage slack allowed in next interval. Then, to make performance aware choices, the configurations with estimated percentage slack greater than  MPS$_{i+1}$ are removed from $E$.  Thus, in effect, the configurations with number of active colors below a certain threshold color are rejected. We call this step as thresholding. 
\item If $E \neq \phi$, then the configuration from $E$ with minimum estimated energy is selected for the next interval $i+1$.
\item If $E=\phi$  then the configuration closest to threshold, viz. $(C_{\star}+L)$ is chosen for next interval. The reason for this is same as explained above. 

\end{enumerate}

  
We now explain the MSM algorithm with a simple example and PSM can be similarly understood. Assume $N$=64 and $L$=8 and in any interval, $C_{\star}$=28. Then, initially, $E=\{20,21...35,36\}$. If MAS$_{i+1}$ is such that the configurations with $C<20$ give an absolute slack value greater than MAS$_{i+1}$, then all configurations in $E$ pass thresholding step and the one with minimum energy is selected for next interval. However, if MAS$_{i+1}$ were such that configurations with  $C<40$ were to be removed, then after thresholding step, $E=\phi$. In such case, the Controller selects the configuration with $36$ (i.e. $C_{\star}+L$) active colors, which is the closest to threshold. In the next interval, $C_{\star}$ becomes 36 and then depending on MAS$_{i+2}$ and threshold-color, a suitable color value can be chosen.

\section{Energy Modeling}\label{sec:energymodel}
 We take into account the energy spent in L2 cache, main memory and the cost of executing the algorithm ($E_{Algo}$), since other components are minimally affected by our approach. Note that for baseline experiments, $E_{Algo}=0$.
\begin{equation}\label{eq:Etot}Energy= E_{L2}+E_{mem}+E_{Algo}\end{equation}
Here energy spent in L2 and memory is composed of both leakage and dynamic energy. 
 Further, we use the symbols $E^{dyn}_{XYZ}$ and $P^{leak}_{XYZ}$ to show the dynamic energy \emph{per access} and leakage energy \emph{per second}, respectively, spent in any component $XYZ$ (e.g. L2, memory, RCE).

To calculate L2 energy, we assume that an L2 miss consumes twice the energy as that of an L2 hit \cite{HanHri02_TVLSI,MitZha12_EnCache}. The leakage energy is proportional to active area of the cache \cite{MitZha12_EnCache,YanPow02_HybridCache}.  Thus,
\begin{equation}E_{L2}= E^{dyn}_{L2}\times(2M_{L2}+H_{L2}) + (P^{leak}_{L2}\times Time\times C)/N\end{equation}

Here $N$ shows the total number of colors and for any interval with $C$ active colors, $M_{L2}$ and $H_{L2}$ show the corresponding number of L2 misses and L2 hits respectively and $Time$ shows time consumed in the interval.  The L2 energy values are obtained using CACTI \cite{cacti_53} for 4-bank, 8-way caches at 45nm technology. For 2MB L2 cache, we get $E^{dyn}_{L2}$=0.985 nJ/access and  $P^{leak}_{L2}$=1.568 Watt. To account for the increased area due to use of gated-V$_{dd}$ technique, we assume 5\% higher value of $P^{leak}_{L2}$ for CASHIER, but not for baseline cache.


To calculate memory energy, we note that $ E^{dyn}_{mem}$=70 nJ and $ P^{leak}_{mem}$=0.18 Watt \cite{ZheLin09_DIMM,MitZha12_EnCache}. Using $A_{mem}$ to denote the number of memory accesses, we get,
\begin{equation}E_{mem}= E^{dyn}_{mem}\times A_{mem}+ P^{leak}_{mem}\times Time \end{equation}

Using $A_{RCE}$ to denote the number of RCE accesses and $E_{Tran}$ to denote  block-transition energy, $E_{Algo}$ is calculated as follows.
\begin{equation}\label{eq:Ealgo}E_{Algo}= E^{dyn}_{RCE}\times A_{RCE} +P^{leak}_{RCE}\times Time + E_{Tran}\end{equation}
To calculate the energy of RCE, we use CACTI. Since CACTI only provides values for power-of-two size caches, we take an upper bound of $S$ as $S=64Z/16R_S$ and estimate energy using CACTI for a single bank structure, with 8B block size (which is minimum data size allowed in CACTI). We only compute energy consumption of tag arrays, since RCE is a tag only structure. For an RCE corresponding to 2MB L2, we get $E^{dyn}_{RCE}$=0.004 nJ/access and  $P^{leak}_{RCE}$=0.007 Watt. Noting that, for every 64 L2 accesses, RCE is accessed only 6 times, we see that RCE energy consumption is a very small fraction of L2 cache energy consumption. Each block transition is assumed to take $0.002$ nJ. Using $Tran$ to denote the total number of blocks transitions, we get 
\begin{equation}E_{Tran}= 0.002\times Tran \text{  nJ}\end{equation}

\section{Energy Saving Results}\label{sec:cashier_energyresults}

We now present the experimental results. Notice that we have used much more strict deadlines than that used by the previous researchers (e.g. \cite{weissel2002process}).

\subsection{Magnitude Slack Method (MSM)} \label{sec:msm}
We test MSM algorithm by assigning slack values in two ways which are  as follows.  

\begin{figure*}[htbp]
 \centering
  \includegraphics [scale=0.50] {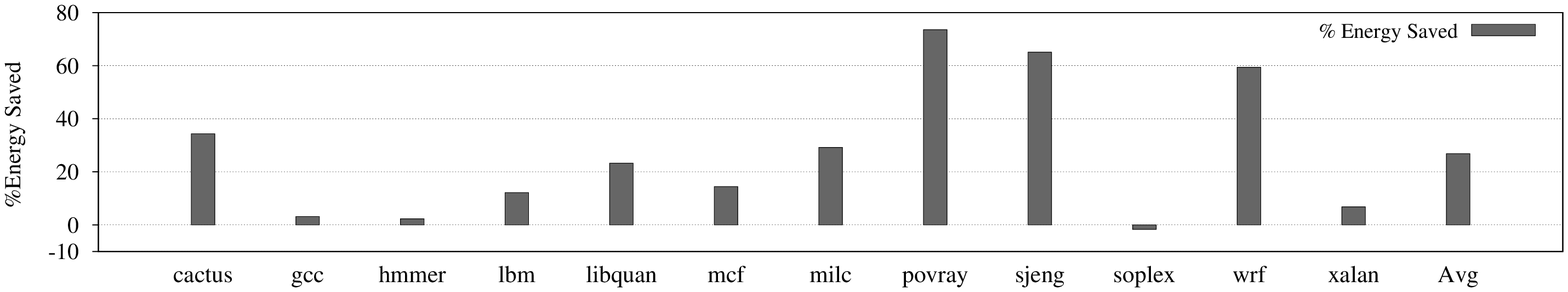}
  \includegraphics [scale=0.50] {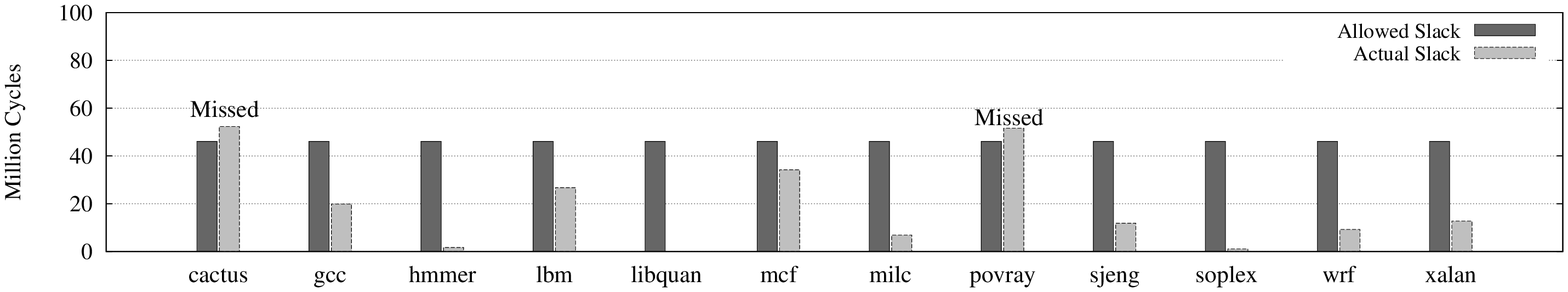}
 \isucaption{Results on Magnitude Slack Method with Uniform Slack Values:  Percentage Energy Saving and Simulation Cycle Increase (\textit{cactus} and \textit{povray} miss their deadlines)   }\label{fig:perfenergyloss_magnitudeuniform}
 \end{figure*}

\textbf{1. Uniform Slack Values: }  We take simulation cycles of baseline experiments of all the benchmarks and sort these values in ascending order. We then find two medians, take their mean, and set 5\% of this value as $T_{slack}$ for \textit{all} the benchmarks. In our experiments, this value was 46.08M cycles.   The results from this experiment are shown in Figure \ref{fig:perfenergyloss_magnitudeuniform}. We observe 26.8\% saving in energy, and two benchmarks (\textit{cactus} and \textit{povray}) miss their deadlines. The results on remaining metrics are presented in Table \ref{tab:cashier_resultsparameter}.

\begin{figure*}[htbp]
 \centering
  \includegraphics [scale=0.50] {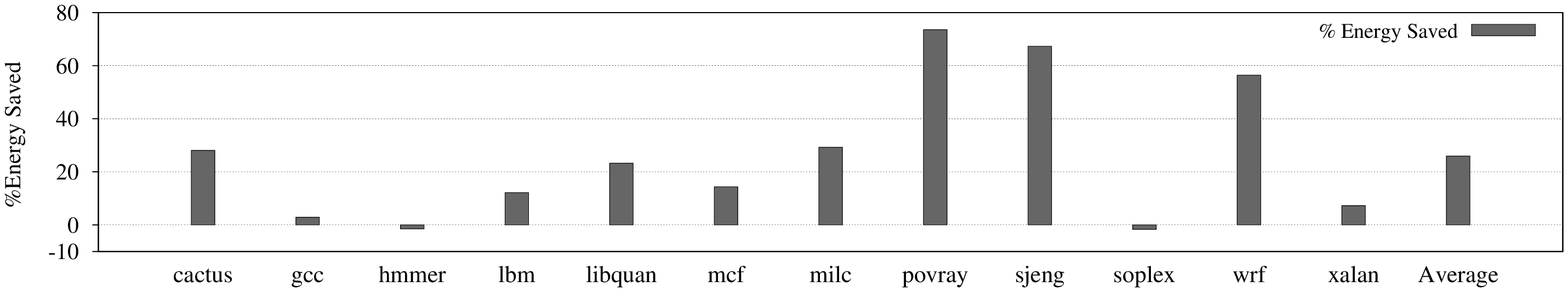}
  \includegraphics [scale=0.50] {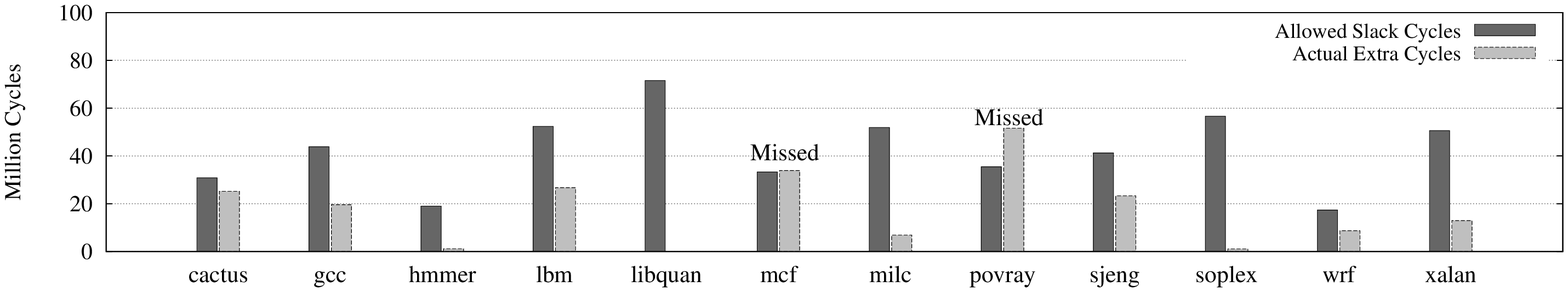}
 \isucaption{Results on Magnitude Slack Method with Different Slack Values: Percentage Energy Saving and Simulation Cycle Increase (\textit{mcf} and \textit{povray} miss their deadlines)   }\label{fig:perfenergyloss_magnitude}
 \end{figure*}

\textbf{2. Different Slack Values: } In this case, we assign different slack values to different benchmarks. For ensuring reasonably strict deadlines and evaluation, we need to randomly choose a slack  which is neither too high, nor too low. Hence, we proceed as follows. We first generated a list $P$ of 12 random numbers in the range of $[0,1]$, using an on-line random number generation utility \cite{random}. We then calculated $(4+p_i) \%$ value  of the baseline simulation cycles, where $p_i \in P$, $i= \{1,2..12\}$. This value is then set as the $T_{slack}$ for MSM algorithm for each of the 12 benchmarks.  Figure \ref{fig:perfenergyloss_magnitude} shows the results for this case. The average saving in energy over the baseline cache is 25.9\%, and two benchmarks, viz. \textit{mcf} and \textit{povray} miss their deadlines.  The values of remaining metrics are shown in Table \ref{tab:cashier_resultsparameter}.

\begin{table}[ht]\footnotesize
\isucaption{Percentage EDP saving, active ratio and MPKI increase}
\label{tab:cashier_resultsparameter}
\centering
\begin{tabular}{|c||c|c|c|}
\hline
Algorithm & EDP saving & Active Ratio & MPKI Increase \\\hline 
MSM, uniform-slack & 25.3\% & 36.5\%&0.44 \\\hline
MSM, different-slack & 24.6\% & 36.0\% & 0.42\\\hline
PSM, $\Upsilon$=5\% & 22.4\% & 45.2\% & 0.38  \\\hline
\end{tabular}	
\end{table}
          
 \subsection{Percentage Slack Method (PSM)}
\begin{figure*}[htbp]
 \centering
  \includegraphics [scale=0.52] {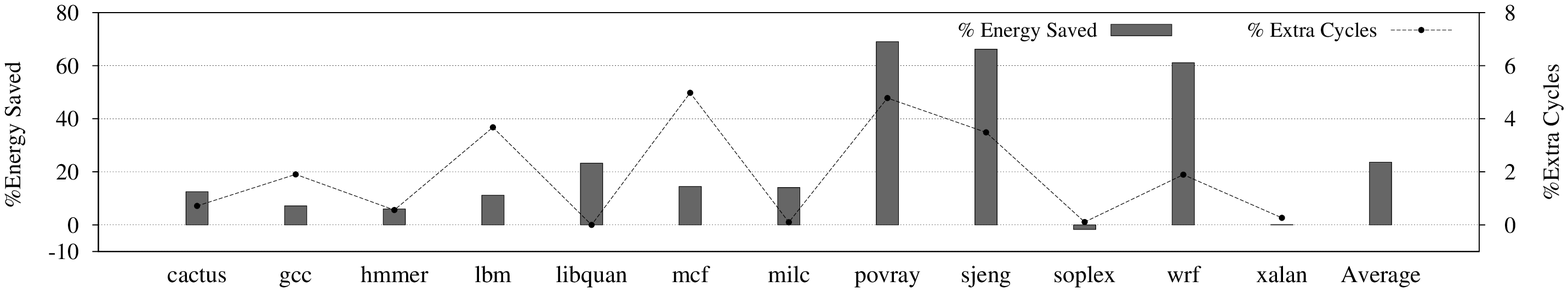}
 \isucaption{Results with Percentage Slack Method: Percentage Energy Saving and Percentage Simulation Cycle Increase for $\Upsilon=$ 5\% (No benchmark misses the deadline) }\label{fig:perfenergyloss}
 \end{figure*}

We tested PSM for percentage slack $\Upsilon =5$\%. Figure~\ref{fig:perfenergyloss} shows the results. The average saving in energy is 23.6\% and none of the benchmark misses its deadline. The values of remaining metrics are shown in Table \ref{tab:resultsparameter}.

\textbf{Discussion: } For both the cases, the MSM algorithm turns off nearly 64\% of the cache and increases L2 misses by less than 0.45 MPKI. Intuitively, the energy saved should increase with the fraction of cache which is turned off and this is confirmed by the results presented in Table \ref{tab:resultsparameter}. The saving in EDP is nearly 25\% and thus, CASHIER keeps a fine balance between performance loss and energy saving. PSM algorithm turns off nearly 55\% of the L2 cache and increases L2 misses by 0.38 MPKI. Clearly, compared to the execution with MSM algorithm, PSM turns off less fraction of cache and as expected, it saves less amount of energy. Still the saving in energy and EDP are significant. These results confirm the effectiveness of CASHIER in saving energy in QoS systems.

\subsection{Parameter Sensitivity Study}\label{sec:parameter}
We now study the sensitivity of CASHIER towards changes in different parameters. For sake of brevity, we omit per-benchmark figures and only present results on percentage energy saving and specify the benchmarks which meet their deadlines. We first test MSM, as explained in Section \ref{sec:msm} above, but this time with $(4+q_i) \%$ of baseline simulation cycles, where $q_i \in Q$, $i= \{1,2..12\}$ and $Q$ is another randomly generated list \cite{random}. We obtain average energy saving of 25.8\% and two benchmarks (\textit{mcf} and \textit{povray}) miss their deadlines.

We then test PSM with $\Upsilon=3$\%. We observe that the average saving in energy is 22.4\% and two benchmarks (\textit{lbm} and \textit{mcf}) miss their deadlines. Further, on testing with $\Upsilon=7$\%, we observe that the average energy saving is 25.0\% and no benchmark misses its deadline. Clearly, CASHIER can adapt itself to save extra amount of energy, for the case when the deadlines are more relaxed.   

Finally, we change the interval length from 5M to 10M instructions and assign slack values as shown in Section \ref{sec:msm}. A larger value of interval length is likely to reduce the overhead of algorithm execution. We observed that for MSM algorithm with uniform slack, 25.0\% energy is saved and only \textit{cactus} misses its deadline. For MSM algorithm with different slack values, 25.2\% energy is saved and only \textit{cactus} misses its deadline. For PSM algorithm 23.8\% energy is saved and no benchmark misses its deadline. Comparing the case of 10M interval size with that of 5M interval size, we see that for MSM algorithm, energy saving is slightly reduced and the number of benchmarks with missed deadlines is also reduced. This can be attributed to reduced reconfiguration overhead with larger interval size.   


\section{Conclusion}\label{sec:cashier_conclusion}
Recent trends of CMOS scaling and increasing cache sizes have made managing the leakage energy consumption of LLC extremely crucial. We have presented Cashier, a cache energy saving approach for QoS systems. Cashier uses dynamic profiling and  reconfiguration to optimize for memory subsystem energy. The experimental results have shown that Cashier intelligently adapts itself according to the available slack to maximize energy saving and outperforms conventional deadline-unaware energy saving techniques.

\chapter{MASTER: A CACHE ENERGY SAVING APPROACH FOR MULTICORE SYSTEMS} \label{chap:master}

\section{Introduction}
In this chapter, we present \textbf{MASTER}, a \underline{m}ulticore c\underline{a}che
energy \underline{s}aving \underline{te}chnique using dynamic cache
\underline{r}econfiguration. Power consumption has been identified as a major threat for future multicore scaling \cite{esmaeilzadeh2011dark} and hence, cache energy saving techniques are extremely important for multicore systems. With increasing number of
cores integrated on a single chip \cite{refIBMPower7,kurd2010westmere}, the
pressure on the memory system is rising and to mitigate this pressure, modern
processors are using large sized LLCs; for example, Intel's 32nm, 8-core Poulson
processor uses 32MB of LLC \cite{riedlinger201132nm}. Further, with each CMOS
technology generation,  leakage energy consumption has been increasing
exponentially
\cite{semiconductor2011international,
rodriguez2006energy} and hence, large LLCs  contribute significantly to the total
processor power consumption \cite{monchiero2008power}. The increased levels of
power consumption necessitate expensive cooling solutions which significantly
increase the overall system cost and design complexity and also restrict
further performance scaling. Further, in several scenarios, the actual number of programs running on a multicore processor are much less than the number of cores and thus, a large amount of cache leakage energy is wasted.  For these reasons, managing the power consumption
of LLCs has become an important research issue in modern processor design.

The conventional cache energy saving techniques face significant challenges when
used for managing energy consumption of shared LLCs in multicore processors. For example, the techniques such
as decay cache \cite{KaxHuz01_CacheDecay} exploit the locality property of memory access
streams and place the `dead' cache lines into low leakage mode for saving
leakage energy. Since single-core workloads typically exhibit high locality,
these techniques are effective in saving energy in single-core systems. However,
in the case of multicore systems with shared LLCs, the independent access
streams from multiple applications are interleaved and thus, the actual memory
access stream exhibits reduced locality.  The techniques which allocate and turn-off cache at way
granularity
\cite{Albonesi99_Selective,Karthik_HPCA12,BarCom2008_Way,wang2011dynamic,
kotera2011power} can only provide few coarse grain partitions (at most, as many
as the number of ways) while drastically reducing the associativity for each
program. Finally, some techniques use offline
profiling or compiler analysis of running applications for saving energy
\cite{Albonesi99_Selective,jiang2011access,YanPow01_IcacheResize,reddy2010cache,
ZhaVah03_Configurable,zhang2002compiler}; however, due to the large number of
possible program combinations in multicore environment, use of offline profiling
becomes increasingly difficult.

 MASTER works by periodically allocating suitable
amount of LLC space to each running application and turning off unused LLC space
to save cache energy. MASTER uses a simple ``cache coloring'' scheme and thus, 
allocates cache at the granularity of a single cache color (Section
\ref{sec:methodology}). For profiling the behavior of running programs under
different LLC cache sizes, MASTER uses a small microarchitectural component,
called ``reconfigurable cache emulator'' (RCE). RCE is a tag-only (data-less)
component and is designed using the set sampling method.  RCE does not lie on
critical access path and because of its small size, its access latency is easily
hidden.  With this lightweight hardware support, MASTER energy saving algorithm
periodically predicts the memory subsystem  energy of running programs for a
small number of color values. Using these estimates, MASTER selects a
configuration with minimum estimated energy and turns off the unused cache
colors for saving leakage energy (Section \ref{sec:EnergySavingAlgorithm}).  For
hardware implementation of cache block switching (i.e. turning-off), MASTER uses
the well-known  gated V$_{\text{dd}}$ technique \cite{PowSeh00_GatedVdd}
(Section \ref{sec:hardwareimplementation}).

We evaluate MASTER using out-of-order simulations with Sniper
\cite{CarHei2011_Sniper}, a state-of-art x86-64 simulator  and multi-programmed workloads from SPEC2006 suite (Section
\ref{experimentResult}). We compare it to decay cache technique (DCT)
\cite{KaxHuz01_CacheDecay} and way adaptable cache technique (WAC)
\cite{BarCom2008_Way}.  The results show that MASTER saves highest amount of
memory subsystem energy (Section~\ref{sec:master_results}). For example, over a shared
baseline LLC, for 2 and 4-core systems (with one program on each core), the average savings in memory subsystem energy by using MASTER are 14.7\% and 11.2\%, respectively.  Using WAC (which, on average, performs better than DCT), these values are only 10.2\% and 6.5\% respectively. Further,  the average value
of weighted speedup and fair speedup using MASTER remain very close to
one ($\ge$0.98) and absolute increase in DRAM APKI (accesses per kilo instructions)
remains less than 0.5. Thus, MASTER does not harm performance or cause
unfairness. Additional
simulation results show that MASTER works well for a wide variety of system parameters.

\section{Background and Related Work}\label{sec:literatureReview}
The energy saving approach of MASTER has two broad steps. In the first
step, the LLC quotas to be allocated to different cores (and to be turned off)
are decided and then these quotas are actually enforced. In the second step,
a leakage control mechanism is used to turn off the cache blocks for saving
energy. In literature, different schemes have been proposed which allocate or
turnoff  cache space at the granularity of  cache colors
\cite{LinLuq08_hpca,LinLuq09_softwarecp,MitZha13_Cashier}, cache ways
\cite{Albonesi99_Selective,Karthik_HPCA12,BarCom2008_Way,QurPat06_UtilityBasedCP,kkedzierski2010power,wang2011dynamic}, cache sets
\cite{PowSeh00_GatedVdd,RanAdv00_Recon,YanPow01_IcacheResize}, both sets and ways (hybrid)
\cite{MitZha12_EnCache,YanPow02_HybridCache} and cache blocks
\cite{KaxHuz01_CacheDecay,FlaKim02_DrowsyCache}. MASTER determines cache quotas
with the goal of optimizing energy efficiency  and enforces it using a cache
coloring scheme.   

The circuit-level leakage control  mechanisms are divided into two types, namely state-destroying \cite{PowSeh00_GatedVdd} and state-preserving
\cite{FlaKim02_DrowsyCache,HanHri02_TVLSI}.  The state-destroying
mechanisms do not retain data in low-leakage mode and hence, access to such a
block incurs a cache miss; however, these mechanisms typically reduce more
leakage power than the state-preserving mechanisms
\cite{LiyPar04_Skadron,PowSeh00_GatedVdd,FlaKim02_DrowsyCache}. The
state-preserving mechanisms retain data in low leakage mode but generally require two supply voltages for each block and also make the cache more
susceptible to noise \cite{ayala2007energy,FlaKim02_DrowsyCache}. Hence, MASTER employs state-destroying leakage control by using gated
V$_{\text{dd}}$ mechanism \cite{PowSeh00_GatedVdd}.
 

Recently, researchers have proposed techniques for saving both leakage and
dynamic energy in caches. With no leakage optimization applied, LLCs spend a
large fraction of their energy in the form of leakage energy
\cite{homayoun2008adaptive,li2002leakage}. Hence, we aim at saving cache leakage energy. Some energy saving techniques work by
statically allocating or turning off a part of cache and do not allow dynamic
runtime reconfiguration
\cite{wang2011dynamic,Albonesi99_Selective,jiang2011access}. However, since the
behavior of applications varies significantly over their execution length,
dynamic cache reconfiguration is important for realizing large energy savings.

 An important difference between MASTER and most existing cache energy saving
techniques (e.g.
\cite{Karthik_HPCA12,PowSeh00_GatedVdd,YanPow02_HybridCache,KaxHuz01_CacheDecay,BarCom2008_Way,kkedzierski2010power})
is that MASTER works to \textit{directly} optimize energy value, while existing
techniques do not directly work to optimize cache energy, rather they aim to
keep the increase in cache misses resulting from cache turnoff small, which
\textit{leads to} energy saving. Due to this feature, MASTER can optimize for
system (or subsystem) energy, instead of only cache energy. Several cache energy
saving techniques proposed in literature (e.g.
\cite{Karthik_HPCA12,reddy2010cache,DroBuy02_Accounting,jiang2011access,
udipi2009non,kotera2011power,li2002leakage}) have been evaluated by considering their effect
on LLC energy  only.  We model both LLC energy
and main memory energy for a more  comprehensive evaluation.



%

\section{System Architecture and Design}\label{sec:methodology}
\begin{figure*}[htbp]
\centering
 \includegraphics [scale=0.53] {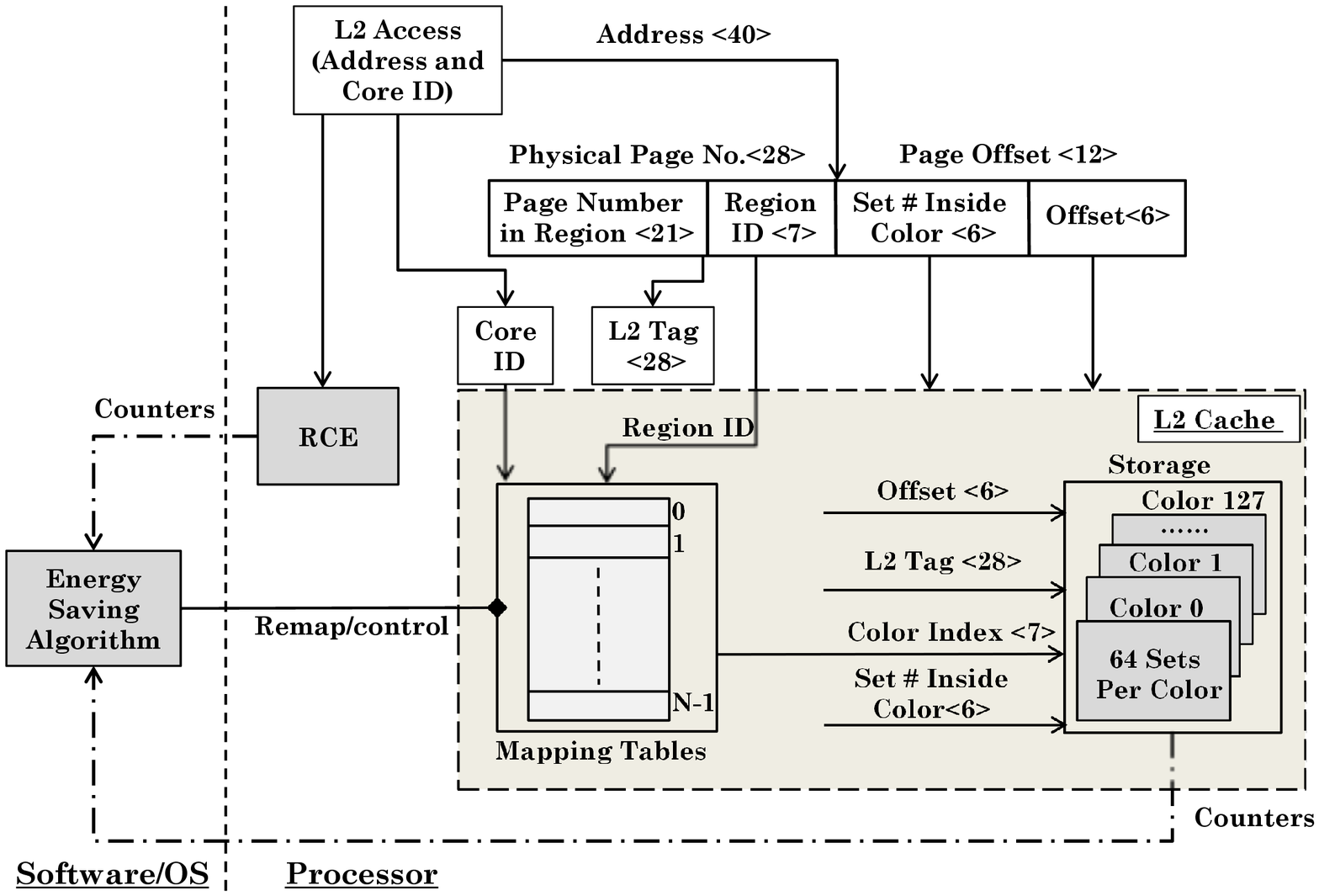}
\isucaption{Flow diagram of MASTER approach (Assuming $M$ = 128, page size = 4KB, cache block size = 64)
}\label{fig:block}
\end{figure*}

MASTER works on the idea that different programs and even different execution
phases of a single program have different working set sizes and hence, by
allocating just suitable amount of cache to the programs, the rest of the cache
can be turned off, with little impact on performance. Figure \ref{fig:block}
shows the flow-diagram of MASTER. In the following, we explain each of the
components of MASTER in more detail.

\textbf{Notations and Assumptions: } We use $N$ to denote the number of cores
and $n$ or $k$ to show core indices. The interval index is shown using $i$. The
maximum number of cache colors is shown as $M$. System page size is taken as 4KB
and all caches use a block size of 64B. The terms ``active''  and ``turned off'' are used
to refer to the cache space (either cache block, color or way), which is in normal
leakage and low leakage mode, respectively. The term ``color value'' denotes the number of colors given to each core and ``configuration'' denotes the
colors given to all the $N$ cores, e.g. a 2-core configuration
$\{37,65\}$ specifies that color values of core 0 and core 1 are 37 and 65,
respectively.  We assume that the LLC is an L2 cache; and the discussion can be extended to the case where LLC is an L3
cache. The baseline cache is taken as shared LLC, as done in several recent works
\cite{LinLuq08_hpca,SanKoz11_Vantage,Karthik_HPCA12}.

\subsection{Cache Coloring Scheme} \label{sec:master_coloring}
For selective cache allocation, MASTER uses cache
coloring scheme \cite{KesHil92_PageColoring,LinLuq08_hpca,LinLuq09_softwarecp},
which is as follows. First, we logically divide the cache into $M$ disjoint
groups, called cache colors, where total number of colors ($M$) is given by
\begin{equation}\label{eq:M}
M =\dfrac{\text{L2CacheSize}} {\text{PageSize} \times \text{L2Associativity}}
\end{equation}

Further, we logically divide the physical pages into disjoint groups, called
\textit{memory regions}. For each core, the number of memory regions is $M$. Thus, a memory region denotes the group of physical pages of a core that share
$\log_2(M)$ least significant bits of the physical page number. A cache color is
given to one or more memory regions of a \textit{single} core and thus, all
physical pages in those memory regions are mapped to the same cache color. For
each core, we use a small \textit{mapping table} of $M$ entries, each
$\log_2(M)$-bit wide, which stores the mapping of memory regions to cache
colors. At any instance, if the number of colors allocated to core $n$ is $c_n$,
then  the mapping table of core $n$ stores the mapping of its $M$ regions to
$c_n$ colors. Thus, cache quotas are enforced by mapping all the memory regions
of a core to only its allocated cache colors. Further, when quota allocation is
such that the sum of allocated colors is less than $M$, the remaining colors
become unused which can be turned off for saving leakage energy. MASTER turns off 
both tag and data arrays of the unused colors, in contrast with some
techniques which only turn off data array and always keep the tag fields active \cite{AbeGon05_IATAC,ZhoTob03_AMC}. 

 Using mapping tables, computation of cache index (set) is done as follows
 (Figure \ref{fig:block}). For any L2 access from core $n$, its memory region ID
 is computed by simple bit-masking. Using memory region ID, the cache color is
 read from the mapping table of core $n$ and the set number inside the color is
 decided by the most significant bits of the page offset.     
 
While previous set level allocation techniques
\cite{PowSeh00_GatedVdd,YanPow02_HybridCache,MitZha12_EnCache} reconfigured the
cache only to power-of-two set-counts, MASTER allocates and turns off cache at
the granularity of a single cache color and hence it reconfigures the cache to
\emph{non}-power-of-two set-counts also; for example, at an instance, it may
keep only 37 colors as active. From Eq. \ref{eq:M}, we find that an 8-way
4MB cache has 128 colors. Thus, with merely an 8-way cache, MASTER provides much
finer granularity of cache allocation than the previous set, way or hybrid (set
and way) level allocation  techniques 
\cite{PowSeh00_GatedVdd,MitZha12_EnCache,Karthik_HPCA12,QurPat06_UtilityBasedCP,YanPow02_HybridCache}.  


Lin et al. \cite{LinLuq08_hpca} present a coloring scheme which does not require hardware
support and can control mapping of every OS page individually. In contrast,
MASTER uses lightweight hardware support and can control the address mapping
only at the level of a memory region which contains multiple pages.
However, the limitation of their scheme is that repartitioning incurs
significant overhead since the data of whole virtual page needs to be copied
from an old physical page to a new physical page. Since MASTER uses mapping
table to add a layer of mapping between physical pages and cache colors, it
avoids the need of page migration and also keeps the reconfiguration overhead
small. Further, as shown in Section \ref{sec:hardwareimplementation}, the overhead of mapping tables is extremely small.  


\subsection{Reconfigurable Cache Emulator
(RCE)}\label{sec:reconfigurablecacheemulator}

For estimating program energy consumption under different color values, the
number of cache misses under them needs to be estimated. A
challenge in obtaining profiling data for color (or set) level  allocation is
that, unlike for way level allocation \cite{QurPat06_UtilityBasedCP}, a single auxiliary
tag structure cannot provide profiling information for different cache
sizes (Note that since MASTER does not dynamically reconfigure
associativity or block size, change in cache size simply means change in the
set-count.). Hence, to estimate performance at multiple cache sizes, these cache
sizes need to be individually profiled. However, since caches have a large
number of colors, profiling for each possible color value would be extremely
costly.
 
To address this issue, MASTER uses RCE, which profiles only a few selected cache
sizes (called profiling points) and uses piecewise linear interpolation to
estimate miss rates for other cache sizes. In this chapter, we use seven profiling
points, each denoted by $(2^{j-1}X)/64$, where $j=\{1,2,3,4,5,6,7\}$ and  $X$
denotes the L2 cache size (or equivalently number of L2 colors). Corresponding
to each profiling point, MASTER uses an auxiliary tag structure, called
\textit{profiling unit} for each core. To keep the overhead of profiling units
small, MASTER leverages ``set sampling'' approach \cite{puzaksampling}. The ratio
of set-counts of L2 and that of a profiling unit is called sampling ratio
($R_s$). 

\begin{figure}[htp]
 \centering  
  \includegraphics [scale=0.56] {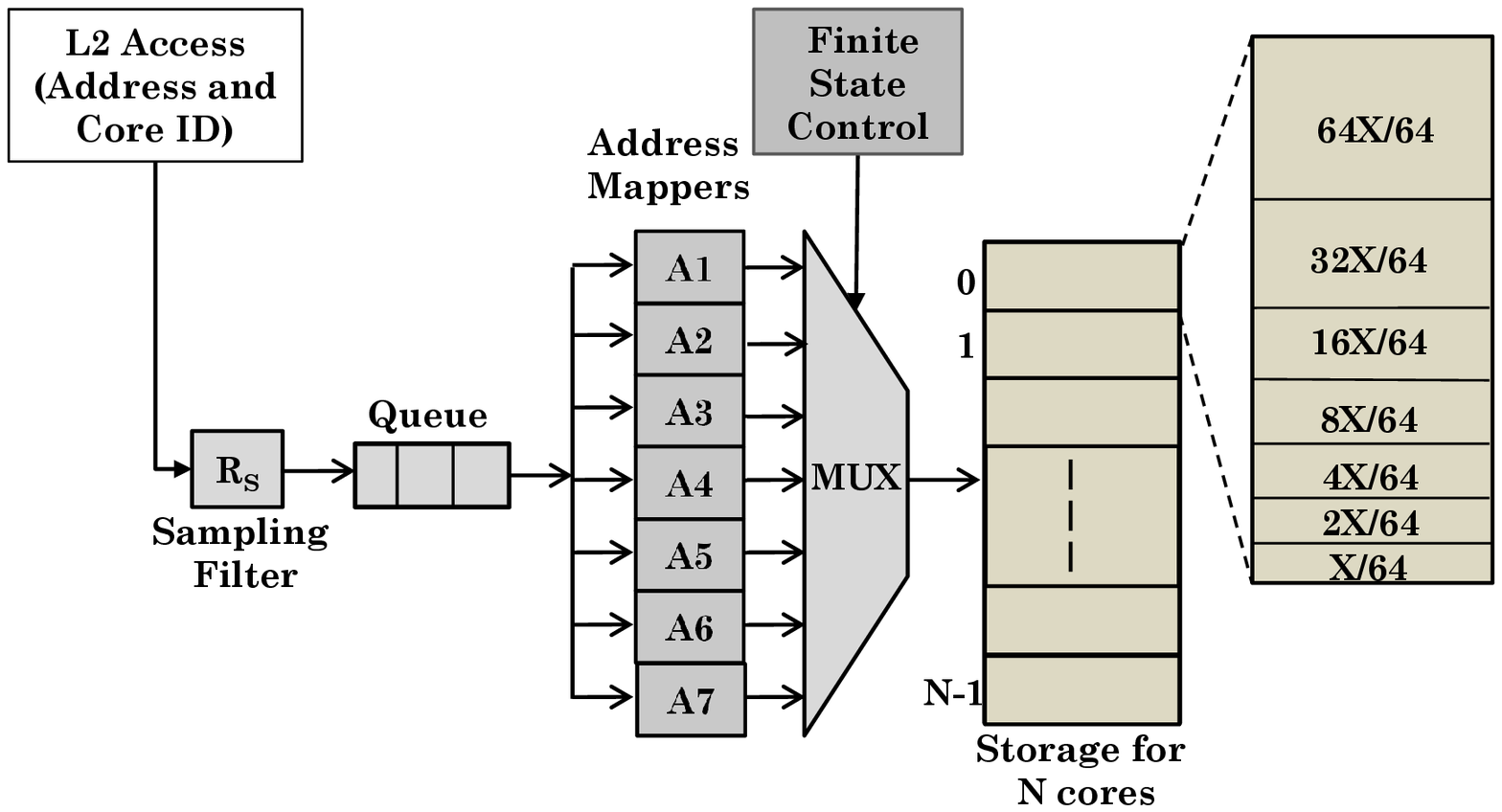}  
 \isucaption{RCE design (Assuming 64 or more
colors)}\label{fig:rcedesign}
 \end{figure}
The RCE works as follows (Figure \ref{fig:rcedesign}). Any L2 access address,
originating from a core (say $n$) is sampled by a sampling filter which removes
block offset bits and uses bit-matching to decide whether the address passes
the filter. An address which passes the filter is further passed through a queue. Then, each 
address mapper (shown as A1 to A7) computes cache tag and
set using traditional set-decoding (and not cache coloring). Also, to map the address to suitable region in the storage, it adds an offset corresponding to its profiling unit and core index of the address.  Afterwards, using a small
multiplexer (MUX), the incoming addresses are sequentially fed to the tag-only
storage region for emulating cache access. 

We now compute the size of RCE. Let $Q$ denote the number of sets in L2 and $S$
denote the number of sets in RCE for all the cores. Further, let $G$ and $L$
denote the size of tag and block size in bits, respectively and $F_{\text{RCE}}$
denote the total size of RCE as a percentage of L2 size. Thus, we get
\begin{align}
S &= \dfrac{(\sum\nolimits_{j=1}^7 2^{j-1})\times N\times Q}{64\times R_s} =  
\dfrac{127NQ}{64R_s} \le \dfrac{2NQ}{R_s} \label{eq:setprofiling}\\
F_{RCE} &= \dfrac{\text{RCESize}}{\text{L2CacheSize}}\times 100=\dfrac{N\times
127G}{64R_s(L+G)}\times 100
\end{align}

In our experiments $R_s$ = 64, $G$ = 28, $L$ = 64$\times$8 and hence, for 2 and 4-core systems, we get $F_{\text{RCE}}$ as 0.3\% and 0.6\%, respectively. To cross-check, we have computed areas of RCE and L2 using CACTI
\cite{cacti_65} for the cache sizes chosen in our experiments (see Section
\ref{sec:Simulation} and \ref{sec:EnergyModel})  and have found values of
$F_{\text{RCE}}$ in the same range. Taking into account both RCE and mapping
tables, we conservatively assume the maximum storage overhead of MASTER as 0.8\%
of L2 which occurs for 4 core systems. Clearly, the overhead of MASTER is small.
The RCE overhead can be further reduced by half by taking the sampling ratio as
128, although it leads to slight reduction in the energy saving achieved (Section
\ref{sec:sensitivity}).

Note that RCE works in parallel to L2 and does not lie at the critical access
path and does not store or communicate data. A miss in RCE does not generate any
request for other caches. Each address mapper is simple, since it only performs  bit-matching and additions. For each sampled address from core $n$, the RCE
storage of core $n$ is accessed seven times. However, due to the use of queue, 
large value of $R_s$ and dataless operation of RCE, no congestion occurs, even
in the case of bursty L2 accesses. RCE design is flexible and can be easily
extended to also profile for sizes such as $X/128$ and $X/256$, although this
also increases the number of profiling units consulted in each RCE access.

\subsection{Marginal Color Utility (MCU) }\label{sec:marginalgain}
In each interval, MASTER computes marginal color utility values which are used
by the energy saving algorithm (Section \ref{sec:EnergySavingAlgorithm}). The
notion of marginal gain has been previously used
\cite{QurPat06_UtilityBasedCP,marginalGain}. In context of MASTER which
uses cache coloring and RCE, we define MCU for the non-uniformly spaced
profiling points for which miss-rate information is available using RCE and use
the unit as a single cache color.  

For each core $n$, at any color value $c_n$, the value of MCU,
$\text{MCU}_n(c_n)$, is defined as the reduction in cache misses per extra unit
cache color.  We assume that between two profiling points, the  number of misses
vary linearly with cache size (piecewise linear approximation) and hence, MCU
remains constant between those profiling points. Let $C_{p}^1=X/64$,
$C_{p}^2=2X/64$ $\ldots$ $C_{p}^7=64X/64$ denote the seven profiling points as
mentioned above. Then, if the number of L2 misses of core $n$ at these profiling
points is denoted by $\text{Miss}_n(C_{p}^j)$ (where $j=\{1,2,3,4,5,6,7\}$ and
$n=\{0,1,\ldots, N-1\}$), then for $C_{p}^{1} \le c_n \le C_{p}^{7}$, 
$\text{MCU}_n(c_n)$ is defined   as follows.     
     \begin{equation}
\text{MCU}_n(c_n) = \begin{cases}
\dfrac{ \text{Miss}_n(C_{p}^j)-\text{Miss}_n(C_{p}^{j+1})}{C_{p}^{j+1}-C_{p}^j} 
& C_{p}^{j} \le c_n < C_{p}^{j+1}\\
\dfrac{ \text{Miss}_n(C_{p}^6)-\text{Miss}_n(C_{p}^{7})}{C_{p}^{7}-C_{p}^6} &
c_n=C_p^7\\
\end{cases}
\end{equation}

 \section{Energy Saving Algorithm (ESA)}\label{sec:EnergySavingAlgorithm}
We now discuss the energy saving algorithm of MASTER which runs after a fixed
interval length (e.g. 5M cycles) and can be a kernel module. Since the future
values are unknown, the algorithm works by using the observed values from
interval $i$ to make predictions about interval $i+1$. Without loss of
generality, we assume single-threaded workloads and hence, use the words `core'
and `application' interchangeably. We discuss generic values of parameters,
along with their specific values for 2 and 4 cores systems. Let $c_n(i)$ denote the color value of core $n$ in interval $i$. The algorithm
has the following two steps.

\textbf{1. Selection of color values: } For each core, ESA intelligently selects $T_{max}$ (= 4 in our experiments) possible
color values. Let ConfigSpace[$n$] be the set of these color values. The selection of color values is done using following criteria. 

\textbf{A.} To avoid application starvation,
ESA allocates at least $Min$ (=$M/64$ in our experiments) colors to
each core. Such color values are termed as `valid' color values.
 
\textbf{B.} To keep
the reconfiguration overhead low and avoid oscillations; `valid' color values
are searched only in close vicinity of $c_n(i)$ (i.e. $c_n(i)\pm 10$).
  
\textbf{C.} Based on intuitive observation, if an application has low MCU, then
reducing its cache allocation does not significantly increase its miss rate but
provides opportunities of turning off the cache or allocating the cache to other
cores. Thus, for applications with low MCU, the color values having smaller
number of active colors are likely to be energy efficient and vice-versa. To
quantify smallness or largeness of MCU, we use four
\textit{application-independent} thresholds, viz. $\lambda_q$ ($q$ = 1, 2, 3,
4), which are heuristically taken as 50, 200, 300 and 1000, respectively in our experiments.
By comparing $\text{MCU}_n(c_n)$ to the threshold values, its range is decided
and thus, the color values for core $n$ are chosen. For example, if for core 3,
$\text{MCU}_3(c_3)$ equals 250 ($\lambda_2 < \text{MCU}_3(c_3)\le \lambda_3$),
then ConfigSpace[$3$] equals $\{c_3-1, c_3, c_3+4, c_3+6\}$, assuming all the color values are
all valid (if not, the invalid color value is replaced by a valid one).   

\textbf{D.} For each of the $T_{max}$ color values in ConfigSpace[$n$], the
contribution of core $n$ in memory subsystem energy is estimated (see next
paragraph). Then, out of these $T_{max}$ color values, $T$ color values with
least energy are selected for each core and the other color values are
discarded. In our experiments, for $N$ = 2, $T$ = $T_{max}$ = 4 and for $N$
$=$ 4, $T$ = 2. Note that for $N$ = 2, $T$ = $T_{max}$ and hence, this step is not required for the 2-core system. For $N$ = 4, the energy computations are done for a maximum of $NT_{max}$ (=16) color values.

For a color value $c_n$, the contribution of core $n$ in memory subsystem energy
is estimated using equation \ref{eq:totalenergy} (Section \ref{sec:EnergyModel}) in the following manner. Since
we are only interested in \textit{comparing} energy for different color values,
and not in their actual magnitudes, we ignore the quantities which are common.
L2 dynamic energy depends on number of L2 misses and hits at $c_n$, which are
estimated using RCE. For a fixed interval length, time consumed is fixed and
hence L2 leakage energy only depends on the active fraction of cache, which is
equal to $c_n/M$. DRAM dynamic energy depends on DRAM accesses and hence on L2
misses and writebacks. L2 miss estimates are already available. The number of
writebacks are assumed to be same for different color values and hence are
ignored. This assumption has only small affect on estimation accuracy since most
applications bring only small number of dirty blocks in L2, not all of which are
expected to be evicted.

\textbf{2. Selection of $N$-core configurations: } ESA now generates all
possible combinations of $N$-core configuration, using color values from
ConfigSpace[$n$] of all $N$ cores.  Out of
these, the configurations with sum of active colors greater than $M$ are
discarded. Depending on the number of remaining configurations, ESA chooses one of
the following steps.
    
\textbf{A.} For the remaining configurations, memory subsystem energy is computed
(procedure is same as above, except that now it is for $N$-core configuration
and not just for a  single core) and the configuration  with minimum energy
(call it $C_{min}$) is selected. Memory subsystem energy for current
configuration (call it $C_{now}$) is also computed. If compared to $C_{now}$,
$C_{min}$ improves energy by at least 0.3\% (chosen arbitrarily), $C_{min}$ is
chosen for the next interval. Otherwise $C_{now}$ is taken for $i+1$.  
 
\textbf{B.} If no configuration remains, $C_{now}$ is taken for $i+1$.

The maximum number of configurations tested is $(T)^N+1$. From above, for both $N$ = 2 and 4, maximum number of configurations tested are always 16+1 = 17.  

 
\textbf{Discussion: } In each execution, ESA only
examines a maximum of 16 color values and 17 configurations and hence, the overhead of ESA is
small. Also, by using MCU values, ESA makes an intelligent prediction about the
configurations which are likely to be most energy efficient. The threshold
values chosen are application-independent and hence, do not require
per-application tuning. As can be seen from the results (Section
\ref{sec:master_results}), our chosen values provide significant energy saving for
almost all the workloads and a designer can further exercise trade-off between
algorithm efficiency and energy saving obtained by choosing a proper value of $T_{max}$,  $T$ and the interval length.  Algorithm implementation is further discussed in Sections
\ref{sec:hardwareimplementation} and \ref{sec:master_compare}.



\section{Implementation} \label{sec:hardwareimplementation}
\textbf{Cache block switching: } For hardware implementation of cache block
switching, MASTER uses gated V$_{\text{dd}}$
scheme \cite{PowSeh00_GatedVdd} which has also been used by several researchers
\cite{KaxHuz01_CacheDecay,HanHri02_TVLSI,LiyPar04_Skadron,Karthik_HPCA12}.  We use a specific
implementation of gated V$_{\text{dd}}$ ( NMOS gated V$_{\text{dd}}$, dual Vt,
wide, with charge pump) which reduces leakage energy by 97\% and results in 5\%
area penalty and 8\% access latency penalty \cite{PowSeh00_GatedVdd}. We account
for these overheads below and in Section \ref{experimentResult}. Also note that mechanism to turn off a subset of LLC is already
provided by the existing commercial processor chips
\cite{kurd2010westmere,naveh2006power}. 


\textbf{Effect on cache access time: } With MASTER, block switching only happens
at the end of an interval and RCE is accessed in parallel to L2 and
hence, these activities do not happen on the critical path. Further, MASTER does
not require use of caches of large associativity which have higher access time
and dynamic energy. 
Hence, the impact of MASTER on cache access time comes due to access to mapping
table and use of gated V$_{\text{dd}}$ scheme. To see the maximum overhead of mapping
tables, which in our experiments, occurs for 4-core system; we take the example
of an 8-way, 8MB cache which has 256 colors. Thus, the total size of mapping
tables of all cores is 8192 bits ($=4\times 256\times 8 $), merely 0.012\% of L2 cache size (tag+data) and hence their access latency and energy consumption are negligible.  
Since mapping tables are changed only during cache reconfigurations, access to them can be folded
into the address decode tree of the cache's tag and data arrays. The gated
V$_{\text{dd}}$  scheme increases  access latency by 8\%. With baseline L2
latency as 12 cycles, we take the L2 latency with MASTER as 13 cycles (Section
\ref{sec:Simulation}).
    
\textbf{Counters: } MASTER uses counters for RCE (recording number of misses in
each profiling point, MCUs etc.) and ESA (recording color values, configurations
and their energy values etc.). Since the energy consumption of counters is much
smaller than that of memory subsystem (LLC+DRAM) and several processors already
have counters for operating system or performance measurement
\cite{KaxHuz01_CacheDecay}, we ignore the overhead of counters in energy
calculations. Also note that MASTER does not require tracking the
application-ownership of each cache block or altering the replacement policy
(unlike \cite{QurPat06_UtilityBasedCP,Karthik_HPCA12}). MASTER works independent
of the replacement policy used (see Section \ref{sec:sensitivity}) and hence,
does not require using a specific replacement policy such as true-LRU which has
higher implementation overhead than the ``approximate LRU'' schemes
\cite{al2004performance}.  Further, MASTER does not require using per-block
counters to monitor cache access intensity (unlike
\cite{FlaKim02_DrowsyCache,KaxHuz01_CacheDecay}) or tables for offline profiling
(unlike \cite{reddy2010cache,wang2011dynamic}).

\textbf{Handling reconfigurations: } L2 reconfigurations are handled in the
following manner. When a color (say $c_n$) is `allocated' to a core (say $n$),
one or more regions of core $n$, which were mapped to some other color, are now
mapped to the color $c_n$ and the blocks of remapped region in the old color are
flushed (i.e. dirty data is written back to memory and other blocks are
discarded). Conversely, when a color (say $c_k$) is `taken away' from a core
(say $k$), the blocks of core $k$ in cache color $c_k$ are flushed  and then,
the regions of core $k$, which were mapped to $c_k$, are now mapped to some
other color(s) of core $k$. Change in mapping is accomplished by using the
mapping table (Section \ref{sec:coloring}). The time taken in running the
algorithm is accounted in Section \ref{sec:master_compare}.

 The existing set level allocation schemes turn off cache at power-of-two set
counts \cite{YanPow02_HybridCache,MitZha12_EnCache} and hence, the change in
set-decoding on reconfigurations necessitates flushing a large number of blocks.
In contrast, with MASTER, cache reconfiguration changes the set locations of
only those addresses which were (or are going to be) stored in the transferred
colors. Thus, MASTER incurs smaller reconfiguration overhead than the previous
schemes. Compared to the lazy reconfiguration  approach
\cite{RanAdv00_Recon,LinLuq08_hpca}, the reconfiguration scheme of MASTER is
simpler, requires less state storage and always maintains consistency. 
Reconfigurations happen only at most once every interval which is of the order
of a few million cycles and hence, the overhead of reconfigurations is amortized over
the interval length.  Indeed, our results (Section \ref{sec:master_results}) show that
MASTER keeps increase in number of DRAM accesses small (less than 0.5 per kilo
instructions) and this confirms that the reconfiguration overhead of MASTER is
small.    

%

 

\section{Experimental Methodology}\label{experimentResult}

\subsection{Simulation Environment and Workload}\label{sec:Simulation}
We conduct out-of-order simulations using interval core model in Sniper x86-64
multi-core simulator \cite{CarHei2011_Sniper}, which has been verified
against real hardware. Each core has a 128-entry ROB, dispatch width
of 4 micro-operations and frequency of 2.8GHz. L1I and L1D caches are private to each core and L2 cache is shared among the
cores. Both L1I and L1D are 32KB, 4-way, LRU caches with 2 cycle latency. The L2
cache is unified 8-way, LRU and its size for 2 and 4-core simulations are
4MB and 8MB respectively. This range of cache sizes are typical in commercial processors \cite{kumar2009family}. L2 latency for baseline simulations is 12 cycles
and for MASTER, DCT and WAC (Section \ref{sec:master_compare}), it is 13 cycles since
they all use gated V$_{\text{dd}}$ scheme. 
 Main memory latency is 196 cycles and memory queue contention is also modeled.
For 2-core configuration, peak memory bandwidth  is 12.8 GB/s and for 4-core
configuration, it is 25.6 GB/s. Interval length is
5M cycles. 
 
We use all 29 SPEC CPU2006 benchmarks with \textit{ref} inputs. For workload construction, the benchmarks
are classified following a methodology similar to Jiang et al.
\cite{jiang2011access}. Based on the change in L2 miss-rate from a 4MB L2 to
64KB L2, benchmarks were sorted and then classified into two groups namely
high-gain (H) and low-gain (L) such that each group has nearly half the
benchmarks. This is shown in Table \ref{tab:classification}.

\begin{table}[htp]\footnotesize
\isucaption{Benchmark classification}
\label{tab:classification}

\begin{center}
\begin{tabular}{|l|l|}
\hline
High(H)
 & astar(As), bzip2(Bz), calculix(Ca), dealII(Dl),  gcc(Gc), gemsFDTD(Gm), gromacs(Gr), lbm(Lb), \\
 & leslie3d(Ls),  omnetpp(Om),  soplex(So), sphinx(Sp), xalancbmk(Xa), zeusmp(Ze)\\\hline
Low(L)
 & bwaves(Bw), cactusADM(Cd), gamess(Ga), gobmk(Gk),  h264ref(H2), hmmer(Hm), libquantum(Lq) \\
 & mcf(Mc), milc(Mi), namd(Nd),  perlbench(Pe), povray(Po), sjeng(Sj), tonto(To), wrf(Wr) \\\hline
\end{tabular}	
\end{center}
\end{table}
\begin{table}[htbp]\footnotesize
\isucaption{Workloads for 2 and 4 core systems. \textmd{H$x$L$y$ shows
that the workload has $x$ high-gain and $y$ low-gain benchmarks}}
\label{tab:workloads}
\begin{center}
\begin{tabular}{|l|l|}
\hline
& 2-core workloads \\\hline
H2L0 &  T1(AsDl), T2(GcLs), T3(GmGr), T4(LbXa), T5(BzLs) \\   \hline
H1L1 & T6(SoMi), T7(ZeCd), T8(CaTo), T9(SpMc), T10(OmLq) \\   \hline
H0L2 &  T11(SjWr), T12(BwNd), T13(HmGa), T14(GkH2), T15(PePo) \\    \hline 
\hline
& 4-core workloads \\\hline
H4L0 &	 F1(SoGrZeLb), F2(OmSpGmGc), F3(BzGrLsGm)\\\hline
H3L1 &	 F4(LsZeOmLq), F5(GmCaLbCd), F6(CaAsXaMc)\\\hline
H2L2 &	 F7(BzDlGaMc), F8(SpGcLqHm), F9(XaLbMiGk)\\\hline
H1L3 &	 F10(SoNdMiBw), F11(DlCdGkGa), F12(AsPeToWr)\\\hline
H0L4 &	 F13(BwPoNdH2), F14(HmSjPoH2), F15(SjToWrPe)\\\hline
\end{tabular}	
\end{center}
\end{table}

Using this classification, multiprogrammed workloads are randomly constructed
with different combinations of H and L benchmarks (Table \ref{tab:workloads}). T1 to T15 are \underline{t}wo-core workloads and F1 to F15 are \underline{f}our-core workloads. Except for completing the left-over groups, each SPEC benchmark is used exactly once for 2-core workloads and exactly twice for 4-core workloads.


 \begin{table}[htbp]\footnotesize
\isucaption{Evaluation Metrics Used}
\label{tab:metrics}
\centering
\begin{tabular}{|l|l|}
\hline
Percent Energy Saved &
$((E(\text{base})-E(\text{scheme}))\times100)/E(\text{base})$\\\hline
Weighted Speedup \cite{LinLuq08_hpca} & $\Sigma_{n}
(\text{IPC}_n(\text{scheme})/\text{IPC}_n(\text{base}))/N$ \\\hline
Fair Speedup \cite{LinLuq08_hpca} & $N/ \Sigma_{n}
(\text{IPC}_n(\text{base})/\text{IPC}_n(\text{scheme}))$ \\\hline
 
\end{tabular}	
\end{table}
The evaluation metrics used are shown in Table \ref{tab:metrics}. Here scheme refers to either MASTER, DCT or WAC. $E$ is
computed as shown in Section \ref{sec:EnergyModel}. Each benchmark was
fast-forwarded for 10B instructions and the workloads were simulated till each core
completes at least 500M instructions. A core that has finished its 500M
instructions is allowed to run, but for computation of fair speedup and weighted
speedup, its IPC is recorded only for 500M instructions, following previous
works \cite{QurPat06_UtilityBasedCP,Karthik_HPCA12,SanKoz11_Vantage}. Energy
values are recorded for entire execution, following \cite{wang2011dynamic},
since this enables us to account for the effect of increased execution time on
energy consumption. Across the workload, average value of fair speedup and
weighted speedups are calculated as geometric means (Gmean) of per-workload
improvements. For all the other quantities reported in the chapter, average values are
calculated as arithmetic means (Amean). To gain insights, we also present
results on the following two quantities. The first is ActiveRatio, which is defined as the active cache area fraction, averaged over the entire simulation length \cite{KaxHuz01_CacheDecay}. The second is absolute increase in DRAM accesses per kilo instructions (APKI)
due to use of a scheme, over baseline. This is calculated as
$(\text{APKI}(\text{scheme})-\text{APKI}(\text{base}))$. Through this, we measure the increase in both L2 misses and writebacks due
to cache turnoff and reconfigurations. We have also checked the increase in L2
misses and writebacks individually and have found similar trends as in DRAM
access increase. We report \textit{absolute} difference values and not the \textit{relative} difference, following previous works \cite{MitZha12_EnCache,TamAzi09_RapidMRC}. 


\subsection{Comparison with Other Techniques} \label{sec:master_compare}
\textbf{Decay Cache Technique (DCT): } DCT \cite{KaxHuz01_CacheDecay} works by
turning off a block which has not been accessed for the duration of `decay
interval' (DI). Following \cite{KaxHuz01_CacheDecay,LiyPar04_Skadron}, DCT is implemented using gated V$_{\text{dd}}$ and hierarchical
counters; both tag and data arrays are decayed and latency of waking up decayed
block is assumed to be overlapped with memory latency. For computing DI, we used
competitive algorithms theory  \cite{KaxHuz01_CacheDecay}. As shown in Section
\ref{sec:EnergyModel}, a 4MB, 8-way L2 cache has a leakage power consumption of
1.39 Watts and dynamic access energy of DRAM is 70nJ. Hence, for 2.8GHz
frequency, the leakage energy per cycle per block for L2 is 1.39
/(2.8$\times$65536) nJ. Thus, the ratio of DRAM access energy and L2 leakage
energy per cycle per block is 9.2M cycles. Hence, we take the value of decay interval as 9.2M cycles.

\textbf{Way Adaptable Cache (WAC) Technique: }  WAC \cite{BarCom2008_Way} saves
energy by keeping only few MRU (most recently used) ways in each set of the
cache active. WAC computes the ratio (call it $Z$) of hits to the least recently
used \textit{active} way and the MRU way. It also uses two threshold values,
viz. $T_1$ and $T_2$. When $Z<T_1$, it is assumed that most cache accesses of the program hit near
MRU ways and hence, if more than two ways are active, a single cache way is
turned off. Conversely, when $Z>T_2$, cache hits are distributed over different
ways and hence, a single cache way is turned on \cite{BarCom2008_Way}. WAC checks for possible reconfiguration after every $K$ cache hits.
Following  \cite{BarCom2008_Way}, we take $T_1 = 0.005$, $T_2= 0.02 $, $K= 100,000$ and use gated V$_{\text{dd}}$ for hardware implementation.

 We have chosen these techniques, since, like MASTER, they both use state-destroying
 leakage control. Also, DCT turns off cache at block granularity (fine
granularity), while WAC turns off cache at way granularity (coarse granularity)
and hence, these techniques help us evaluate MASTER against different energy
saving mechanisms.  Time overhead of running MASTER, DCT and WAC algorithms is taken as 500, 300 and 20 cycles, respectively. When cache is reconfigured,
all techniques incur additional 600 cycles average overhead.  

We have also experimented with the statically, equally-partitioned cache. On average, for 2 and 4 core configurations, this scheme leads
to nearly 2\% and 4\% \textit{loss} in energy compared to the shared
baseline, respectively. Hence, on taking this scheme as the baseline, the savings of MASTER will be even larger. For sake of brevity, we omit these results.
 
  
 \subsection{Energy Modeling}\label{sec:EnergyModel}
 We model the energy spent in L2 cache, DRAM and the energy cost of algorithm
execution ($E_{\text{Algo}}$). We use the following notations. In any interval (or entire execution),
$E$ denotes the total energy consumed. $E_{\text{L2}}$ and $E_{\text{DRAM}}$
show the energy spent in L2 cache and DRAM respectively.
$P^{\text{Leak}}_{\text{xyz}}$ and $E^{\text{Dyn}}_{\text{xyz}}$ show the
leakage energy \emph{per second} and the dynamic energy \emph{per access},
respectively, in a component xyz (e.g. L2, DRAM and RCE). $G_f$ and $D_f$ ($=1-G_f$) show the fraction of $E^{\text{Dyn}}_{\text{L2}}$, which is spent in accessing data array and tag array, respectively. $DE_{\text{L2}}$  and 
$LE_{\text{L2}}$ show the total dynamic and leakage energy consumed in L2. 
$E_{\text{tran}}$ shows the total energy consumed in block transitions and
$E_{\chi}$ shows the energy consumed in a single block transition. Tran shows
the number of block transitions. 
 In an interval, $F_A$, $W$, $M_{\text{L2}}$ and $H_{\text{L2}}$ show the active
fraction of cache, number of active ways, L2 misses and L2 hits respectively.
Assoc shows L2 associativity. Time denotes the time length of an interval in
seconds.  $A_{\text{DRAM}}$ and $A_{\text{RCE}}$ show the number of DRAM
accesses and RCE accesses, respectively. $\Upsilon$ shows the area overhead of
gated V$_{\text{dd}}$ cell as a fraction of area of the normal cell.
$P_{\text{off}}$ shows the leakage power consumption at low leakage as a
fraction of normal leakage power, $P^{\text{Leak}}_{\text{L2}}$. 

For computing L2 leakage energy, we account for the consumption of both active
and low-leakage portion of the cache and also assume that the increase in area
due to the use of gated V$_{\text{dd}}$ leads to an increase in leakage energy
in the same proportion.
The L2  dynamic energy in accessing data array is assumed to scale with the number of active ways
\cite{Albonesi99_Selective,kotera2011power,selectiveCache} and an L2 miss is assumed to consume
twice the dynamic energy as that of an L2 hit
\cite{HanHri02_TVLSI,MitZha12_EnCache}.  Thus, we get
\begin{align}
 \label{eq:totalenergy}E&= E_{\text{L2}}+E_{\text{DRAM}}+E_{\text{Algo}} \\
 E_{\text{L2}} &= LE_{\text{L2}} + DE_{\text{L2}}  \\   
 LE_{\text{L2}} &= P^{\text{Leak}}_{\text{L2}} (1+\Upsilon)\times (F_{A}
+ (1-F_{A}) P_{\text{off}}) \times \text{Time} \\
DE_{\text{L2}} &= E^{\text{Dyn}}_{\text{L2}}\times(2
M_{\text{L2}}+H_{\text{L2}})\times (G_f+\dfrac{D_f\times W}{\text{Assoc}})  \\
E_{\text{DRAM}} &= P^{\text{Leak}}_{\text{DRAM}}\times \text{Time} +
E^{\text{Dyn}}_{\text{DRAM}}\times A_{\text{DRAM}} \\
E_{\text{Algo}} &= E_{\text{tran}}+E^{\text{Dyn}}_{\text{RCE}}\times
A_{\text{RCE}} +P^{\text{Leak}}_{\text{RCE}}\times \text{Time} \\
E_{\text{tran}}&= E_{\chi}\times \text{Tran}
\end{align}

Note that for baseline experiments, $E_{\text{Algo}}=0$, $\Upsilon$ = 0 and
$F_A$ = 1 and $P_{\text{off}}$ value is not required. RCE energy cost is only incurred in MASTER. For MASTER, DCT and baseline,
$W$ = Assoc, since these techniques do not turn off the cache ways. Based on CACTI, we take $G_f $= 0.03 and $D_f $= 0.97 for all cache sizes. For MASTER,
DCT and WAC, $F_{A}$ represents the fraction of active colors, active blocks and
active ways, respectively. For gated V$_{\text{dd}}$ scheme, $P_{\text{off}}$ =
0.03 and $\Upsilon$ =  0.05  \cite{PowSeh00_GatedVdd}, which applies to MASTER,
DCT and WAC.  The values of $P^{\text{Leak}}_{\text{L2}}$ and
$E^{\text{Dyn}}_{\text{L2}}$ are obtained using CACTI \cite{cacti_65} assuming
8-bank, 8-way at 32nm and they are shown in Table~\ref{tab:L2cacheEnergy}.
$P^{\text{Leak}}_{\text{DRAM}}$and $ E^{\text{Dyn}}_{\text{DRAM}}$ are taken as
0.18 Watt and 70 nJ, respectively \cite{ZheLin09_DIMM,MitZha12_EnCache} and
$E_{\chi}$ is taken as 2 pJ \cite{MitZha12_EnCache}.

 The energy values of RCE are computed using CACTI \cite{cacti_65} and Eq.
\ref{eq:setprofiling}. Since RCE only stores tags, we take the energy values of
tag arrays only. These values act as upper bounds of RCE energy consumption,
since without data arrays, dirty bits etc., RCE can be implemented even more
efficiently. The values of $P^{\text{Leak}}_{\text{RCE}}$ and
$E^{\text{Dyn}}_{\text{RCE}}$ are shown in   Table~\ref{tab:L2cacheEnergy},
assuming 8B block size and a single bank structure. Noting that for every 64 L2
accesses,  RCE is accessed only 7 times, we conclude that the energy consumption of RCE is a very small fraction of L2 energy consumption.

\begin{table}[htbp]\footnotesize

\centering
\isucaption{Energy values for L2 Cache and Corresponding $N$-core RCE }
\label{tab:L2cacheEnergy}
\begin{center}
{
\begin{tabular}{|c|c|c||c|c|c|}
\hline
 & \multicolumn{2}{c||}{L2 cache} &  \multicolumn{3}{c|}{RCE} \\ \cline{2-6}  
\rule{0pt}{10pt}Cache & $E^{\text{Dyn}}_{\text{L2}}$& 
$P^{\text{Leak}}_{\text{L2}}$ & Number of &$E^{\text{Dyn}}_{\text{RCE}}$
&$P^{\text{Leak}}_{\text{RCE}}$ \\
Size & (nJ/access) & (Watt) & cores ($N$) &(nJ/access) & (Watt) \\\hline
  
  4MB& 0.289& 1.39 &2 & 0.005 &0.006\\\hline
  8MB &0.438& 2.72 &4 &  0.016 &0.023\\\hline

\end{tabular}
}
\end{center}
\end{table}

%
%





\section{Results and Analysis} \label{sec:master_results}
\subsection{Comparison of Energy Saving Techniques} \label{sec:comparison}
\begin{figure*}[htbp]
 \centering
  \includegraphics [scale=0.53] {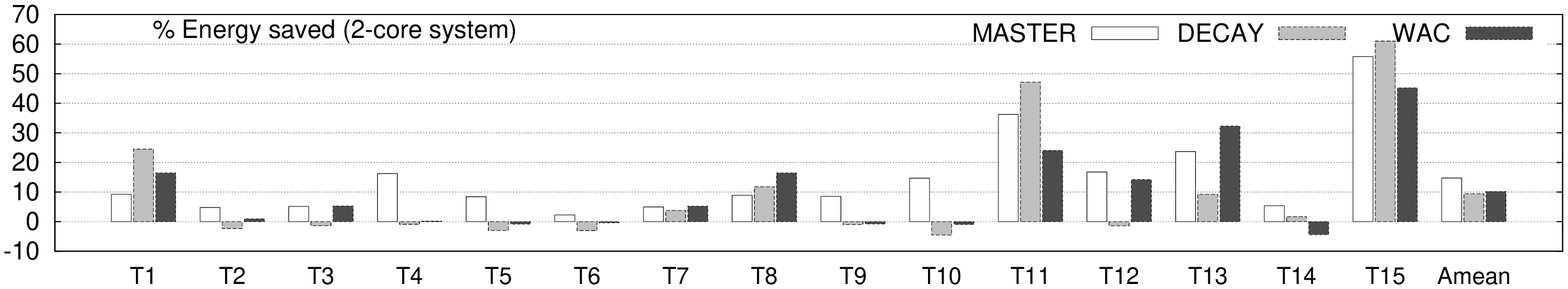}
   \includegraphics [scale=0.53] {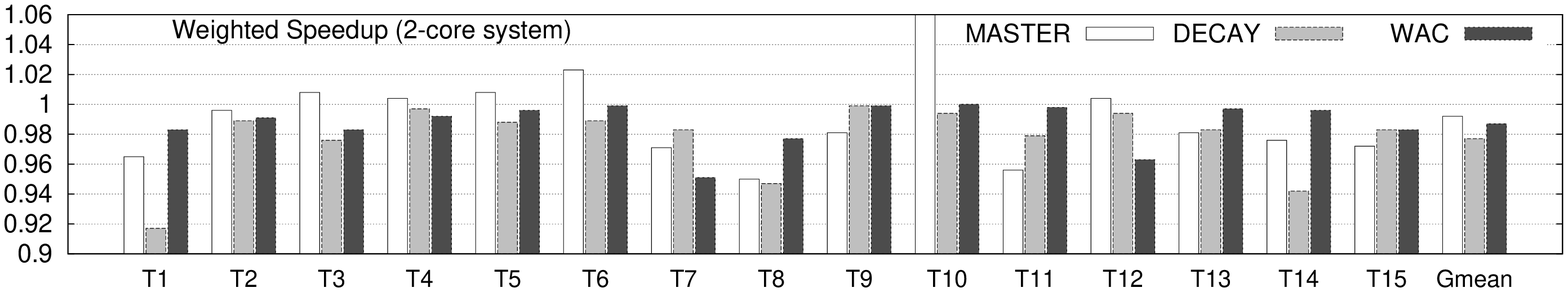}
 \isucaption{Results on percentage energy saved and weighted speedup for 2 
core system}
\label{fig:ensaved248master}
 \end{figure*}

\begin{figure*}[ht]
 \centering
   \includegraphics [scale=0.53] {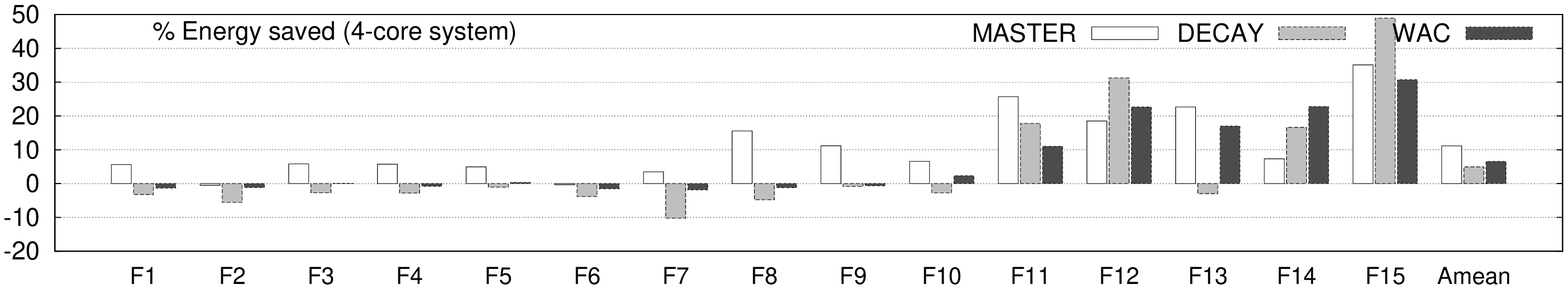}
   \includegraphics [scale=0.53] {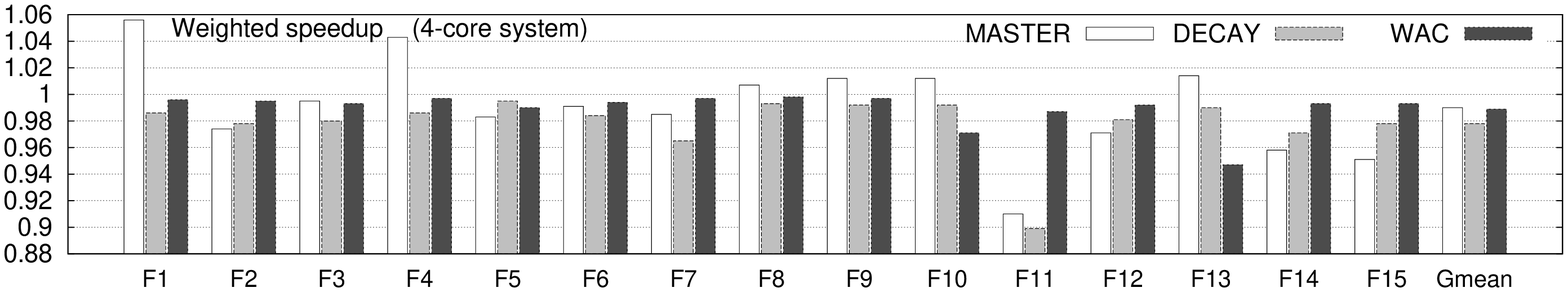}
     
 \isucaption{Results on percentage energy saved and weighted speedup for  4 core system}
\label{fig:ensaved248mastern}
 \end{figure*}

Figure \ref{fig:ensaved248master} and \ref{fig:ensaved248mastern} show the energy saving and weighted speedup
results.

Other quantities are summarized in Table
\ref{tab:othermetrics} and figures for them are omitted for brevity. For 2 and 4-core system, energy savings of MASTER (DCT and WAC) are 14.72 (9.43 and 10.18) and  11.16 (4.92 and 6.55), respectively.  

\begin{table}[htbp]\footnotesize
\isucaption{Results on fair speedup, active ratio and DRAM APKI increase}
\label{tab:othermetrics}
\centering
\begin{tabular}{|c|c|c||c|c||c|c|}
\hline
     & \multicolumn{2}{c||}{Fair speedup} &\multicolumn{2}{c||}{Active Ratio } &\multicolumn{2}{c|}{APKI Increase }\\\hline
         &N=2 & N=4 & N=2 & N=4 & N=2 & N=4 \\\hline
MASTER & 0.99 & 0.99  & 0.53 & 0.52 & -0.51 & 0.17 \\\hline
DCT    &  0.98 & 0.98  & 0.69 & 0.80 & 0.55 & 0.53   \\\hline
WAC    &   0.99 & 0.99  & 0.74 & 0.81 & 0.16 & 0.23  \\\hline
\end{tabular}	
\end{table}

Clearly, MASTER provides largest improvement in energy efficiency, weighted
speedup and fair speedup. With increasing $N$, intra-application
interference increases and locality of memory access stream decreases and hence,
the energy saving achieved by application-insensitive techniques such as DCT and WAC decreases. This fact is  confirmed by the results on ActiveRatio which show that the average ActiveRatio with DCT and WAC are more than 0.69.  In contrast, MASTER turns off a large fraction of cache while keeping DRAM access increase low and this translates into large energy savings.

With MASTER, fair speedup values are close to one. Thus, by allocating cache in proportion to the cache demand of individual applications, MASTER maintains fairness and does not affect QoS (quality of service) or cause thread starvation. Further, despite turning off and flushing a portion of L2, MASTER \textit{reduces} the
DRAM APKI for many workloads, such as T4, T6, T10, F9 etc. In fact, for 2-core system, on average, DRAM APKI is reduced by 0.51. This is because, by managing the cache quota of different applications and containing the
thrashing applications, MASTER reduces the number of L2 misses and writebacks.  DCT and WAC increase DRAM APKI more than MASTER.

 Looking into the essential energy saving mechanisms of different techniques, we
observe that DCT considers the access intensity to cache block as a measure of
its usefulness or liveliness and uses this information to turn off the cache.
However, for many benchmarks and especially for streaming ones such as
libquantum and milc, access intensity shows up to be a poor measure of data
reuse and usefulness of a block and hence, for most workloads, DCT does not save
large amount of leakage energy. With increasing $N$, the intra-application
interference reduces the opportunity to turnoff cache reduces even
further. The advantage of DCT is that it turns off cache at block granularity and hence, achieves larger energy saving for some workloads such as T15.  
 
WAC works by using ratio of hits in MRU and LRU positions as a measure of locality present in the memory access stream and turns off cache at way granularity, while always keeping at least 2 ways active. Clearly, due to way level allocation approach, WAC turns off cache only at coarse granularity and reduces the associativity of the cache. The advantage of WAC is that it always turns off least recently used blocks in the LRU chain which are less likely to be reused in the future. Further, by turning off ways, it also reduces the dynamic energy consumed in accessing data array of the cache.        

MASTER works by estimating energy consumption of a few configurations and choosing a configuration with highest energy efficiency. It takes into account the cache demands of each application and hence, can easily account for streaming or
non-streaming applications. Further, MASTER enforces strict cache quotas and alleviates inter-application interference, which also helps in maintaining performance and fairness. MASTER allocates cache at color granularity and hence it does not hurt associativity.  With MASTER, the contribution of $E_{\text{Algo}}$ in total memory subsystem
energy consumption for 2 and 4 core systems is 0.25\% and 0.39\%,
respectively. Given the large energy saving achieved by MASTER, its small
overhead is justified. A limitation of MASTER is that it allocates at least $M/64$ colors to each application and hence, for applications with very small working set size, it may lose the opportunity to turnoff the cache further. This limitation can be easily addressed by reducing the lower limit (see Section \ref{sec:reconfigurablecacheemulator}), depending on the typical working set size of the applications and
acceptable RCE overhead. Based on our experiments, we have observed that $M/64$ color limit is reasonable since it enables significant energy savings and also avoids any possibility of performance degradation.

The aggressiveness with which an energy saving technique should turn off the cache depends, not only on the application behavior, but also on the factors such as relative energy consumption of cache and other processor components. While DCT and WAC cannot directly take other components into account, their effect is implicitly seen in the choice of decay interval in DCT and $K$, $T_1$ and $T_2$ in WAC. Thus,  statically choosing the optimal (or best) value(s) of the parameters in these techniques is likely to require significant efforts and the values may also vary for different platforms and optimization targets.  In contrast, MASTER is capable of accounting and directly optimizing for system (or subsystem) energy at runtime and it can easily adjust its aggressiveness of cache turnoff depending on the trade-off between energy saving and performance loss from cache turnoff. In fact, the energy saving approach of MASTER presented here can be easily extended to optimize for overall system energy by merely
including the energy model of other processor components.

\subsection{Sensitivity To Different Parameters} \label{sec:sensitivity}
We henceforth focus exclusively on MASTER. We study its sensitivity for different system parameters. In each case, only a single parameter is changed from the default configuration and the results are shown in Table \ref{tab:resultsparameter}. 
Wherever applicable, for changed
parameters, the energy values such as $E^{\text{Dyn}}_{\text{RCE}}$ etc. were
computed as shown in Section \ref{sec:EnergyModel}. For sake of
brevity, we omit these values. In all cases, the average fair speedup is more than 0.97 and hence, these results are also omitted. 

The following two parameters apply to MASTER technique. 

\textbf{Interval length: } To see the possiblity of reducing reconfiguration overhead, we change the interval length to 10M cycles. As shown in Table \ref{tab:resultsparameter}, this slightly reduces energy saving and slightly improves performance, which is expected. Thus, MASTER can work at coarse interval sizes and is not  very sensitive to the choice of a specific interval length.
 
\textbf{Sampling ratio ($R_s$): } We change $R_s$ in RCE to 128. From Table \ref{tab:resultsparameter}, we observe a small reduction in energy saving, which is due to reduced accuracy in profiling information, although the energy savings are still large. Thus, at the cost of slightly reduced energy saving, the overhead of RCE can be further reduced.    
\begin{table}[ht]\footnotesize
\isucaption{Energy saving, weighted speedup (WS) and APKI Increase for different parameters.
\textmd{Default parameters: interval length = 5M cycle, $R_s=64$, Assoc = 8, LRU policy. Results with default parameters are also shown.}}
\label{tab:resultsparameter}
\centering
\begin{tabular}{|c|c|c|c|c|c|c|}
\hline
     & \multicolumn{2}{c}{\% Energy Saved} &\multicolumn{2}{|c|}{WS} &\multicolumn{2}{|c|}{APKI Increase} \\\hline
     & N=2 & N=4  &N=2 & N=4 &N=2 & N=4\\\hline
Default        & 14.7 & 11.2& 0.99 & 0.99  & -0.51  &  0.17 \\\hline  
Interval=10M   & 14.0 & 11.6& 1.00 & 1.00  & -0.68  & -0.17 \\\hline
$R_s$ = 128    & 12.9 & 10.3& 0.99 & 0.99  & 0.04   &  0.61 \\\hline
Assoc = 16     & 15.8 & 13.9& 0.99 & 0.99  & -0.51  &  0.23\\\hline
FIFO policy    & 12.9 & 12.8& 0.99 & 1.00  &  -0.54 & -0.18\\\hline
PLRU policy    & 14.3 & 12.0& 0.99 & 0.99  & -0.45  &  0.14 \\\hline
\end{tabular}	
\end{table}

The following two parameters apply to both baseline and MASTER.

\textbf{Cache associativity: } On changing L2 associativity (Assoc) to 16, while keeping the size same, we observe that MASTER still offers large energy savings  (Table \ref{tab:resultsparameter}).

\textbf{Replacement policy: } We first change the replacement policy to FIFO (first-in, first-out) and then to MRU bits
based pseudo-LRU (PLRU) \cite{al2004performance}. The large value of energy savings (Table \ref{tab:resultsparameter})  show that MASTER works independent of the replacement policy used.


%
 
\subsection{The Case When The Number Of Programs Is Less Than The Number of Cores}
As discussed before, in several cases the actual number of programs running on a processor are much less than the number of cores. This is especially expected to be true for future processors which would have a large number of cores. To test the effectiveness of MASTER in such cases, we simulate 4-core configuration with 2-core workloads (shown in Section \ref{sec:Simulation}). We run one program each on the first two cores while the other two cores remain idle.  Using MASTER, we observe an energy saving of 25.3\%, weighted speedup of 0.96, fair speedup of 0.96, DRAM APKI increase of 1.29 and active ratio of 32.6\%. Clearly, since in this case, the cache size available to each core is large, MASTER aggressively reconfigures the cache to provide large energy saving.


%

%

 \section{Conclusion}\label{conclusion}
In this chapter, we have presented  MASTER, a cache leakage energy saving approach
for  multicore caches. MASTER uses coloring scheme to partition cache space at
the granularity of a single cache color. By using low-overhead RCE for estimating
performance and energy of running applications at multiple cache sizes, MASTER
periodically reconfigures the LLC to most energy efficient configuration. Out-of-order simulations performed
using SPEC06 workload have shown that MASTER is effective in saving memory subsystem energy and does not harm performance or cause unfairness.


\chapter{MANAGER: A CACHE ENERGY SAVING APPROACH FOR MULTICORE QOS SYSTEMS}\label{chap:manager}
\section{Introduction}
In this chapter, we present \textbf{MANAGER}, a \underline{m}ulticore shared c\underline{a}che e\underline{n}ergy s\underline{a}vin\underline{g} techniqu\underline{e} for quality-of-se\underline{r}vice systems. As cache energy consumption becomes an increasing fraction of processor power consumption \cite{lahiri2004power,monchiero2008power}, cache energy saving techniques have become extremely important for multicore QoS systems.   
 Several recent trends motivate this shift.  Since LLC is the last line of defense against the memory wall and the QoS which a program gets from the platform is crucially affected by the behavior of shared LLC \cite{hsu2006communist,chandra2005predicting,guo2007chaos}, modern processors use large LLC, e.g. Intel's 32nm Westmere processor uses 12MB LLC \cite{kurd2010westmere}. With each CMOS technology generation, leakage energy consumption has been drastically increasing \cite{rodriguez2006energy} and thus, energy consumption of large LLCs is on rise. Hence, effective management of LLC in multicore processors is important for achieving both QoS and energy efficiency. 
 
The existing cache energy saving techniques have several limitations when used in multicore QoS systems. Some techniques aim to aggressively save energy \cite{MitZha12_EnCache,KaxHuz01_CacheDecay} and hence, for QoS systems, they may either fail to meet QoS requirement or lose the opportunity to save energy. Further, modern multicore processors run arbitrary combinations of benchmarks and hence, the techniques which require offline profiling (e.g. \cite{Albonesi99_Selective,YanPow01_IcacheResize}) become infeasible to use.  Several energy saving techniques are application-insensitive and only rely on locality of memory access streams \cite{KaxHuz01_CacheDecay}. Since the memory access streams from different applications exhibit different locality properties and memory sensitivity; a co-scheduled program can make it difficult to meet QoS of one program or trying to meet QoS of one program may lead to starvation of co-scheduled program.
Thus, to address the challenges of achieving energy efficiency in multicore QoS systems, novel techniques are required. 

  MANAGER aims to optimize memory subsystem energy, while ensuring QoS for one program (called ``target'' program) in best-effort manner (Section \ref{sec:qosformulation}). In several scenarios, different programs have different importance, for example a data-critical program has higher priority than  a program performing system backup. Similarly, in usage models such as server consolidation,  SLAs (service level agreements) motivate performance isolation for some applications. MANAGER is a useful technique for such systems. Further, MANAGER uses software control to ensure QoS, which makes it effective since the relative priorities of running programs are best known in the operating environment. 
 

 
MANAGER uses a small reconfigurable cache emulator (RCE) to dynamically predict energy efficiency of multiple configurations (Section \ref{sec:systemdesign}). Also, by comparison with the miss-rate estimates obtained from RCE, the minimum amount of cache which needs to be allocated to the target program is decided, such that its QoS target can be met. Among configurations fulfilling this criterion, the most energy efficient configuration is chosen, which is used for the next interval (Section \ref{sec:manager_esa}). The overhead of MANAGER is small.  For a 2-core system, MANAGER adds an overhead of less than 0.4\% of the LLC cache size (Section \ref{sec:manager_implementation}).

 
 We evaluate MANAGER using out-of-order simulations with Sniper x86-64 simulator, and dual-core workloads from SPEC2006 suite. The results show that MANAGER saves large amount of memory subsystem energy, while ensuring QoS for most workloads. For example, for 5\% allowed performance loss of target program,  4MB LLC and 29 dual-core workloads, the average energy saving over statically, equally-partitioned baseline LLC is 13.5\% and only one workload misses its QoS deadline.  


 



\section{Related Work}
As we move to the exascale era, the applications running on modern processors are  presenting increasingly higher resource demands \cite{PanMit2009_Baywave,raju2012high,agrawal2008new,MitGup2013_QA,raju2009domain}. Several e-learning and multimedia applications present QoS demands \cite{pande2007network,sood2006novel}. To address this challenge, several studies have proposed cache-partitioning methods which use either QoS \cite{LinLuq08_hpca,iyer2004cqos,nesbit2007virtual} or performance \cite{QurPat06_UtilityBasedCP} as the optimization target.  Iyer \cite{iyer2004cqos} discuss techniques to assign and enforce priority  for the applications and then allocate desired amount of cache using methods such as way-partitioning.  Iyer et al. \cite{iyer2007qos} propose QoS-aware cache partitioning which aims to improve the performance of the high priority application in the presence of other applications. In contrast to these works, our work aims to minimize memory subsystem energy while ensuring a pre-defined QoS for a target (high priority) program (see Section \ref{sec:qosformulation}).

Herdrich et al. \cite{herdrich2009rate} propose \textit{rate-control} techniques (e.g. clock-modulation, DVFS) for addressing cache QoS issues and managing power dissipation. Their approach works by throttling the processing rate of a core running a low-priority task, if its execution 
is interfering with a high priority task due to platform resource contention. In contrast, our work uses \textit{resource-control}, like several previous works \cite{iyer2007qos,iyer2004cqos}. The resource-control techniques work by partitioning the resources (e.g. cache, memory bandwidth) among running programs to achieve desired QoS.


Most of the existing cache energy saving techniques (e.g. \cite{MitZha12_EnCache,KaxHuz01_CacheDecay}) aim to aggressively save energy and hence, may not ensure QoS. A few other QoS-based energy saving techniques (e.g. \cite{MitZha13_Cashier}) only work for single-core systems  and hence, cannot be directly used for ensuring QoS in multicore systems. Further, a 4MB, 8-way cache has 128 cache colors and thus, the cache coloring technique used in our work provides much finer granularity of reconfiguration than that provided by several existing techniques e.g. \cite{MitZha12_EnCache,YanPow01_IcacheResize,Albonesi99_Selective}.


\section{Notations and QoS Formulation}\label{sec:qosformulation}
We assume single-threaded cores and hence, use the words core and program interchangeably. We assume that the LLC is L2 cache, although the techniques presented here can be extended to the case where LLC is L3 cache. $M$ denotes the number of cache colors and $N$ denotes the number of cores. An arbitrary core index is shown as $n$ and interval index is shown as $i$. $L$, $W$ and $G$ denote the cache block size, tag size and system page size, respectively. In our experiments, we assume $L$ = 512 bits (=64B), $W$ = 24 bits and $G$ = 4KB. 

In this chapter, the QoS requirement is formulated as follows.  For a two-core workload, the first program is termed as the ``target'' program and the second one as the ``partner'' program. The QoS guarantee is to ensure that compared to baseline execution, the performance loss of the target program is no more than $\Omega$\% \cite{LinLuq08_hpca}, while the objective is to save overall memory subsystem energy. 

Let $IPC[n_t]$ refer to the $IPC$ of core $n_t$ which runs the target program. Baseline refers to the statically partitioned cache of the same total cache size with half of the cache capacity allocated to each program (i.e. target and partner). Then, the QoS target is met if 
\begin{equation}
\dfrac{IPC_{\text{baseline}}[n_t]- IPC_{\text{MANAGER}}[n_t]}{IPC_{\text{baseline}}[n_t]}\times 100 \le \Omega
\end{equation}

The technique proposed by Lin et al. \cite{LinLuq08_hpca} requires offline specification of baseline IPC. In contrast, our technique estimates baseline IPC \textit{also} during runtime by using RCE, thus, providing significant improvement over existing techniques.  Note that our QoS formulation differs from other works (e.g. \cite{chang2007cooperative}), where QoS summarizes the behavior of an entire workload. Further, Varadrajan et al. \cite{varadarajan2006molecular} define QoS requirement in terms of miss-rate goal. We define QoS requirement in terms of IPC, which has been more widely used.

\section{System Architecture}\label{sec:systemdesign}
\begin{figure}[htp]
\centering

 \includegraphics [scale=0.60] {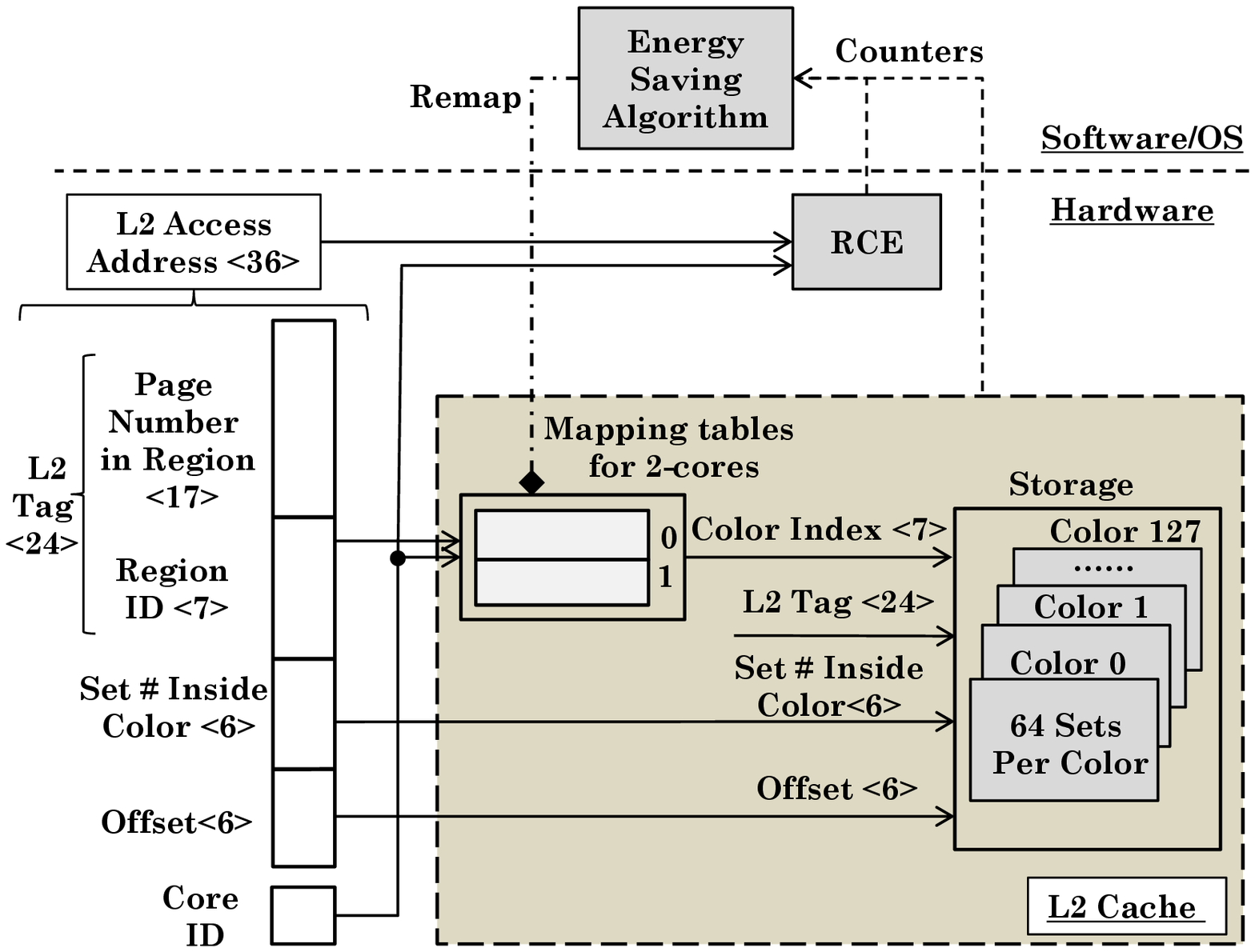}
\isucaption{Overall Flow Diagram of MANAGER ($N$=2, $M$=128)}\label{fig:manager_flowdiagram}
\end{figure}

The overall architecture of MANAGER is shown in Figure \ref{fig:manager_flowdiagram}. We now describe each component of MANAGER in detail. 

\subsection{Cache Coloring}
For selective cache allocation, MANAGER uses cache coloring technique \cite{MitZha13_Cashier,LinLuq08_hpca}, which works as follows. We logically divide the cache into $M$ parts called ``cache colors''. Here $M$ is given by 
\begin{equation}\label{eq:manager_numColor}
M = \dfrac{\text{Size}_{\text{L2}}}{G\times \text{Associativity}_{\text{L2}}}
\end{equation} 

 We further logically divide the physical pages into groups such that the physical pages of a core that share $\log_2(M)$ least significant bits of the physical page number are in the same ``memory region''. Thus, the number of memory regions for each core is $M$. Cache coloring technique allocates a given cache color to one or more memory regions of a \textit{single} core, such that all physical pages in those memory regions are mapped to the same cache color. To record the mapping of memory regions to colors, for each core, a \textit{mapping table} is used, which has $M$ entries, each $\log_2(M)$-bit wide. At any instance, if core $k$ has $c_k$ colors, then its mapping table would store the mapping of its $M$ regions to $c_k$ colors. In this way, the cache quota of different cores can be enforced and the unused colors can be turned off for saving leakage energy. From Eq. \ref{eq:manager_numColor}, a typical 4MB, 8-way cache has 128 cache colors, thus MANAGER provides fine granularity of reconfiguration.


\subsection{Reconfigurable Cache Emulator (RCE)}\label{sec:manager_rce}
To estimate cache miss-rate under different cache configurations\footnote{Since MANAGER does not dynamically change the associativity and block size, change in configuration simply refers to change in set-count.}, we use  auxiliary tags for each core. Each such unit is referred to as a profiling unit. To keep the size of the profiling unit small, we use set-sampling technique \cite{QurPat06_UtilityBasedCP,MitZha12_EnCache}.

A single profiling unit cannot provide profiling information for different number of sets or colors, hence, for each cache configuration which is profiled, a different profiling unit would be required. To keep the overhead of profiling small, we use only six profiling units for each core and estimate miss-rate for other configurations using interpolation. The profiling units chosen are $2^{(j-1)}X/32$ , where $X$ refers to the L2 cache size and $j=\{1, 2,\ldots, 6\}$. The complete profiling structure, consisting of all profiling units of all the cores is referred to as  ``reconfigurable cache emulator'' (RCE).   
   
 \begin{figure}[htp]
\centering
 \includegraphics [scale=0.60] {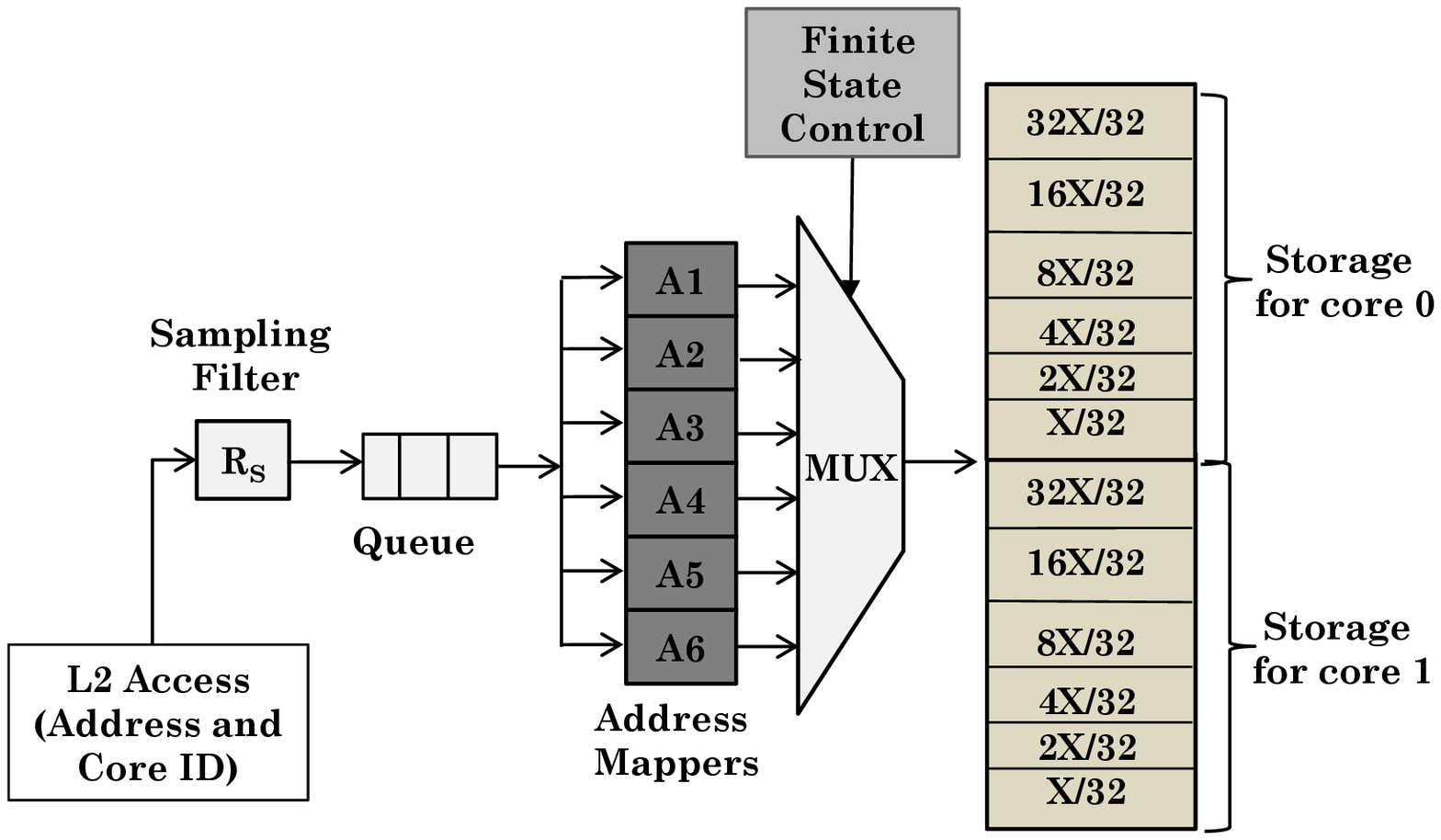}
\isucaption{RCE  Design in MANAGER}\label{fig:manager_rcediagram}
\end{figure}
 The RCE works as follows (Figure \ref{fig:manager_rcediagram}). Each L2 address is first sampled using a sampling filter, which has a sampling ratio ($R_S$) of 64. The addresses which pass the filter are passed through queue to avoid congestion. Then, the addresses are fed to address mappers, which compute the tag and set (index) location and also add an offset to map the address to suitable profiling unit. Afterwards, using a MUX, the incoming addresses are fed to the profiling units of the originating core.
       
       
 We now compute the size of RCE. Let $Z$ and $S$ be the number of sets in L2 and RCE, respectively. Let  $\Theta$ show the percentage overhead of RCE, compared to L2. Then, we have 
 \begin{eqnarray}
 S &=& \dfrac{(\sum\nolimits_{j=1}^6 2^{j-1})\times N\times Z}{32\times R_S} =  
\dfrac{63NZ}{32R_S} \le \dfrac{2NZ}{R_S} \label{eq:manager_setprofiling}\\
 \Theta &=&  \dfrac{\text{Size}_{\text{RCE}}\times 100}{\text{Size}_{\text{L2}}} = \dfrac{N\times
2W}{R_S(L+W)}\times 100      
 \end{eqnarray}

Substituting values, we obtain $\Theta$ = 0.28\%. To cross check, we have used CACTI 6.5 \cite{cacti_65}  to compute the area of RCE and L2 for the cache sizes used in experiments (see Section \ref{sec:manager_simulation} and \ref{sec:manager_energymodel}) and have observed the value of $\Theta$ in the same range. Thus, the overhead of RCE is small. We account for the energy consumption of RCE in our energy model in Section \ref{sec:manager_energymodel}.

\subsection{Execution Time Estimation}\label{sec:manager_cpistack}
To estimate the effect of cache misses on program execution time, MANAGER uses CPI stack technique \cite{CarHei2011_Sniper}. The CPI stack shows the contribution of different components to overall performance. It shows base CPI and lost cycles due to events such as instruction interdependencies, memory stalls etc., taking into account the possible overlaps between execution and miss events.  

Cache misses affect program execution time through memory stall cycles. We assume that, in a given interval, memory stall cycles depend linearly on the number of load misses, and hence, their ratio, called SPM (Stall cycles Per load Miss), is same for different configurations (i.e. different number of load misses). This assumption holds reasonably well, since in an interval, ESA only searches for configurations which differ from existing configuration in a small number of active colors (Section \ref{sec:manager_esa}). 

The RCE uses extra counters to estimate load misses under different configurations and by multiplying these values with SPM, the stall cycles under any cache configuration can be estimated. Using stall cycles and base CPI obtained from CPI stack, total execution cycles (and hence execution time) can be easily estimated. These values are used for computing memory subsystem energy, as shown in Section \ref{sec:manager_energymodel}. Also, for the target program, the estimated execution time under different configurations is used to meet its QoS target (Section \ref{sec:manager_esa}).

\subsection{Marginal Gain}\label{sec:manager_marginalgain}
In each interval, ESA selects configurations using marginal gain values. Marginal gain (MG$_n(x)$) for core $n$ with color value $x$ is defined as the reduction in cache misses per extra unit cache color. We assume that between two profiling points, the  number of misses vary linearly with cache size and thus, MG$_n$ remains constant between those profiling points. Thus, MG$_n$ is defined as: 
 
 \begin{equation}
\text{MG}_n(c_n) = \begin{cases}
\dfrac{ \text{Miss}_n(D_j)-\text{Miss}_n(D_{j+1})}{D_{j+1}-D_j} 
& D_{j} \le c_n < D_{j+1}\\
\dfrac{ \text{Miss}_n(D_5)-\text{Miss}_n(D_{6})}{D_{6}-D_5} &
c_n=D_6\\
\end{cases}
\end{equation}
Here $D_1$ to $D_6$ refer to the 6 profiling points mentioned above (viz. $D_1 = X/32 \ldots D_6=32X/32$) and Miss$_n(D_j)$ refer to the cache misses of core $n$ at color value $D_j$. We show the use of marginal gain in configuration selection in the next section.

\section{Energy Saving Algorithm (ESA) }\label{sec:manager_esa}

We now describe our energy saving algorithm (ESA), which can be part of a kernel module. The decision to start ESA is taken as follows. A Boolean flag is initially reset. After every $K$ (e.g. 10M) instructions of target program, the flag is set. After every 1000 cycles, the flag is checked and whenever the flag is found to be set, ESA starts working. At the end of ESA execution, the flag is reset. 

We use ``color-value'' to refer to the colors of each core and configuration to refer to the color-value combination for 2 cores. Let $c_n^{\star}$ denote the current color value of core $n$ in interval $i$. At the end of an interval $i$, the algorithm executes the following steps. 

\textbf{Step 1: } We first define a quantify $t_{i}$, which is useful in understanding the algorithm. At the end of each interval, the algorithm estimates the extra time (called $\tau$) that the current configuration of \textit{target program} has taken over and above its baseline configuration, i.e. $M$/2 colors, for that interval (Notice that estimates for baseline configuration are also obtained in runtime from 16X/32 profiling unit of RCE and not from offline profiling). Further, over all the intervals, ESA accumulates $\tau$ values to get $t_i$. At the end of interval $i$, $t_i$ gives the estimate of increased execution time due to working of ESA,  \textit{till that interval}.    
 
\textbf{Step 2: } At the end of an interval $i$, if the actual execution time is $T_{i}$, then $(T_{i}- t_{i})$ shows the estimate of baseline execution time for the same execution window. Let $\beta _i$ be the current percentage loss in performance of target program over baseline, then we have    

\begin{equation}
\beta _{i} = \dfrac{t_{i}\times 100}{(T_{i}- t_{i})}
\end{equation}

ESA always attempts to conservatively keep  $\beta _{i}$ below the actual allowed percentage slack ($\Omega$), by a small margin $\chi$ (0.4\% in our experiments). Thus, $\beta _{i} \leq \Omega - \chi$.


\textbf{Step 3: } To ensure that the target program meets its QoS requirement, in each interval, ESA allows a certain percentage loss (say $\Delta_i$) in performance to save energy, such that the overall performance loss of target program is less than $\Omega \%$. The value of $\Delta_i$ is chosen based on $\beta _{i}$ and $\Omega$. 
Since our technique controls cache allocation, specifying $\Delta_i$, in turn, specifies the minimum amount of cache size (i.e. number of cache colors) that must be allocated to target program. Let $Min$ denote this color limit.    

\textbf{Step 4: } For both target program and partner program, four candidate color-values are selected using marginal color values, as follows. Intuitively, for a program with large marginal gain, color values with smaller number of active colors are likely to be energy efficient and vice versa. Hence, ESA uses four application-independent thresholds (viz. 50, 200, 300, 100) to decide the range of MG$_n$ and then a suitable color value is chosen in vicinity of the current color value (c$_n^{\star}$). These candidate color values should also fulfill the following criterion.
\begin{enumerate}
\item[C1] To avoid thrashing, each core receives at least $M/32$ colors; thus a candidate color value must have at least $M/32$ colors.
\item [C2] In any interval, at most 12 colors can be given to a program or taken from it.
\item[C3] For the target program, all color values should have $Min$ or more colors. 
\end{enumerate}

Note that if allocating at-least $Min$ colors to target program requires transferring more than 12 colors in an interval (which may happen due to sudden change in working set size of the target program), condition C3 is relaxed.  This avoids oscillation and high reconfiguration overheads. Moreover, since ESA aims to meet a global (and not per-interval) QoS requirement; a positive or negative deviation from the allowed slack is compensated by feedback adjustment.       

\textbf{Step 5: } From per-core color values, sixteen (=4$\times$4) combinations are formed which represent 2-core configurations. Of these, the configurations with sum of active colors greater than $M$ are discarded.
 
\textbf{Step 6:} For the remaining configurations, the memory subsystem energy is estimated. From these configurations, the one with minimum energy consumption is selected and is chosen for the next interval.

Note that in each interval, ESA examines at most 16 color values and thus, its overhead is small. The use of marginal gain values helps in quickly finding the suitable cache size for a program and use of application-independent thresholds avoids the need of per-application tuning. ESA allocates at least $M$/32 colors to each program and thus, may provide coarser granularity of cache allocation than other schemes which allocate cache at block granularity, e.g. \cite{KaxHuz01_CacheDecay}. However, our choice helps in keeping reconfiguration overhead and performance loss small. Further, large energy savings obtained in the experiments (Section \ref{sec:manager_results}) have confirmed that our choice works well in practice.   
Also, if desired, this limit can be further reduced by adding extra profiling levels in the RCE. 

\section{Implementation}\label{sec:manager_implementation}
For hardware implementation of cache block switching, we use a specific implementation of gated V$_{\text{dd}}$ (NMOS gated V$_{dd}$, dual V$_t$, wide, with charge pump), which reduces leakage power by 97\%, while increasing the access latency by 8\% and cell area by 5\% \cite{PowSeh00_GatedVdd}. We account for these overheads below and in Section \ref{sec:energymodel}. Note that the hardware functionality to turn off a portion of cache is already provided by the existing commercial processor chips \cite{kurd2010westmere,naveh2006power}. 
 
 MANAGER does not require caches of large associativity (which have higher access time),  changes to replacement policy (unlike \cite{QurPat06_UtilityBasedCP}) or offline profiling (unlike \cite{YanPow01_IcacheResize,wang2011dynamic}). With MANAGER, block switching happens only at the end of an interval and hence, change in mapping tables happens infrequently. For a 2-core system with 8-way, 4MB L2, the total size of mapping tables is merely 1792 bits (=$2\times128\times7$), which is merely 0.005\% of the L2 size (tag+data). Thus, the size and access time of mapping tables are negligible and access to them can be folded into the address decode tree of the cache's tag and data arrays. Also, RCE is accessed in parallel to L2. Thus, these activities do not lie on critical access path. Gated V$_{\text{dd}}$ scheme increases access time by 8\%. Hence, with baseline L2 latency as 12 cycles, the L2 latency with MANAGER is taken as 13 cycles. 

L2 cache reconfigurations are handled as follows. When a color is taken away from a core, the blocks of its owner core are flushed from it (i.e. dirty data are written back to memory and other blocks are discarded). When a color is allocated to a core (say $Q$), one or more regions of $Q$, which were mapped to some other color, are mapped to the new color. The blocks of remapped regions in previous colors are flushed. Change in region mapping is achieved using the mapping table.    

Time overhead of running algorithm is taken as 500 cycles and when the L2 is reconfigured, an additional 600 cycles average overhead is incurred. Reconfigurations happen only at the end of a large interval length and thus, the reconfiguration cost is amortized over the interval length. As shown in Section \ref{sec:manager_results}, the average increase in DRAM access on using MANAGER is small or even \textit{negative}. This confirms that the reconfiguration overhead is small.  



\section{Experimentation Methodology}\label{sec:manager_experimental}

\subsection{Simulation Platform and Workload}\label{sec:manager_simulation}

We perform out-of-order simulations using interval core model from Sniper x86-64
multi-core simulator \cite{CarHei2011_Sniper}. Each core has a frequency of 2.8GHz, an 128-entry ROB and a dispatch width of 4 micro-operations. L1I and L1D caches are private to each core and L2 cache is shared. Both L1D and L1I are 32KB, 4-way, LRU caches and have 2 cycle latency. The unified L2
is 4MB, 8-way, LRU. L2 latency for baseline simulations is 12 cycles and for our technique, it is 13 cycles (Section \ref{sec:manager_implementation}). Main memory latency is 196 cycles; peak memory bandwidth is 12.8 GB/s and memory queue contention is also modeled. $K$ is taken as 10M instructions.

We use all 29 SPEC CPU2006 benchmarks with \textit{ref} input. We constructed 29 two-core multiprogrammed workloads by randomly combining different benchmarks. Each benchmark is a target program and partner program in exactly one workload. The workloads are shown in Table \ref{tab:manager_workloads}.

\begin{table}[htbp] 
\isucaption{Workloads Used For Experimentation}
\label{tab:manager_workloads}
\centering
\begin{tabular}{|l|l|l|l|}
\hline
T1& astar dealII & T2& bwaves bzip2  \\\hline 
T3& bzip2 povray & T4& cactusADM gemsFDTD \\\hline
T5&  calculix tonto & T6& dealII cactusADM \\\hline
T7& gamess astar  &T8& gcc leslie  \\\hline
T9& gemsFDTD gromacs  & T10& gobmk  omnetpp  \\\hline
T11& gromacs gamess & T12& h264ref wrf \\\hline
T13&  hmmer mcf  & T14& lbm hmmer \\\hline
T15& leslie sjeng& T16& libquantum soplex \\\hline
T17&  mcf gobmk&T18& milc calculix \\\hline
T19& namd zeusmp  & T20& omnetpp libquantum\\\hline
T21&perlbench lbm & T22&  povray perlbench  \\\hline
T23& sjeng, gcc & T24&soplex milc\\\hline
T25& sphinx  xalan& T26& tonto h264ref\\\hline
T27& wrf bwaves & T28& xalan namd \\\hline
T29& zeusmp  sphinx & &  \\\hline

\end{tabular}
\end{table}

\subsection{Evaluation Metrics}   
We use the following metrics. The first one is the percentage saving in memory subsystem energy, which is computed as shown in Eq. \ref{eq:totalenergy}. Further, we use weighted speedup (WS) \cite{LinLuq08_hpca} and fair speedup (FS) \cite{LinLuq08_hpca}, which are defined as 
\begin{eqnarray}
WS &=&(\Sigma_{n} (\text{IPC}_n(\text{MANAGER})/\text{IPC}_n(\text{baseline})))/N \\
FS &=&N/ (\Sigma_{n} (\text{IPC}_n(\text{baseline})/\text{IPC}_n(\text{MANAGER})))
\end{eqnarray}

Also, we show cache active ratio \cite{KaxHuz01_CacheDecay} (active cache area fraction, averaged over the entire simulation length) and absolute increase in DRAM access per kilo instruction (APKI), which is computed as $(\text{APKI}(\text{MANAGER})-\text{APKI}(\text{base}))$. We present absolute change in APKI and not percentage change, following \cite{MitZha12_EnCache}. 


Across the workload, weighted speedup and fair speedup are averaged using geometric mean and all the other quantities are averaged using arithmetic mean. We fast-forwarded each benchmark for 10B instructions. The simulation is run till each benchmark in the workload completes its 500M instructions \cite{QurPat06_UtilityBasedCP}. IPC of a program is only computed for its first 500M instructions \cite{QurPat06_UtilityBasedCP}. Energy is computed for the entire simulation length \cite{wang2011dynamic} since this allows us to comprehensively account for the effect of loss of performance due to cache turnoff on energy consumption.    
 
 

\subsection{Energy Model}\label{sec:manager_energymodel}
We model the energy spent in L2 cache ($E_{\text{L2}}$), DRAM ($E_{\text{Mem}}$) and the energy cost of algorithm
execution ($E_{\text{Algo}}$), since other components are minimally affected by our technique. Our notations are as follows. For any interval,
$E$ denotes the total energy consumed and $T$ shows the time length in seconds. 
For  a component xyz (e.g. L2, DRAM and RCE), $P^{\text{Leak}}_{\text{xyz}}$ and $E^{\text{Dyn}}_{\text{xyz}}$ show the
leakage energy \emph{per second} and the dynamic energy \emph{per access},
respectively. $DE_{\text{L2}}$  and 
$LE_{\text{L2}}$ denote the total dynamic and leakage energy consumed in L2, respectively. $E_{\chi}$ shows the energy consumed in a single block transition. Tran shows the number of block transitions. In an interval, $F_A$,  $M_{\text{L2}}$ and $H_{\text{L2}}$ show the active
fraction of cache, L2 misses and L2 hits respectively. $A_{\text{Mem}}$ and $A_{\text{RCE}}$ denote the number of accesses to DRAM and RCE, respectively. The area overhead of a gated V$_{\text{dd}}$ cell as a fraction of area of the normal cell is shown as $\Upsilon$ and the fraction of normal leakage power, which is still consumed at low-leakage is shown as $P_{\text{off}}$.

For computation of L2 leakage energy, we account for the consumption of both active
and turned-off (i.e. low-leakage) fraction of the cache. Also, we assume that the increase in cell area due to the use of gated V$_{\text{dd}}$ leads to an increase in leakage energy in the same proportion. Further, we assume that an L2 miss consumes
twice the dynamic energy as that of an L2 hit \cite{HanHri02_TVLSI,MitZha12_EnCache}.  Thus, we get
\begin{align}
 \label{eq:manager_totalenergy}E&= E_{\text{L2}}+E_{\text{Mem}}+E_{\text{Algo}} \\
 E_{\text{L2}} &= LE_{\text{L2}} + DE_{\text{L2}}  \\   
 LE_{\text{L2}} &= P^{\text{Leak}}_{\text{L2}} \times (1+\Upsilon)\times (F_{A}
+ (1-F_{A}) P_{\text{off}}) \times T \\
DE_{\text{L2}} &= E^{\text{Dyn}}_{\text{L2}}\times(2
M_{\text{L2}}+H_{\text{L2}}) \\
E_{\text{Mem}} &= P^{\text{Leak}}_{\text{Mem}}\times T +
E^{\text{Dyn}}_{\text{Mem}}\times A_{\text{Mem}} \\
E_{\text{Algo}} &= E_{\chi}\times \text{Tran}+E^{\text{Dyn}}_{\text{RCE}}\times
A_{\text{RCE}} +P^{\text{Leak}}_{\text{RCE}}\times T 
\end{align}

For baseline experiments, $E_{\text{Algo}}=0$, $\Upsilon$ = 0,
$F_A$ = 1 and $P_{\text{off}}$ value is not required. For MANAGER, $P_{\text{off}}$ = 0.03 and $\Upsilon$ =  0.05  \cite{PowSeh00_GatedVdd}. Using CACTI 6.5 \cite{cacti_65}, for 4MB, 8-way L2 at 32nm, we obtain  $P^{\text{Leak}}_{\text{L2}}$ = 1.39 Watt and $E^{\text{Dyn}}_{\text{L2}}$ = 0.289 nJ/access. The  RCE energy values are computed using CACTI 6.5 \cite{cacti_65} and Eq.
\ref{eq:setprofiling}. We assume 8B block size and only account for energy consumption of tags, since RCE is a tag-only (data-less) component.  For a 2-core system and $R_S$=64, an RCE corresponding to 4MB L2 has  $P^{\text{Leak}}_{\text{RCE}}$ = 0.006 Watt and $E^{\text{Dyn}}_{\text{RCE}}$= 0.005 nJ/access.  Clearly, the energy consumption of RCE is a very small fraction of L2 energy consumption.

 We assume that the DRAM uses aggressive power saving mode as allowed in DDR3 DRAM, and hence $ P^{\text{leak}}_{\text{Mem}}$ = 0.18 Watt, when there is no memory access \cite{ZheLin09_DIMM,MitZha12_EnCache}. Also, $ E^{\text{Dyn}}_{\text{Mem}}$ = 70 nJ \cite{ZheLin09_DIMM,MitZha12_EnCache} and
$E_{\chi}$ = 2 pJ \cite{MitZha13_Cashier}. The energy consumption of counters is negligible compared to that of memory subsystem and hence, is ignored.

\section{Results}\label{sec:manager_results}
\subsection{Main Results}
\begin{figure*}[ht]
 \centering
   \includegraphics [scale=0.50] {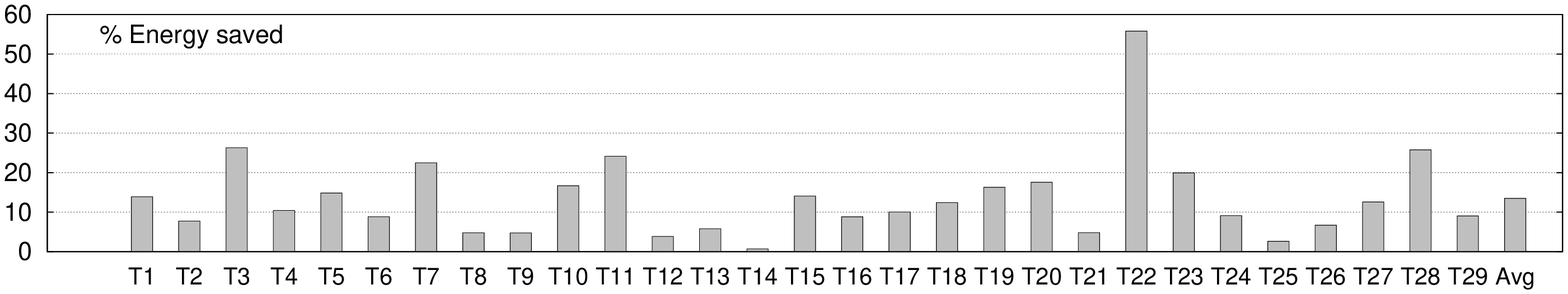}
   \includegraphics [scale=0.50] {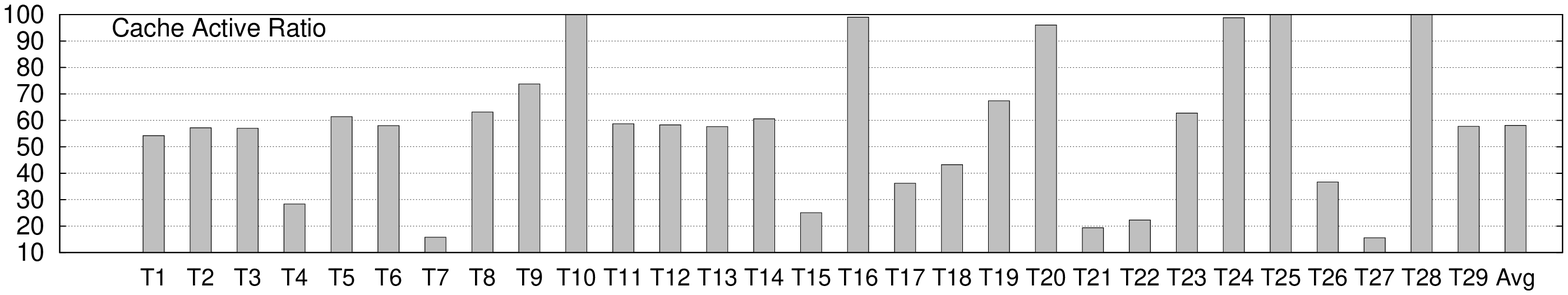}
   \includegraphics [scale=0.50] {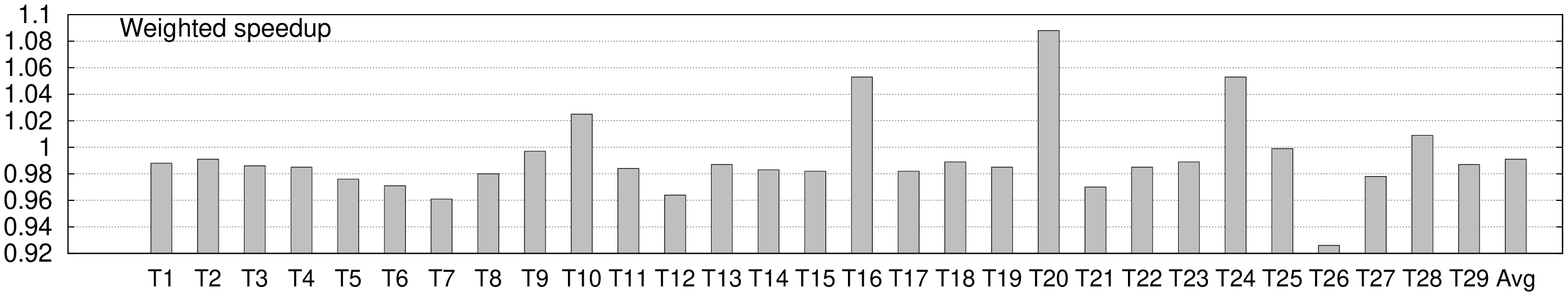}   
 \isucaption{Results on percentage energy saved, active ratio and weighted speedup}
\label{fig:ensaved2astern}
 \end{figure*}

Figure \ref{fig:ensaved2astern} shows the results on percentage energy saving, active ratio and weighted speedup. For brevity, we omit the per-workload values for remaining metrics and only state the average. The average fair speedup is 0.99 and average increase in DRAM APKI is -0.35. Only one workload, viz. T7  misses the QoS deadline.

We now analyze the results. First, MANAGER achieves large energy savings, while keeping the performance almost same as baseline as shown through the value of weighted speedup which is close to one. Further, the average fair speedup is close to one, which indicates that  MANAGER does not cause unfairness or thread-starvation. Only one workload misses its QoS target, which shows that MANAGER ensures meeting most QoS deadlines.  Also, on average DRAM APKI is \textit{reduced} and thus, despite turning off the cache, DRAM accesses do not increase. This is because by partitioning the cache according to demands of different programs, MANAGER contains thrashing programs (e.g. libquantum ) and increases the quota of cache intensive programs (e.g. soplex, omnetpp).

\subsection{Parameter Sensitivity Study}
We now study the sensitivity of MANAGER for different parameters. In each case, we only change a single parameter from the default configuration and summarize the results in Table \ref{tab:manager_resultsparameter}. The value of FS are omitted for brevity, since it is nearly same as that of WS.
 
\begin{table}[ht] 
\isucaption{MANAGER results for different parameters. \textmd{Default parameters: $\Omega$=5\% and $K$= 10M. Results with default parameters are also shown for comparison.}}
\label{tab:manager_resultsparameter}
\centering
\begin{tabular}{|c|c|c|c|c|c|c|}
\hline
        & Energy &WS  & $\delta$APKI & Active & Missed \\
        & Saving &    &              & Ratio & QoS \\\hline
Default & 13.5\% & 0.99 & -0.35 & 58.1\% & T7 \\\hline
$\Omega$=3\% & 13.1\% & 0.99 & -0.37 & 60.3\% & T7,T14,T26 \\\hline
$\Omega$=7\% & 13.4\%  & 0.99  & -0.34 & 57.9\% & none \\\hline
$K$= 5M & 13.3\% & 0.98 & -0.29 & 53.3\% &  T12\\\hline
$K$= 15M & 12.4\% & 0.99 & -0.40 & 64.0\% & none \\\hline
\end{tabular}	
\end{table}
 
\textbf{Change in $\Omega$:} On changing $\Omega$ to 3\%, the active ratio is increased and thus, energy saving is slightly reduced. Due to more strict deadline, three workloads miss their QoS. On changing $\Omega$ to 7\%, the cache active ratio and energy saving remain almost same as that for $\Omega$ = 5\%. Further, due to more relaxed deadline, no workload misses its deadline. For both cases, WS remains close to 1 and DRAM APKI is reduced. 
   
\textbf{Change in $K$:} Changing $K$ to 5M increases the aggressiveness of cache turnoff, as shown in reduced value of active ratio, but it also increases DRAM APKI, and due to their interaction, the energy saving is slightly reduced. Only T12 misses the QoS deadline. On increasing $K$ to 15M, less cache is turned off, which leads to reduced energy saving. No workload misses its QoS. 
     
The results presented in this section show that MANAGER works well for wide range of parameters and achieves a right balance between energy saving and performance loss.    
     

%
%
%
%
%
%
%
%

\section{Conclusion}\label{sec:manager_conclusion}
In this chapter, we presented MANAGER, which uses dynamic profiling with cache reconfiguration to save energy in multicore LLCs. MANAGER uses software control with lightweight hardware support. The simulation results have confirmed that MANAGER is a useful technique for saving energy in memory subsystem, and does not harm performance or cause unfairness, while also meeting QoS of most programs. Our future work will focus on synergistically integrating MANAGER with DVFS (dynamic voltage/frequency scaling) techniques to save even larger amount of energy and provide better quality-of-service to programs.



\chapter{FULL-SYSTEM SIMULATION ACCELERATION USING SAMPLING TECHNIQUE} \label{chap:simulation}

\section{Overview of Our Approach}
Simulation plays a vital role in the study and analysis of proposed architecture designs \cite{khaitan2008multifrontal,mittal_surveyOpnet,khaitan2009fast}. Recently, efforts have been directed towards the development of simulators and the techniques for accelerating simulations, such as benchmark truncation, processor warm-up, simulation sampling etc. However, these development efforts have remained isolated and hence, these approaches lack one or the other desired feature. For example, full-system simulators (e.g. \cite{gems05}) allow detailed modeling and higher accuracy compared to processor-only simulators. However, their extremely slow speed severely restricts their utility. This forces the designers to use simulators which do not model the details of hardware with sufficient closeness and thus lead to a large modeling error or to use truncated and hence inaccurate benchmarks. To address these issues, several simulation acceleration techniques have been proposed; but they have been generally implemented in uni-core simulators only. Thus, efforts such as using simulation acceleration technique to benefit full-system simulators etc. can bring together the best of both and make an efficient simulator available to the architecture community.

As a step towards addressing this need, we present our work on integrating SMARTS (Sampling Microarchitecture Simulation) simulation acceleration technique into the GEMS (General Execution-driven Multiprocessor Simulator) simulator modules (Ruby and Opal). Figure \ref{7a} shows the flow-diagram of our approach. Our approach leads to a fast, full-system simulation platform with detailed memory system and detailed processor simulator. We discuss the challenges faced in integration and the design choices made to address them. Further, we make recommendations for improving specific components, which can further speedup this simulation platform. The experiments performed with SPEC2K benchmarks show that using the sampling approach results in a significant increase in simulation speed of the detailed processor simulator, with very small error in estimating CPI (Cycle-per-instruction). Specifically, across our workload the geometric mean of the speed-up obtained over detailed full-system simulation is 28$\times$. Further, the average error (arithmetic mean) in estimating CPI is only 0.73\% with the minimum being as low as 0.1\%. This shows the effectiveness of our approach.

An architecture simulator must have high simulation speed, to be able to execute a large-enough execution length of any application  without requiring months of simulation-hours. However, many of the existing simulators provide much smaller simulation speed than desired.   As an example, the average speed of out-of-order module of GEMS simulator (called Opal) is 69 KIPS compared to 740 KIPS speed of \emph{sim-outorder} \cite{parallelSim}. Further, the slowdown factor of Opal has been reported  as nearly 140,000 compared to real-hardware \cite{gemstutorial}, although the exact slow-down varies depending on the protocol choice, configuration and workload etc. Thus, a time of 10 seconds in real-hardware would require 1,400,000 seconds or more than 16 days of simulation when run using out-of-order module of GEMS. This motivates us to use the approach of sampling for accelerating GEMS modules to make their use feasible in simulation studies.

Currently the SMARTS technique is implemented in \emph{sim-outorder} simulator from simplescalar, which is not a full-system simulator. Thus, the potential of accelerating cycle-accurate full-system simulators using SMARTS technique has not been utilized. Full-system simulators are known to be more accurate for OS-intensive
workloads than those simulators that omit the OS \cite{Cain02}.
The simulations performed for cache design usually require monitoring of much larger number of instructions than those performed for pipeline or branch predictor. Hence the current out-of-order simulators with their slow execution rate are quite inefficient in facilitating experimentation with various design alternatives, especially when the number of design choices is large. Thus, we believe that our approach of integrating simulation acceleration into a full-system simulator would be quite useful for the computer architecture research community for cache design research.
Simple-scalar simulates the DEC Alpha ISA (instruction-set-architecture), while Simics can simulate a variety of instructions sets. However, Opal in particular is tied to the SPARC ISA. Thus, our work implements the SMARTS sampling technique on another ISA.

\section{Related Work}
 Simulation holds a vital role in the field of computing systems \cite{mitpande11_simplex,mittalref13,gupta_eureqa,GupMit08_MIMO}. In recent years, several simulation methods and acceleration techniques have been proposed \cite{khaitan2008optimization,summary06,mittalref14,khaitan2010class}. 

Simple-scalar is a uniprocessor microarchitectural simulator suite which provides different simulators such as functional, detailed; execution and trace-driven simulators etc \cite{simpleScalar}. Simics \cite{simics02} is a full system simulation platform, capable of running operating systems and commercial workloads.  It is an efficient,  system-level instruction set simulator and supports several targets and host architectures. It provides the facility of configuration checkpointing through which the entire state of simulation can be stored on the disk, ported or loaded any time. Wisconsin Multifacet General Execution-driven Multiprocessor Simulator (GEMS) \cite{gems05} is a set of timing simulator modules which run over Simics.  Zesto is a detailed-timing simulator which models x86 microarchitecture \cite{LohSub_Zesto}. Zesto models many x86-specific features, which are not implemented in other state-of-art simulators. However, the increased modeling-accuracy comes at the cost of reduced simulation speed and thus its simulation speed is in tens of KIPS (kilo-instructions-per-second) \cite{LohSub_Zesto}.

Compared to the real-hardware, architectural simulators show a large slow-down in simulation speed. This obstructs the architects from doing a complete evaluation of their proposed architectural modification. To address this issue, several techniques have been proposed to reduce the simulation time.  Kleinosowski and Lilja \cite{KleLil02_BenchmarkReduction}  propose benchmark reduction technique, which aims at creating benchmarks of reduced length, that tries to mimic the characteristics of the original benchmark. The reduced inputs enable the architects to run simulations in reasonable amount of time. Despite this, benchmark reduction technique has several limitations.  It works only for few programs and produces large errors for other benchmarks. Moreover, the reduction process is very time consuming, since such reduction needs to be applied for every new benchmark individually. Thus it is not scalable.

Sherwood et al. propose SimPoint technique for selecting representative subsets of benchmark traces by offline analysis of basic blocks \cite{simpoint02} . This technique has also been extended to other platforms \cite{PatCoh04_MPKI}. SimPoint technique works on the assumption that dynamic instances of basic block sequences with similar
profiles have the same behavior. Thus, by measuring a particular sequence only once and weighting it appropriately to represent all remaining instances, SimPoint captures the characteristic of entire execution stream of the benchmark program. 


Wunderlich et al. \cite{smarts}  discuss a Sampling Microarchitecture Simulation (SMARTS) methodology for simulation acceleration and implement it in \emph{sim-outorder}.  Wenisch, Wunderlich, Falsafi and Hoe \cite{live} replace the functional warming approach used in \cite{smarts} with checkpointed warming (using live-points). This modification improves the speed at the cost of limiting the re-usability since it imposes limits on some aspects of microarchitecture parameters, such as maximum size or associativity of cache and hence is inappropriate for the applications requiring flexibility in design choice. An additional space overhead for storing the live-point library and the time overhead of gzip-compression further restrict the utility of this approach. This overhead increases with increasing cache sizes and more realistic simulators. Barr et al. discuss their approach of accelerating multi-processor simulation using memory timestamp record (MTR) and evaluate it using Bochs full-system simulator \cite{barr2005accelerating}. Wenisch, Wunderlich, Ferdman et al. \cite{simflex06} implement sampling technique for full-system timing-accurate simulation of uni-processor and multiprocessor systems, using their simulators which hook into Simics. 
\section{Review of SMARTS Sampling Acceleration Technique} \label{sec:method}
The basis of the sampling methodology is developed in \cite{smarts}. We briefly review it here.  SMARTS uses systematic sampling approach where sampling units are selected from an ordered population at a fixed sampling interval $k$, such that $n=N/k$, where $N$ = size of population, $n$ = Number of samples and $k$ = sampling interval.

By using the coefficient of variation, the optimal sampling interval ($k$) is selected which captures a benchmark's variation and promises a certain confidence level in the estimates. By measuring only certain chosen sections (a.k.a. sampling units) out of the full benchmark stream, the simulation sampling approach estimates the cumulative property of a population, with quantifiable accuracy and confidence level for the error on estimates. Before the actual measurement, a short window (of size $W$ instructions) of detailed warming is introduced to remove the effect of bias due to stale microarchitectural states. For rest of the instructions, fast-forwarding using the functional warming approach (maintaining large microarchitectural state such as branch predictors and cache hierarchy) is found to be superior and cost-effective compared to functional simulation.

For SPEC2K benchmarks, the authors found that a sampling unit size(U) of 1000 instructions provides sufficient accuracy. Thus simulation rate of SMARTS is insensitive to the speed of detailed simulation, but mainly depends on the speed of functional warming.

Since the validity of SMARTS sampling approach is well-established, and GEMS is also widely used , in this work we do not focus on  testing or establishing their validity. Rather, we focus more on integrating detailed simulator and simulation acceleration technique (i.e. simulation component) to bring the best of two together.

\section{Design Methodology and Proposed Speed Optimizations} \label{subsec:optimization}
A key requirement of SMARTS is the availability of functional warming mode (also known as fast-forwarding mode) that updates cache state while fast-forwarding the program.

Thus Simics with GEMS module should provide these simulation modes. Using the terminology in \cite{smarts} we observed this correspondence:
\begin{enumerate}
\item Simics only: functional simulation
\item Simics with Ruby: functional warming (fast-forwarding with cache update)
\item Simics + Opal + Ruby: detailed simulation.
\end{enumerate}

\begin{figure*}[htpb]
\centering
\includegraphics [scale=0.4] {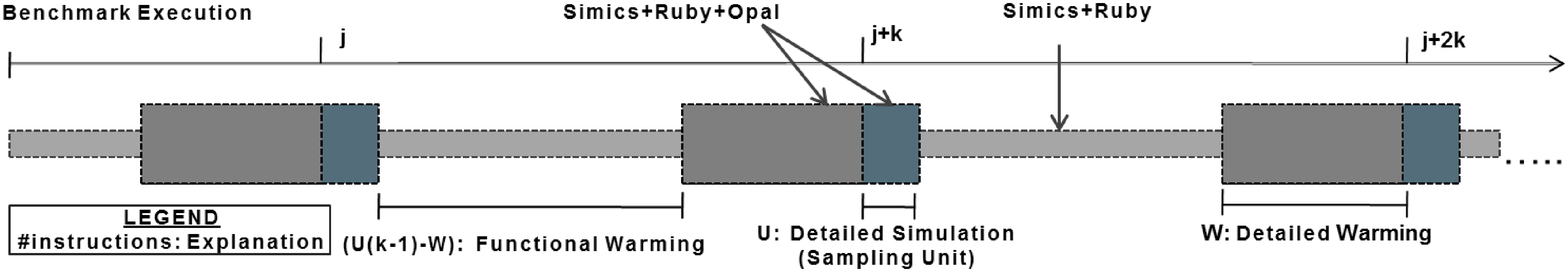}
\isucaption{Simulation Acceleration Approach}\label{7a}
\end{figure*}

Based on this our overall approach is shown in Figure \ref{7a}.

The speed-up of SMARTS comes from running a large fraction of the instructions in functional mode and running only a fraction of instructions in detailed mode. For simplescalar, the authors in \cite{smarts} observed the values of $S_F=1$, $S_{FW}=0.55$, $S_D=1/60=0.0167$.  Here $1/S_D$ and $1/S_{FW}$ show relative slow-downs of detailed simulation and functional warming simulation respectively over functional simulation. The simulation rate of SMARTS with functional warming stays close to $S_{FW}$, i.e. the rate of functional-warming itself. For GEMS, we found the values to be $S_F=1$, , $S_D= 1/440=0.0023$ and $S_{FW}=1/90=0.011$.  It is clear that simulation with Ruby is relatively much slower than its counterparts in simplescalar. It seems that GEMS module will not benefit too much from SMARTS approach.

To address this issue, we make use of data and instruction STCs (Simulation Translation Caches) in Simics \cite{simics02} during the functional-warming phase. The STC is a pure software cache which targets virtual address to host address translations. STC stores the information about ``harmless'' memory addresses, which means the addresses where an access would not cause any device state change or side-effect. Thus, a particular memory address is mapped by the STC only if the given  logical-to-physical mapping is valid; the access would not affect the MMU (TLB) state and there are no breakpoints, callbacks, etc. associated with that address. STC is a pure optimization technique and does not affect the simulation, by more than one memory access per million, less so in UltraSPARC architecture. The use of STCs greatly accelerates the simulation of Ruby module and makes application of SMARTS simulation with GEMS possible. 

Moreover, based on experiments, we also found that not enabling magic-breakpoint leads to some improvement in the speed of execution of Ruby module. The magic-instruction is special NOOP (no-operation) instruction, which has been selected for every simulated processor architecture. When the simulator executes such a magic instruction, it triggers a hap and calls all the callbacks functions registered on this hap. Simics uses \emph{magic-break-enable} command, which changes the way Simics handles the execution of the magic-breakpoint instruction. On enabling the magic-breakpoint in simics, the magic instruction is treated as a Simics breakpoint. On disabling the magic-breakpoint in simics, it will simply generate a hap. If nothing is listening to the hap then nothing will happen and execution will continue as if the instruction were a NOOP. Thus, our simulations do not make use of magic-instructions and hence we disable magic-breakpoints to gain simulation speed.

Thus, the use of the above mentioned optimizations (namely use of STC and disabling magic-breakpoint) is referred to as ``optimized'' case (abbreviated as $Op$), and we observed its simulation rate as , $S_{FW}^{Op}=1/10=0.1$. The former case is referred to as ``un-optimized'' (abbreviated as $NOp$) case and, as shown above, its simulation rate is $S_{FW}^{NOp}=1/90=0.011$. It is clear, that the speed-optimizations proposed above open the opportunity of achieving simulation acceleration in GEMS.

\section{Addressing Challenges Faced in Implementing Simulation Acceleration}
 A few issues need to be addressed for enabling implementation of SMARTS in GEMS.  Simics does not allow unloading of Opal module; thus currently a switching between sampling and non-sampling phase is only possible by using checkpoint method. This method involves storing checkpoint (including cache state) at the end of measurement phase, exiting Simics, removing the references to Ruby and Opal (manually or through a text-processing program) and then reloading execution by using the checkpoint. Since checkpoints are stored in an incremental order and are dependent on the previous checkpoints, recording hundreds or thousands of checkpoints this way slows down the execution considerably since Simics would have to keep going through the checkpoints in reverse order for collecting total information. A typical application may show hundreds or thousands of phase changes between sampling and non-sampling phase and in such a case, the last checkpoint would depend on all the previous checkpoints to properly generate the architectural state of the execution. Thus, this approach has high time and space overhead. An alternative for this could be to store all the information in a single checkpoint, thus making them independent of the previous checkpoints. This approach, however, incurs a much larger space overhead (since each checkpoint becomes very large) and is thus, infeasible for real-life applications.

 We alleviate this overhead by implementing \textbf{\emph{suspend}} and \textbf{\emph{reconnect}} commands for enabling switching between different phases. At the first instance of switching to sampling phase, Opal module is loaded into Simics by using \emph{load-module}. Consequently in each interval, at the end of sampling phase, Opal module issues \textbf{\emph{suspend}} command; this disconnects its Ruby interface and then Ruby is attached as timing model interface for Simics for beginning functional warming phase and Simics acts as the processor. At the beginning of sampling phase, Opal module issues \textbf{\emph{reconnect}} which automatically disconnects Ruby from timing interface of Simics and connects it to Opal. In this phase, Opal acts as the processor and uses Simics to verify its functional correctness. Thus, compared to the naive method of switching using checkpoints, our implementation provides a space and time efficient method of switching.

The switching method, inherent in sampling approach poses some unique challenges, which are absent in either of the functional-alone or detailed-alone simulation.  It may be possible that at the time of switching from sampling-phase to non-sampling phase, the Opal processor may have a cache demand outstanding. Using a sharp boundary of 1000 instructions for sampling measurement phase may lead to forcing the miss request to be flushed; this is likely to lead to inaccuracy and error. To address this issue, we modified the sampling scheme to allow more than 1000 instructions for measurement phase. The switch to non-sampling phase is made only when the outstanding requests are completed. We have found that, in practice, this leads to a negligible increase in the average number of instructions beyond 1000. Moreover, since we are actually measuring CPI (cycles per instruction), the variation is averaged out.

\section{Experimental Results}
\begin{figure*}
  \begin{center}

    \subfigure[CPI values: From Detailed, Sampling with Optimized case and Sampling with Un-optimized case respectively]{\label{fig:CPIValue} \includegraphics[scale=0.3]{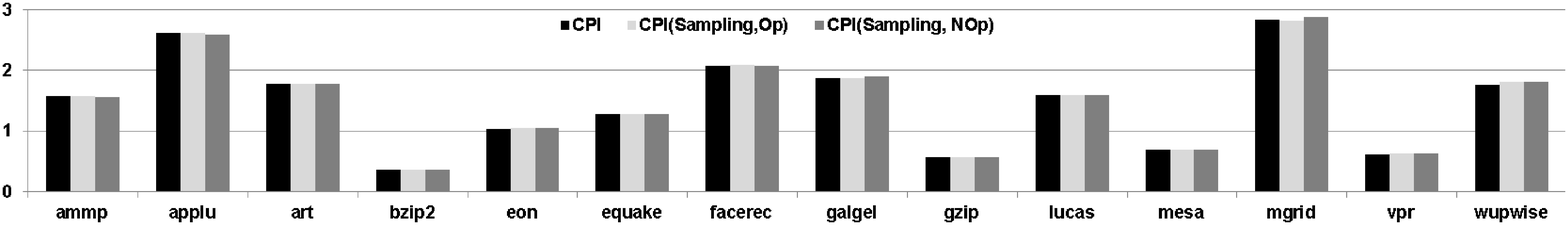}}
    \subfigure[Magnitude of Percent Relative Error, compared to CPI from Detailed Simulation]{\label{fig:CPIError} \includegraphics[scale=0.3]{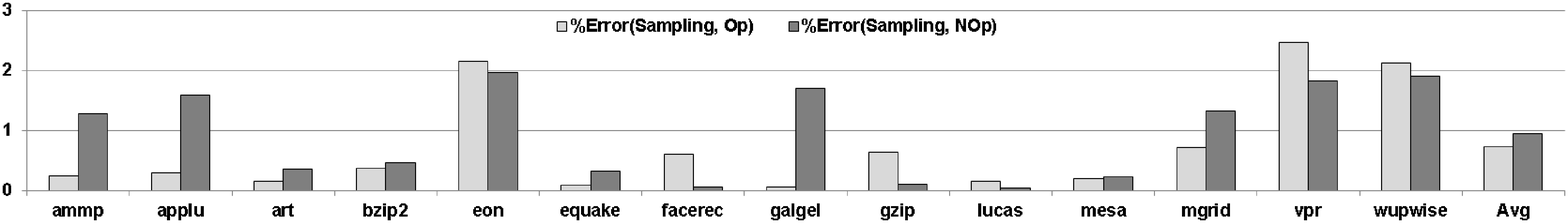}}
  \end{center}
  \isucaption{Simulation Acceleration Experimental Results: CPI Values and Errors in CPI Estimation  }
  \label{fig:simulation_result}
\end{figure*}

As for SMARTS methodology parameters, we conduct experiments for confidence level of 99.7\% and confidence interval of $\pm$3\%. We take initial value of $n_{init}$ as 1000 for all simulations. This is due to the fact that the number of instructions in `test' inputs is much smaller compared to `ref' inputs. Since $n_{init}$ is a compromise between simulation rate and likelihood of meeting confidence requirement, hence, if $n_{init}$ is found to be insufficient for some benchmarks, a second simulation is run using $n_{tuned}$ calculated from $\hat{V_x}$ of the initial run.

\begin{table}[htbp]
  \centering
  \isucaption{Simulation times (in minutes) and Speedups.}
\begin{tabular}{|r|r|r|r|r|r|}\hline
&\multicolumn{3}{|c|}{Simulation Times}&\multicolumn{2}{|c|}{Speedups}\\
\cline{2-6}
Program&$T_D$	&$T^{Op}_S$	&$T^{NOp}_S$ &$T_D/T^{Op}_S$		&$T_D/T^{NOp}_S$\\\hline
ammp	&2591.0	&92.6	&596.1&28.0		&4.3\\\hline
applu	&124.8	&4.4	&27.3&28.4		&4.6\\\hline
art	&311.9	&15.5	&85.1&20.1		&3.7\\\hline
bzip2	&3253.0	&28.0	&745.9&116.2		&4.4\\\hline
eon & 44.4     & 16.4 &40.0&2.7   &1.1\\\hline
equake	&221.5	&22.0	&95.8&10.1		&2.3\\\hline
facerec	&1524.1	&33.95	&286.2&44.9		&5.3\\\hline
galgel	&1748.0	&17.1	&317.2&102.2		&5.5\\\hline
gzip	&771.0	&38.56	&238.4&20.0		&3.2\\\hline
lucas	&1849.7	&20.3	&385.8&91.1		&4.8\\\hline
mesa	&853.8	&66.6	&284.2&12.8		&3.0\\\hline
mgrid	&6829.0	&260.3	&1515.9&26.2		&4.5\\\hline
vpr	&323.7	&6.6	&59.7&49.0		&5.4\\\hline
wupwise	&4156.5	&149.9	&876.6&27.7		&4.7\\\hline

\end{tabular}

\label{tab:simulation_result}%
\end{table}%

\section{Results}\label{sec:simulation_results}
Figure~\ref{fig:CPIValue} shows the CPI values obtained from detailed simulation, sampling simulation with optimizations (as shown in Section~\ref{subsec:optimization}) and sampling simulation without these optimizations.  Here $T_D$ shows detailed simulation time. $T^{Op}_S$ and $T^{NOp}_S$ show the time of sampling simulation with and without proposed speed optimizations respectively.
As shown in Figure~\ref{fig:simulation_result}, the differences found in the CPI with this approach and that with ``optimized'' case (STCs enabled/magic-break disabled) are negligibly small and only affect the CPI value in the second or third place after decimal. However, the large speed-ups obtained in optimized case over un-optimized case far outweigh the loss in accuracy and justify the use of optimized case.

To mathematically quantify the magnitude of error in estimating CPI, we use magnitude of percentage relative error (MPRE), which is defined as 
\begin{equation}
MPRE_{Method}=\left|\frac{CPI-CPI_S^{Method}}{CPI}\times 100\right|
 \end{equation}
Here $CPI_S^{Method}$ refers to cycle-per-instruction value obtained using sampling simulation, using either $Method$ of 1. sampling simulation with optimization or 2. sampling simulation without optimization. Also, $CPI$ refers to the cycle-per-instruction value obtained from detailed simulation. Figure~\ref{fig:CPIError} shows the MPRE values for our workload. It is clear that average value of MPRE is found as 0.73\%, with the minimum being as low as 0.1\%. 

Table~\ref{tab:simulation_result} summarizes the results on simulation times using different techniques. Across our workload, the geometric mean of the speed-up is found as nearly 28$\times$ and arithmetic mean of speed-up is 41$\times$. The speed of detailed simulation with GEMS is 69KIPS \cite{parallelSim} and by providing a speed-up of nearly 28$\times$, our work enables effective simulating speed of nearly 1.93 MIPS using a detailed, full-system simulator. The difference in the speed-ups for different benchmarks can be attributed to their different coefficients of variation, which affects their sampling interval and hence the relative number of instructions simulated in functional warming mode and detailed mode.

Thus the closeness of estimated CPI values with actual CPI values, along with the acceleration achieved, prove the validity of our approach of integration.
\section{Conclusion}

The contributions of and advantages from our work are as follows:
\begin{enumerate}
\item We suggest several speed optimizations for functional warming in GEMS and also implement \textbf{\emph{suspend}} and \textbf{\emph{reconnect}} commands to alleviate the need of taking Simics checkpoints. These extensions greatly reduce simulation time with SMARTS sampling approach.
\item This work enables the advantages of SMARTS sampling method to be used with detailed full system simulators. Moreover, through simulation acceleration, the benefits of GEMS+Simics can be further utilized.

\item We discuss the issues faced in implementing SMARTS on GEMS and by addressing them, we validate the SMARTS sampling methodology for GEMS simulator. 

\end{enumerate}

The integration effort has additionally helped us in getting many insights and finding issues related to portability. Based on our experiences with the implementation, we make an important recommendation for speed enhancement of this simulation framework. Since the simulation speed with SMARTS approach mainly depends on the speed of functional warming \cite{smarts}, any improvement in the speed of Ruby will especially accelerate this simulation framework.  This, in turn, also requires speeding up Simics simulator, because currently, GEMS spends most of its time switching between Simics and Ruby. Because it is the Simics/Ruby switch, that is the dominant source of overhead, thus by Amdahl's law, speeding up Ruby alone will be insufficient. GEMS modules currently work only with Simics 3.0 version and transporting GEMS modules to the faster version of Simics would enhance their speed. Supporting module-unloading in these versions will also help in easy switching between different phases.
In summary, efforts need to be directed to accelerate both Simics simulators and GEMS module (Ruby). This suggestion is also significant for the design of any future memory system or processor simulator.

\chapter{CONCLUSION AND FUTURE WORK} \label{chap:conclusionfuturework} 
In this research, we have made important contributions to the development of algorithms and architectures for improving energy efficiency of caches in high-performance processors. We have proposed specific techniques for single-core and multi-core systems, single-tasking and multi-tasking systems, real-time and QoS systems. The following table summarizes the characteristics of different techniques.

\begin{table*}[htbp]  \small
  \centering
  \isucaption{A Comparison and Overview of Different Cache Energy Saving Techniques Proposed In This Thesis}
    \begin{tabular}{|c|c|c|c|}
    \hline
   
    & \multirow{2}{*}{Single-core/Multi-core} & Real-time and  & \multirow{2}{*}{Cache Allocation Mechanism} \\
    & &  QoS System& \\\hline
EnCache & Single, multicore(shared cache) & No & Selective sets and selective-ways \\\hline 
Palette & Single, multicore(shared cache) & No & Cache Coloring \\\hline
CASHIER & Single, multicore(shared cache) & Yes & Cache Coloring\\\hline
MASTER & Multicore (uses cache partitioning) & No & Cache Coloring\\\hline
MANAGER & Multicore (uses cache partitioning)   & Yes & Cache Coloring\\\hline    
    \end{tabular}
    \label{tab:overallsummary}
  
\end{table*}

From Table \ref{tab:overallsummary}, it is clear that EnCache, Palette and CASHIER do not use cache partitioning and hence, they are suitable for single-core systems or multi-core systems with shared caches. MASTER and MANAGER use cache partitioning. CASHIER and MANAGER are suitable for QoS and real-time systems, while other techniques are suitable for the systems with no deadline. Apart from EnCache, all other techniques use cache coloring method.

The proposed techniques use dynamic profiling and dynamic cache reconfiguration and do not require offline profiling or tuning of their parameters. Due to this feature, these techniques can be easily scaled to processors with large number of cores. These techniques directly optimize for energy and hence, are capable for optimizing for system or subsystem energy. Especially in the context of multicore systems, hardly few techniques exist which enable the designers to save cache leakage energy. Thus, our techniques are extremely useful for multicore systems.

Use of our techniques provides energy savings which also gives headroom for performance scaling, since within the same power budget extra computations can be run. Also, saving of energy leads to reduced cooling cost and chip temperature.

 Extensive simulation results have confirmed that our techniques are effective in saving energy in memory subsystem. We have also evaluated our techniques for different simulation parameters and have found that they are robust towards change in their parameters. We have also evaluted their overheads, both using CACTI and through numerical evaluation and have found that their overheads are small. Further, they do not harm performance and provide higher energy saving than conventional energy saving techniques. The techniques proposed for multicore systems do not cause unfairness and the techniques proposed for QoS systems meet the QoS requirement of most programs. These features make the techniques extremely useful for the product systems.

We believe that the insights gained from our techniques will be highly useful for researchers to designing ``green'' processors of tomorrow. Our research also leads several open dimensions for future researchers to develop them into working ideas. In what follows, we list some possible future works.

\begin{enumerate}
\item The techniques proposed here can be implemented and evaluted on real-processors.
\item The energy saving techniques such as MASTER, can be extended to processors with tens of cores.
\item The energy saving techniques can be synergistically integrated with other methods of saving energy, such as DVFS etc. Also, the leakage energy saving techniques proposed in this thesis can be combined with the techniques for saving dynamic energy to further increase the energy savings achieved.
\item These techniques can be extended to other techniques which aim to improve performance and reduce cache miss-rate so that the increase in miss-rate caused by our technique can be offset and in the case where energy saving opportunities are not present, the algorithm can aim to maximize performance.   
\item The dynamic cache reconfiguration idea presented here can be extended to the GPUs (graphics processing units) also. Since GPUs use much smaller sized caches than the CPUs, the contribution of caches in the overall power consumption of GPUs is small. However, given that the overall power consumption of GPUs is still large (e.g. the high-end GPUs consume up to 300 Watts of peak power), any energy saving achieved in GPU caches can be highly useful in improving their energy efficiency. In GPUs, access to global memory causes stall and during this time, caches can be transitioned to state-preserving low-leakage mode for saving energy.  
\end{enumerate}



\begingroup
\renewcommand \bibname{PUBLICATION AND HONORS}

\unappendixtitle
\phantomsection
\addcontentsline{toc}{chapter}{PUBLICATION AND HONORS}
\textbf{Honors: } 
\begin{enumerate}
 \item ECpE Fellowship of \$2500.
 \item Peer Research Award of \$200 from ISU. 
 \end{enumerate}
\endgroup

\renewcommand{\bibname}{\centerline{BIBLIOGRAPHY}}
\unappendixtitle
\newpage
\phantomsection
\addcontentsline{toc}{chapter}{BIBLIOGRAPHY}
\bibliographystyle{IEEEtran}
\bibliography{PhDReferences}

\end{document}